\magnification1095
\input amstex
\hsize32truecc
\vsize44truecc
\input amssym.def
\font\ninerm=cmr9 at9truept
\font\twbf=cmbx12
\font\twrm=cmr12

\footline={\hss\ninerm\folio\hss}

\def\section#1{\goodbreak \vskip20pt plus5pt
\noindent {\bf #1}\vglue4pt}
\def\eq#1 {\eqno(\text{\rm2.#1})}
\def\rf#1 {(\text{\rm#1})}
\let\al\aligned
\let\eal\endaligned
\let\ealn\eqalignno
\def\cL{\Cal L}
\def\gH{\frak h}
\def\gG{\frak g}
\def\gM{\frak m}
\def\hi{{\hat\imath}}
\def\hj{{\hat\jmath}}
\let\o\overline
\let\ul\underline
\let\a\alpha
\let\b\beta
\let\d\delta
\let\D\varDelta
\let\e\varepsilon
\let\f\varphi
\let\g\gamma
\let\G\varGamma
\let\k\kappa
\let\la\lambda
\let\La\varLambda
\let\om\omega
\let\O\varOmega
\let\mPi\varPi
\let\P\varPsi
\let\pa\partial
\let\t\widetilde
\let\u\tilde
\def\Sh{\operatorname{sh}}
\def\Ch{\operatorname{ch}}
\def\ctgh{\operatorname{ctgh}}
\def\tgh{\operatorname{tgh}}
\def\arctg{\operatorname{arctg}}
\def\ar{\operatorname{arccos}}
\def\ars{\operatorname{arsh}}
\def\ns{\operatorname{ns}}
\def\sc{\operatorname{sc}}
\def\sh{\operatorname{sh}}
\def\sn{\operatorname{sn}}
\def\cn{\operatorname{cn}}
\def\dn{\operatorname{dn}}
\def\sd{\operatorname{sd}}
\def\cd{\operatorname{cd}}
\def\nd{\operatorname{nd}}
\def\tov#1{\widetilde{\overline#1}}
\def\gag{\overset\text{\rm gauge}\to\nabla}
\def\br#1#2{\overset\text{\rm #1}\to{#2}}
\def\ddt{\frac d{dt}}
\def\({\left(}\def\){\right)}
\def\[{\left[}\def\]{\right]}
\def\kl{\left\{}\def\kr{\right\}}
\def\h#1#2{\,{\vphantom F}_#1F_#2}
\def\dr#1{_{\text{\rm#1}}}
\def\ur#1{^{\text{\rm#1}}}
\def\uP{\ul{\t P}}

\def\GR{General Relativity}
\def\cf{configuration}
\def\ct{constant}
\def\co{cosmological}
\def\dt{dependent}
\def\ef{effective}
\def\ev{evolution}
\def\ex{exponential}
\def\il{inflation}
\def\lg{lagrangian}
\def\gr{gravitational}
\def\pt{perturbation}
\def\q{quint\-essence}
\def\U{Universe}
\def\Mp{M\dr{pl}}
\def\rAB{\text{AB}}
\def\Int{\operatorname{int}\nolimits}
\def\crt{\operatorname{crt}\nolimits}

{\twbf
\advance\baselineskip4pt
\centerline{M. W. Kalinowski}
\centerline{\twrm (Higher Vocational State School in Che\l m, Poland)}
\centerline{Cosmological models}
\centerline{in the Nonsymmetric}
\centerline{Kaluza--Klein Theory}}

\vskip20pt plus5pt

{\bf Abstract.} We consider a dynamics of Higgs' field in the framework
of \co\ models involving a scalar field $\varPsi$ from
Nonsymmetric Kaluza--Klein (Jordan--Thiry) Theory. 
The field $\varPsi$ plays here a r\^ole of a \q\ field. We consider phase
transition in \co\ models of the second and of the first order due to \ev\ of
Higgs' field. We developed \il ary models including calculation of an amount
of \il. We match some \co\ models, calculating a Hubble parameter and an age
of the \U.

\section{1. Introduction}
We develop in the paper \co\ consequences of the Nonsymmetric Kaluza--Klein
(Jordan--Thiry) Theory. The theory has been extensively described in the first
point of Ref.~[1]. We refer to the book on nonsymmetric field theory for all
the details of our theory and we present only \co\ applications of the theory.

Let us give a short description of the theory.

We  develop  a  unification  of  the  Nonsymmetric
Gravitational Theory and gauge fields  (Yang--Mills'  fields)  including
spontaneous symmetry breaking and  the  Higgs'  mechanism  with  scalar
forces  connected  to  the  gravitational  constant.   The  theory  is 
geometric and unifies tensor-scalar gravity with massive  gauge  theory 
using a multidimensional manifold in a Jordan--Thiry manner.
We use a nonsymmetric version of this theory.
The general scheme is the following.  We  introduce
the principal fibre bundle over  the  base $V = E \times  G/G_{0}$  with  the
structural group $H$, where $E$ is a space-time, $G$ is a compact  semisimple
Lie group, $G_{0}$ is its compact subgroup and $H$  is  a  semisimple  compact
group.  The manifold $M = G/G_{0}$ has an interpretation as a ``vacuum states
manifold" if $G$ is broken to $G_{0}$ (classical vacuum states).  We define on
the space-time $E$, the nonsymmetric tensor $g_{\alpha\beta}$  from  N.G.T.,  which  is
equivalent to the existence of two geometrical objects
$$\eqalignno{\noalign{\vskip2pt}
\o{g}&= g_{(\alpha \beta )}\overline{\theta }^\alpha \otimes \overline{\theta}^\beta 
&(1.1)\cr
\noalign{\vskip2pt}
\ul{g}&= g_{[\alpha \beta ]}\overline{\theta }^\alpha \wedge \overline{\theta}^\beta
&(1.2)\cr\noalign{\vskip2pt}}
$$
the symmetric tensor $\o{g}$ and the  2-form $\ul{g}$.  Simultaneously we  introduce
on $E$ two connections from N.G.T. $\o{W}_{\beta \gamma}^\alpha $ and 
$\widetilde{\overline{\varGamma}}_{\beta \gamma}^\alpha $. On the homogeneous space $M$
we define the nonsymmetric metric tensor 
$$
g_{\tilde{a}\tilde{b}} = h^{0}_{\tilde{a}\tilde{b}} 
+ \zeta  k^{0}_{\tilde{a}\tilde{b}} \eqno(1.3)
$$
where $\zeta $ is the dimensionless constant, in  a  geometric  way.  Thus  we
really have the nonsymmetric metric tensor on or $V = E \times 
G/G_{0}$. 
$$
\gamma _{\rAB} =
\left(\vcenter{\offinterlineskip\tabskip0pt
\halign{\strut\tabskip5pt plus10pt minus10pt
\hfill#\hfill&\vrule height10pt#&\hfill#\hfill\tabskip0pt\cr
$\vrule height0pt depth6pt width0pt g_{\alpha \beta }$ && $0$\cr 
\noalign{\hrule}
$\vrule height9pt depth4pt width0pt 0$ &&
$r^2g_{\tilde{a}\tilde{b}}$\cr}}\right) 
\eqno(1.4)
$$
$r$ is a parameter which characterizes the size   of  the  manifold $M =
G/G_{0}$.  Now on the principal bundle $\ul{P}$ we define the connection $\omega $,  which
is the 1-form with values in the Lie algebra of $H$.

After  this  we   introduce   the   nonsymmetric   metric   on $\ul{P}$
right-invariant with respect to the action of the group $H$,  introducing
scalar field $\rho $ in a Jordan--Thiry manner. 
The only  difference  is  that
now  our  base  space  has   more   dimensions   than   four.   It   is
$(n_{1}+4)$-dimensional, where $n_{1} = \dim (M) = \dim (G)-\dim (G_{0})$.  In  other
words,  we  combine  the  nonsymmetric  tensor $\gamma _{\rAB}$  on $V$   with   the
right-invariant nonsymmetric tensor on the group $H$ using the connection
$\omega $ and the scalar field $\rho $.  We suppose that the factor $\rho $  depends  on  a
space-time point only.  This condition can be abandoned and we consider
a more general case where $\rho =\rho (x,y)$, $x\in E$, $y\in M$ resulting in  a  tower  of
massive scalar field $\rho _{k}, k=1,2\ldots .$  This  is  really  the Jordan--Thiry
theory  in  the  nonsymmetric  version  but   with $(n_{1}+4)$-dimensional
``space-time".  After this we act in the classical manner.
We introduce the linear  connection
which is compatible with this nonsymmetric metric. This  connection  is
the multidimensional analogue of the connection
$\widetilde{\overline{\varGamma}}{}^{\alpha}_{\beta \gamma }$ on the  space-time
$E$.  Simultaneously  we  introduce  the  second  connection   $W$.    The
connection $W$ is the multidimensional analogue of the $\o{W}$-connection  from
N.G.T. and Einstein's Unified Field Theory.  
Now we  calculate  the  Moffat--Ricci
curvature scalar $R(W)$ for the connection $W$ and  we  get  the  following
result. $R(W)$ is equal to the sum of the Moffat--Ricci curvature on  the
space-time $E$ (the  gravitational  lagrangian  in  Moffat's  theory  of
gravitation), plus $(n_{1}+4)$-dimensional lagrangian  for  the  Yang--Mills'
field from the Nonsymmetric Kaluza--Klein Theory plus  the  Moffat--Ricci
curvature scalar on the homogeneous  space $G/G_{0}$  and  the Moffat--Ricci
curvature scalar on the group $H$ plus  the  lagrangian  for  the  scalar
field~$\rho $.  The only difference is that our Yang--Mills' field is  defined
on $(n_{1}+4)$-dimensional  ``space-time"   and   the   existence   of   the
Moffat--Ricci\vadjust{\vskip-.6pt} 
curvature scalar of the connection on the homogeneous space
$G/G_{0}$.\vadjust{\vskip-.6pt}  
All of these terms (including $R(\o{W})$)  are  multiplied  by  some
factors depending on the scalar field~$\rho $.

This lagrangian depends on the point of $V = E \times  G/G_{0}$ i.e.\ on  the
point of the space-time $E$ and on the point 
of $M = G/G_{0}$.  The  curvature
scalar on $G/G_{0}$ also depends on the point of $M$.

We now go to the group structure  of  our  theory.   We  assume $G$
invariance of the connection $\omega $ on the principal fibre bundle $\ul{P}$, the so
called Wang-condition.   According  to  the
Wang-theorem the  connection $\omega $  decomposes  into  the
connection $\widetilde{\omega }\dr{E}$  on  the  principal  bundle $Q$  over  space-time $E$  with
structural group $G$ and the multiplet of scalar fields ${\varPhi} $.  Due  to  this
decomposition the multidimensional  Yang--Mills'  lagrangian  decomposes
into: a 4-dimensional Yang--Mills' lagrangian with  the  gauge  group $G$
from the Nonsymmetric Kaluza--Klein Theory, plus  a  polynomial  of  4th
order with respect to the fields ${\varPhi} $, plus a term which is quadratic with
respect to the gauge derivative of ${\varPhi}$  (the gauge derivative with respect
to the connection $\widetilde{\omega }\dr{E}$ on a space-time $E$) plus a new term which is of 2nd
order in the ${\varPhi} $, and is linear with respect  to  the Yang--Mills'  field
strength.  After this we perform the  dimensional  reduction  procedure
for the Moffat--Ricci scalar curvature on the manifold $\ul{P}$.   We  average
$R(W)$ with respect to the homogeneous space $M = G/G_{0}$.
In this way we get the lagrangian of our theory.  It is the
sum  of  the  Moffat--Ricci  curvature  scalar   on $E$ (gravitational
lagrangian) plus a Yang--Mills' lagrangian with gauge group $G$  from  the
Nonsymmetric Kaluza--Klein Theory plus a  kinetic  term  for
the scalar field ${\varPhi} $, plus a potential $V({\varPhi} )$ which is of  4th  order  with
respect to ${\varPhi} $,  plus ${\Cal L}_{\Int}$  which  describes  a  nonminimal  interaction
between the scalar field ${\varPhi} $ and the Yang--Mills' field, plus cosmological
terms, plus  lagrangian  for  scalar  field $\rho $.   All  of  these  terms
(including $\o{R}(\o{W})$) are multiplied of course by some factors depending on
the scalar field $\rho $.  We redefine tensor $g_{\mu \nu }$ and $\rho$
and pass from scalar field $\rho $ to ${\varPsi} $
$$ 
\rho  = e^{-{\varPsi} }. \eqno(1.5)
$$
After this  we  get  lagrangian  which  is  the  sum  of  gravitational
lagrangian,   Yang--Mills'   lagrangian,   Higgs'   field    lagrangian,
interaction  term ${\Cal L}_{\Int}$  and  lagrangian  for  scalar  field ${\varPsi} $   plus
cosmological terms.  These terms depend now on the scalar field ${\varPsi} $.   In
this way we have in our theory a multiplet of scalar fields $({\varPsi} ,{\varPhi} )$.   As
in  the  Nonsymmetric-Nonabelian   Kaluza--Klein   Theory   we   get   a
polarization  tensor  of  the  Yang--Mills'   field   induced   by   the
skewsymmetric part of the metric on the space-time and on the group $G$.
We get an additional term in the Yang--Mills' lagrangian induced by  the
skewsymmetric part of the metric $g_{\alpha \beta }$.  We get also ${\Cal L}_{\Int}$,
which is absent in the dimensional reduction procedure known up to  now.
Simultaneously,  our  potential  for  the
scalar--Higgs' field really differs from the analogous potential.
Due to the skewsymmetric part of the metric  on $G/G_{0}$
and on $H$ it has a more complicated structure.   This  structure  offers
two kinds of critical points for the minimum of  this  potential: ${\varPhi} ^{0}_{\crt }$
and ${\varPhi} ^{1}_{\crt }$.  The first is known in the classical,  symmetric  dimensional
reduction procedure and  corresponds  to  the  trivial
Higgs' field (``pure gauge").  This is the ``true" vacuum  state  of  the
theory.  
The second, ${\varPhi} ^{1}_{\crt }$, corresponds to a more complex configuration.
This is only a local (no absolute) minimum  of~$V$. It is  a  ``false"
vacuum. The Higgs' field is not a ``pure" gauge here.  In the first case
the unbroken group is always $G_{0}$.  In the second case, it is in  general
different and strongly depends on the details of the theory: groups $G_{0}$, 
$G$, $H$, tensors $\ell _{ab}$, $g_{\tilde{a}\tilde{b}}$ and  the  constants $\zeta$, $\xi $.   It  results  in  a
different spectrum of mass for intermediate bosons.  However, the scale
of the mass is the same and it is fixed by a constant $r$ (``radius"  of
the manifold $M = G/G_{0})$.  In the first case $V({\varPhi} ^{0}_{\crt }) =0$, in  the  second
case it is, in general, not zero $V({\varPhi} ^{1}_{\crt }) \neq 0$.  Thus, in the first case,
the cosmological constant is a sum of the scalar  curvature  on $H$  and
$G/G_{0}$, and in the second case, we should  add  the  value $V({\varPhi} ^{1}_{\crt })$.   We
proved that using the constant $\xi$ we are able in some cases to make  the
cosmological constant as small as we want (it  is  almost  zero,  maybe
exactly zero, from the observational data point of view).  Here we  can
perform the same procedure for the  second  term  in  the  cosmological
constant using the constant $\zeta $.  In the first case we are able  to  make
the cosmological constant sufficiently small but this is  not  possible
in general for the second case.

The transition from ``false" to ``true" vacuum occurs  as  a  second
order phase transition.  We discuss this transition  in
context of the first order phase transition in models of  the  Universe.
The interesting  point  is  that  there
exists an effective scale of masses, which depends on the scalar 
field~${\varPsi} $.

Using Palatini variational principle we get an equation for fields
in our theory.  We find  a  gravitational  equation  from  N.G.T.  with
Yang--Mills', Higgs' and  scalar  sources  (for  scalar  field ${\varPsi} )$  with
cosmological terms.  This gives us  an  interpretation  of  the  scalar
field ${\varPsi} $ as an effective gravitational constant
$$
G\dr{eff} = G_{N}e^{-(n+2){\varPsi} }. \eqno(1.6)
$$
We get an equation for this scalar  field ${\varPsi} $.   Simultaneously  we  get
equations for Yang--Mills' and Higgs' field. We also discuss the  change
of the effective scale of mass, $m\dr{eff}$ with a relation to the  change  of
the gravitational constant~$G\dr{eff}$.

In the ``true" vacuum case we  get  that  the  scalar  field ${\varPsi} $  is
massive  and  has  Yukawa-type  behaviour.   In  this  way   the   weak
equivalence principle is satisfied.  In the  ``false"  vacuum  case  the
situation is more complex.  It  seems  that  there  are  possible  some
scalar forces with infinite range.  Thus  the  two  worlds  constructed
over the ``true" vacuum and the ``false" vacuum  seem  to  be  completely
different: with different unbroken groups, different mass spectrum  for
the broken gauge and Higgs' bosons,  different  cosmological  constants
and with different behaviour for the scalar field ${\varPsi} $.   The  last  point
means that in the ``false" vacuum case the  weak  equivalence  principle
could be violated and the gravitational  constant  (Newton's  constant)
would increase in distance between bodies.

We explore Einstein $\la$-transformation for a connection $W^{\u A}{}_{\u B}$
($(m+4)$-dimensional) in order to find an interpretation of ${\Bbb R}_+$ gauge
invariance of $\t W_\mu$ field.
We discuss ${\Bbb R}_{+}$ and ${\bold U}(1)\dr{F}$ invariance.
We  decide  that
${\bold U}(1)\dr{F}$  invariance  from  G.U.T.  is  a  local  invariance.   Due  to  a
geometrical construction we are  able  to  identify  $\o{W}_{\mu }$  from  Moffat's
theory of gravitation with the four-potential $\widetilde{A}\ur{F}_{\mu }$ corresponding  to  the
${\bold U}(1)\dr{F}$ group (internal rotations connected to fermion charge).  In  this
way, the fermion number is conserved and plays the role of  the  second
gravitational charge.  Due to the Higgs' mechanism
$\widetilde{A}\ur{F}_{\mu }$  is  massive  and
its strength, $\widetilde{H}\ur{F}_{\mu \nu }$, is of short range with Yukawa-type behaviour.   This
has  important  consequences.   The   Lorentz-like   force   term   (or
Coriolis-like force term) in the equation of motion for a test particle
is of short range with Yukawa-type behaviour.  The range of this  force
is smaller than the  range  of  the  weak  interactions.   Thus  it  is
negligible in the equation of motion for a test particle.   We  discuss
the possibility of the cosmological origin of the mass  of  the  scalar
field ${\varPsi} $ and\vadjust{\eject} geodetic equations on $\ul{P}$.  We consider an infinite tower  of
scalar fields ${\varPsi} _{k}(x)$ coming from the expansion of the  field ${\varPsi} (x,y)$  on
the manifold $M=G/G_{0}$ 
into harmonics of  the  Beltrami--Laplace  operator.
Due to Friedrichs' theory we can 
diagonalize  an  infinite  matrix  of
masses for ${\varPsi} _{k}$ transforming them into new  fields
${\varPsi}'_k$.   The  truncation 
procedure means here to take a zero mass mode 
${\varPsi} _{0}$ and equal it to ${\varPsi} $.

We discuss in the paper \co\ models involving field $\varPsi$ which plays a
r\^ole of a \q\ field. We find \il ary models of the \U\ and discuss a
dynamics of the Higgs field. Higgs' field dynamics undergoes a second order
phase transition which causes a phase transition in an \ev\ of the \U. This
ends an \il ary epoch and changes an \ev\ of the field~$\varPsi$. Afterwards
we consider the field~$\varPsi$ as a \q\ field building some \co\ models with
a \q\ and even with a $K$-essence.
A~dynamics of a Higgs field in several approximations gives us an amount of
an \il. We consider also a fluctuation spectrum of primordial fluctuations
caused by a Higgs field and a speed up factor of an \ev. We speculate on a
future of the \U\ based on our simple model with a special behaviour of a \q.

The paper is divided into three sections. In the second section we discuss
\il ary models and phase transitions obtained due to dynamics of Higgs'
fields. In the third section we present an \ev\ of Higgs field and
quintessential models of the \U. We consider several approaches to get an
amount of \il. Eventually we calculate an amount of an \il\ in a simplified
model of Higgs' field \ev\ and a power spectrum of primordial fluctuations
caused by a Higgs field.

\section{2. Inflationary models and phase transitions}

According to new observational data [2] concerning distances of type Ia
supernovae it seems that we need some kind of ``dark energy'' which drives
the evolution of the Universe. This ``dark energy'' can be considered as a
\co\ constant or more general as \co\ terms in field equations (in the
lagrangian). In the case of \co\ model this type of ``dark
energy''---``vacuum energy'' is a cause to accelerate the evolution of the
Universe (a~scale factor $R(t)$) (see Ref.~[3]). The \co\ constant is
negligible on the level of the Solar System and on the level of the Galaxy.
Moreover it can be important if we consider even nonrelativistic movement of
galaxies in a cluster of galaxies (see Ref.~[4]). In some papers considered
\co\ terms result in changing with time of a \co\ constant (see Ref.~[5]).
Some of them introduce additional scalar field (or fields) in order to give a
field-theoretical description of such an evolution of a \co\ ``constant''.
Those scalar fields are independent in general of the additional scalar
fields in inflationary models.

Thus the inflation field (or fields in multicomponent inflation, which can be\break
the same as some of Higgs' fields from G.U.T.-models) can be different from
those\break fields. In particular considering scalar-tensor theories of gravitation
results in so called quintessence models (see Ref.~[6]). Moreover in such
theories there is a natural field-theoretical background for an inconstant
``gravitational constant''. In such a way this quintessence field can be used
in twofold ways. First as a source of change in space and time of a
gravitational constant. Secondly as a source of \co\ terms leading to the
model of quintessence and a change in time of a \co\ ``constant''. In our
theory we have a natural occurrence of these phenomena due to the scalar
field~$\varPsi$ (or~$\rho$). Let us consider the \lg\ of our theory paying a
special attention to the part involving the scalar field~$\varPsi$, i.e.:
$$
\al
L&=\o R(\o W)+8\pi G_N\biggl(e^{-(n+2)\varPsi}\cL\dr{YM}(\t A)+
\frac{e^{-2\varPsi}}{4\pi r^2}\cL\dr{kin}(\gag\varPsi)\\
&-\frac{e^{(n-2)\varPsi}}{8\pi r^2}V(\varPhi)-\frac{e^{(n-2)\varPsi}}{2\pi r^4}\cL
\dr{int}(\varPhi,\t A)\biggr)\\
&-8\pi G_N\cL\dr{scal}(\varPsi)+e^{n\varPsi}\(\frac{e^{2\varPsi}\a^2_s}{l^2\dr{pl}}
\t R(\t \G)+\frac{\ul{\t P}}{r^2}\),
\eal \eq1
$$
$$
\cL\dr{scal}(\varPsi)=\(\o M\t g^{(\g \nu )}+n^2g^{[\mu \nu ]}g_{\d\mu}
\t g^{(\d\g)}\)\varPsi_{,\nu }\varPsi_{,\g}\, .
$$
We put $c=\hbar=1$.

Now we rewrite the \lg~(2.1) in the following form.
$$
\al
L&=\o R(\o W)-8\pi G_N e^{-(n+2)\varPsi}L\dr{matter}\\
&-8\pi G_N\cL\dr{scal}(\varPsi)+e^{n\varPsi}\(\frac{e^{2\varPsi}}{l^2\dr{pl}}
\a^2_s\t R(\t \G)+\frac{\ul{\t P}}{r^2}\),
\eal \eq1a
$$
where in $L\dr{matter}$ we include all the terms from Eq.~(2.1) with
Yang-Mills' fields, Higgs' fields, their interactions and coupling to the
scalar field~$\varPsi$.

The effective \gr\ constant is defined by
$$
G\dr{eff}=G_N e^{-(n+2)\varPsi}
$$
in such a way that the \lg\ of the Yang-Mills field in $L\dr{matter}$ is
without any factor involving scalar field~$\varPsi$. If the scalar field~$\varPsi$
is constant (e.g. $\varPsi=0$) we can redefine all the fields in such a way that
we get ordinary (standard) \lg s for these fields.

Let us consider the situation after a spontaneous symmetry breaking and
simplify to the case of $g_{[\mu \nu ]}=0$. We get
$$
L=\t{\o R} - 8\pi G\dr{eff}L\dr{matter} -8\pi G_NL\dr{scal}(\varPsi)
+8\pi G_N U(\varPsi) \eq2
$$
where 
$$
L\dr{scal}=\o M g^{\g \nu }\varPsi_{,\nu }\cdot \varPsi_{,\g},\quad \o M>0 \eq3
$$
$$
8\pi G_N U(\varPsi)=-\frac{V(\varPhi^K\dr{crt})}{r^4}l^2\dr{pl}e^{(n-2)\varPsi}
+e^{(n+2)\varPsi}\(\frac{\a^2_s \t R(\t\G)}{l^2\dr{pl}}\)+
\frac{\ul{\t P}}{r^2}e^{n\varPsi}. \eq4
$$

We get the following equations
$$
\t{\o R}_{\mu \nu }-\frac12 \t{\o R}g_{\mu \nu }=8\pi G\dr{eff}
\br{matter}T_{\kern-9pt \mu \nu }+8\pi G_N\br{scal}T_{\kern-5pt\mu \nu }, \eq5
$$
where $\t{\o R}_{\mu \nu }$ and $\t{\o R}$ are a Ricci tensor and a scalar
curvature for a Riemannian geometry generated by $g_{\mu \nu }=g_{(\mu \nu )}$,
$$
\br{matter}T^{\mu\nu}=(p+\rho)u^\mu u^\nu - pg^{\mu \nu } \eq6
$$
is an energy-momentum tensor for a matter considered as a radiation plus a
dust. 
$$
8\pi G_N\br{scal}T_{\kern-5pt\mu \nu }=8\pi G_N \(\frac{\o M}2 g_{\mu\nu} \cdot(g^{\a\b}
\varPsi_{,\a}\cdot\varPsi_{,\b})-\o M\varPsi_{,\mu}\cdot
\varPsi_{,\nu }\)+g_{\mu \nu }\o\la_{cK}
\eq7
$$
where
$$
\o\la_{cK}=2e^{(n-2)\varPsi}\frac{m^4_{\t A}}{\a^4_s} l^2\dr{pl}V(\varPhi^K\dr{crt})
-e^{(n+2)\varPsi}\frac{\a^2_s}{2l^2\dr{pl}}\t R(\t\G) - e^{n\varPsi}\frac{m^2_{\t A}}
{2\a^2_s}\ul{\t P}, \eq8
$$
where $m_{\t A}$ is a scale of a mass for massive Yang-Mills' fields (after a
spontaneous symmetry breaking for a true vacuum case). It is convenient to
write 
$$
\o\la_{cK}=\frac{e^{(n-2)\varPsi}}2 \a_K - \frac{e^{n\varPsi}}2 \g -
\frac{e^{(n+2)\varPsi}}2 \b \eq9
$$
$$
\al
\a_K&=\a_K(\xi,\zeta,m_{\t A},\a_s)\\
\b&=\b(\xi,\a_s,m_{\t A})\\
\g&=\g(\zeta,m_{\t A},\a_s)
\eal \eq9a
$$
$K=0,1$ corresponds to true and false vacuum case, i.e.\ 
$$
V(\varPhi^0\dr{crt})=0,\ V(\varPhi^1\dr{crt})\ne 0, \eq10
$$
$$
\a_0=0,\ \a_1\ne0. \eq10a
$$
For a scalar field $\varPsi$ we have the following equation:
$$
\al
16\pi G_N \o M g^{\a\b}\(\tov\nabla_\a(\pa_\b\varPsi)\)&+(n-2)e^{(n-2)\varPsi}
\a_K-(n+2)e^{(n+2)\varPsi}\b\\
&- n e^{n\varPsi}\g - (n+2)G\dr{eff}T=0
\eal
\eq11
$$
where $T=\rho-3p$ is a trace of an energy-momentum tensor for a matter field.
For we are interested in \co\ models we take for a metric tensor a
Robertson-Walker metric:
$$
ds^2=dt^2 - R^2(t)\left[ \frac{dr^2}{1-kr^2} + r^2d\theta^2 + r^2\sin^2\theta
d\f^2\right], \quad k=-1,0,1 \eq12
$$
and we suppose that $\varPsi,\rho,p$ are functions of $t$ only.

One gets
$$
\al
\frac1{M^2\dr{pl}}\ddot \varPsi&=\frac3{M^2\dr{pl}} H\dot \varPsi
-\(\frac{n-2}{\o M}\)\a_Ke^{(n-2)\varPsi}+
\frac{(n+2)}{\o M}\b e^{(n+2)\varPsi}\\
&+\frac n{\o M}e^{n\varPsi}\g+\frac{(n+2)}{M^2\dr{pl}}e^{-(n+2)\varPsi}\cdot
(\rho-3p)
\eal
\eq13
$$
$$
M\dr{pl}=\(\sqrt{8\pi G_N}\)^{-1}=\frac{m\dr{pl}}{\sqrt{8\pi}}\,.
$$

In this case we get standard equations for a \co\ model adapted to our theory
$$
\frac{3\ddot R}R = -8\pi G\dr{eff}\(\frac12(\rho+3p)\)
-8\pi G_N\(\frac12(\rho_\varPsi+3p_\varPsi)\), \eq14
$$
$$
R\ddot R+2\dot R+2k= 8\pi G\dr{eff} \(\frac12(\rho-p)\)R^2+
8\pi G_N\(\frac12(\rho_\varPsi-p_\varPsi)\)R^2. \eq15
$$

Using Eqs (2.14--15) one easily gets
$$
H^2+k = \frac{8\pi G\dr{eff}}3 \rho + \frac{8\pi G_N}3 \rho_\varPsi \eq16
$$
where $H=\dfrac{\dot R}R$ is a Hubble constant
$$
\al
8\pi G_N\rho_\varPsi& =8\pi G_N \frac{\o M}2 {\dot \varPsi}^2+
\frac12\o\la_{cK}\\
8\pi G_N p_\varPsi& =8\pi G_N \frac{\o M}2 {\dot \varPsi}^2-
\frac12\o\la_{cK}.
\eal
\eq17
$$

Let us consider a \co\ model for a ``false vacuum'' case, i.e.\ without matter
and only with a scalar $\varPsi$ and a vacuum energy $V(\varPhi^1\dr{crt})\ne0$. In
this case $\rho=p=0$ and we get
$$
\eqalignno{
&2\o M\ddot \varPsi - \frac{6\o M\dot R}R \dot\varPsi - M^2\dr{pl}\frac{d\la_{c1}}{d\varPsi}=0
&(2.18)\cr
&{\dot R}^2+k = \frac13 \(\frac12\,\frac{\o M}{M^2\dr{pl}}{\dot\varPsi}^2
+\frac12\la_{c1}\)R^2 &(2.19)\cr
&\frac{3\ddot R}R = -\frac{\o M}{M^2\dr{pl}} {\dot \varPsi}^2+\frac12\la_{c1}.
&(2.20)
}
$$
Let us take
$$
\varPsi=\varPsi_1=\text{const.} \eq21
$$
Thus we get
$$
\frac{d\la_{c1}}{d\varPsi}(\varPsi_1)=0 \eq22
$$
and
$$
\eqalignno{
&(n+2)\b x^4_1 + n\g x^2_1 - (n-2)\a_1=0 &(2.23)\cr
&{\dot R}^2+k = \frac16 \la_{c1}(x_1)R^2 &(2.24)\cr
&\frac{3\ddot R}R = \frac12 \la_{c1}(x_1) &(2.25)
}
$$
$$
x_1=e^{\varPsi_1}. \eqno(*)
$$

One gets from Eq.\ (2.25)
$$
\o R(t)=R_0 e^{H_0 t} \eq26
$$
where
$$
H_0=\sqrt{\frac{\la_{c1}(x_1)}6} \eq27
$$
is Hubble constant (really constant). From Eq.~(2.23) we obtain
$$
x_1=\sqrt{\frac{-n\g+\sqrt{n^2\g^2+4(n^2-4)\a_1\b}}{2(n+2)\b}}. \eq28
$$
For $\a_1>0$, $\b>0$ we get
$$
\sqrt{n^2\g^2+4(n^2-4)\a_1\b} > n|\g| \eq29
$$
$$
\la_{c1}(x_1)=\frac{x_1^{n-2}}{2(n+2)^2\b}
\left[n\g^2 - \g\sqrt{n^2\g^2
+4(n^2-4)\a_1\b}+4\a_1\b(n+2)\right]
\eq30
$$
$$
H_0=\frac{x_1^{(n-2)/2}}{2(n+2)\sqrt{3\b}}
\left[n\g^2 - \g\sqrt{n^2\g^2
+4(n^2-4)\a_1\b}+4\a_1\b(n+2)\right]^{1/2}. \eq31
$$
Thus we get an exponential expansion of the Universe. Using Eq.~(2.24) we
get also $k=0$ (i.e.\ a flatness of a space).

In this way we get de Sitter model of the Universe. This is of course a very
special solution to the Eqs~(2.18--20) with very special initial conditions
$$
\left.
\al
\o R(0)&=R_0\\
\frac{d\o R}{dt}(0)&=H_0R_0\\
\varPsi(0)&=\varPsi_1\\
\frac{d\varPsi}{dt}(0)&=0
\eal
\right\} \eq32
$$
Let us disturb the solution by a small perturbation and examine its
stability. Let
$$
\al
&\varPsi=\varPsi_1+\f,\quad |\f|,|\dot\f|\ll \varPsi_1\\
&R=\o R+\d R,\quad |\d R|,|\d \dot R|\ll \o R.
\eal \eq33
$$
One gets in a linear approximation
$$
\eqalignno{
&2\o M\ddot\f + 6\o MH_0\dot \f+M^2\dr{pl} \frac{d^2\la_{c1}}{d\varPsi^2}
(\varPsi_1)\f=0
&(2.34)\cr
&\frac{3\d\ddot R}{\o R}=-\o M\dot\f^2 &(2.35)\cr
&2\dot{\o R}\d\dot R+k= \frac16\o M\dot\f^2({\o R}^2+\o R\d R).&(2.36)
}
$$
One gets
$$
\f=\f_0e^{-\frac32H_0t}\sin\(\frac{\sqrt p}2t + \d\) \eq37
$$
where (we take for simplicity $M=\dfrac{\o M}{M^2\dr{pl}}$)
$$
p=-\D=\frac1{2M}\,\frac{d^2\la_{c1}}{d\varPsi^2}(\varPsi_1)-9H_0^2>0
\eq38
$$
if 
$$
\al
&M<M_0=\frac43\,\\
&\times\frac
{n(n+2)\g\sqrt{n^2\g^2+4(n^2-4)\a_1\b}-4(n+2)^2(n-1)\a_1\b-n^2(n+2)\g^2}
{n\g^2+4(n+2)\a_1\b-\g\sqrt{n^2\g^2+4(n^2-4)\a_1\b}}\,.
\eal
\eq39
$$

Thus we get an exponential decay of the solution $\f$, i.e.\ a damped 
oscillation around $\varPsi=\varPsi_1$. It means the solution $\varPsi=\varPsi_1$ is
stable against small \pt s of initial conditions for~$\varPsi$, i.e.\ 
$$
\eqalignno{
\varPsi(0)&=\varPsi_1+\f_0\sin\d &(2.40)\cr
\frac{d\varPsi}{dt}(0)&=\f_0\(-\tfrac32 H_0\sin\d+\cos\d\).& (2.41)
}
$$
Making $\f_0$ and $\d$ sufficiently small we can achieve smallness of \pt s
of initial conditions. The exponential decay of the solution to Eq.~(2.34)
can be satisfied also for some different conditions than Eq.~(2.39) if we
consider aperiodic case. However, we do not discuss it here.

Thus we get
$$
0\le \frac12 \o M\dot\f^2 \le a^2e^{-3H_0t} \eq42
$$
where $a^2$ is a constant.

From Eq.~(2.35) one gets
$$
0\le \d \ddot R\le a^2e^{-4H_ot}, \eq43
$$
which means that $\d\ddot R\sim0$ and $\d \dot R=O(\d R(0))$. It means 
$\d R=\d R(0)+\d \dot R(0)t$.
Moreover from Eq.~(2.36) we get that $k=0$ due to the fact that $\d R$ is a
linear function of~$t$. In this way we get
$$
R(t)\cong \o R(t)+\d R(0)+\d \dot R(0)t=
\(R_0+\frac{\d R(0)+\d \dot R(0)t}{e^{H_0t}}\)e^{H_0t}. \eq44
$$

This means that the small \pt s for initial conditions for~$R$ result in a
\pt\ of~$R_0$ in such a way that $R_0$ is perturbed by a quickly decaying
function. Thus our de Sitter solution is stable under small \pt s and is an
attractor for any small perturbed initial data. One can think that this
evolution will continue forever. However, we should remember that we are in a
``false'' vacuum regime and this configuration of Higgs' fields is unstable.
There is a stable configuration---a ``true'' vacuum case for which $\a_K=0$
($K=0$). For $\a=0$ the considered de Sitter evolution cannot be continued.
Thus we should take under consideration a second order phase transition 
in the configuration
of Higgs' fields from metastable state to stable state, from ``false'' vacuum
to ``true'' vacuum, from $\varPhi^1\dr{crt}$ to $\varPhi^0\dr{crt}$. In this case
the Higgs fields play a r\^ole of an order parameter. Let us suppose that the
scale time of this phase transition is small in comparison to $\dfrac1{H_0}$
and let it take place locally. We suppose locally a conservation of a density
of an energy. Thus
$$
\br{scal}T_{\kern-4pt 00}=\br{scal}T_{\kern-4pt 00}+\br{matter}T_{\kern-9pt 00}. \eq45
$$
We suppose also that the scalar field $\varPsi$ will be close to the new
equilibrium (new minimum for \co\ terms) and that the matter will consist of
a radiation only:
$$
\ealn{
\br{scal}T_{\kern-4pt 00}&=\frac12 \la_{c1}(x_1) &\text{(2.46a)}\cr
\br{scal}T_{\kern-4pt 00}+\br{matter}T_{\kern-9pt 00}&=\frac12 \la_{c0}(x_0)+8\pi G_N
\frac{\rho_r}{x_0^{n+2}}. &\text{(2.46b)}
}
$$

For $x_0=e^{\varPsi_0}$ is a new equilibrium point we have:
$$
\ealn{
&\frac{d\la_{c0}}{d\varPsi}(\varPsi_0)=0 &(2.47)\cr
&\la_{c0}=-e^{(n+2)\varPsi}\b - e^{n\varPsi}\g &(2.48)\cr
&\frac{d\la_0}{d\varPsi}=-e^{n\varPsi}\((n+2)\b e^{2\varPsi}+n\g\). &(2.49)
}
$$

One gets
$$
e^{\varPsi_0}=x_0=\sqrt{\frac{n|\g|}{(n+2)\b}} \eq50
$$
and supposing 
$$
\g<0 \eq51
$$
thus
$$
\la_{c0}(\varPsi_0)=\frac{2|\g|^{\frac n2+1}n^{\frac n2}}{\b^{\frac n2}
(n+2)^{\frac n2+1}}\,. \eq52
$$
From Eqs (2.46ab) one gets
$$
\al
\rho_r&=\frac1{16\pi G_N} \biggl\{
\frac{x_1^{n-2}}{(n+2)^2\b} \Bigl[n\g^2 - \g\sqrt{n^2\g^2+4(n^2-4)\a_1\b}
+2\a_1\b(n+4)\Bigr]\\
&-\frac{2|\g|^{\frac n2+1}}{\b^{\frac n2}(n+2)^{\frac n2+1}}\biggr\}
\frac{n^{\frac{n+2}2}\g^{\frac{n+2}2}}{(n+2)^{\frac{n+2}2}\b^{\frac{n+2}2}}\,.
\eal
\eq53
$$
This gives us matching condition for a second order phase transition and
simultaneously it is an initial condition for a new epoch of an evolution of
the Universe plus a condition $\dot\varPsi(t_r)=0$, $R(t_r)=\o R(t_r)$, where
$t_r$ is a time for a phase transition to occur. Notice we have simply
$\varPsi_0\ne \varPsi_1$. Thus we have to do with a discontinuity for a
field~$\varPsi$. Let us consider Hubble's constants for both phases of the
Universe 
$$
\al
H_0^2&=\frac{\la_{c1}(x_1)}6 \\
H_1^2&=\frac{\la_{c0}(x_0)}6 =\frac{|\g|^{\frac n2+1}n^{\frac n2}}{3\b^{\frac
n2}(n+2)^{\frac n2+1}}\\
H_0^2&\ne H_1^2
\eal
\eq54
$$
Summing up we get
$$
\al
R(t_r)&=\o R(t_r)=R_0 e^{+H_0t_r}\\
\dot \varPsi_1(t_r)&=\dot \varPsi_0(t_r)=0 \\
\varPsi_1&\ne \varPsi_0\\
H_0^2&\ne H_1^2
\eal
\eq55
$$

Thus we see that the second order phase transition in the \cf\ of Higgs'
fields results in the first order phase for an evolution of the Universe. We
get discontinuity for Hubble constants and values of scalar field before and
after phase transition.

Let us calculate deceleration parameters before and
after phase transition
$$
\ealn{
&q=-\frac{R\ddot R}{{\dot R}^2} &(2.56)\cr
&q=-1 &(2.57)
}
$$
before phase transition and
$$
q=-1 \eq58
$$
after phase transition. 

Let us come back to the Eqs (2.14), (2.16), (2.13). Let us rewrite
Eq.~(2.13) in the following form
$$
\o M\ddot \varPsi+\frac{6\o M\dot R}R \dot\varPsi + M^2\dr{pl}\frac{d\la_{c0}}
{d\varPsi}=0. \eq59
$$
(Remember we now have to do with a radiation for which the trace of an
energy-momentum tensor is zero.)

Let us suppose that $\dot\varPsi\approx0$. Thus Eq.~(2.59) simplifies 
$$
\frac1{M^2\dr{pl}} M\ddot \varPsi + \frac{d\la_{c0}}{d\varPsi}=0 \eq60
$$
and we get the first integral of motion
$$
\ealn{
& \frac{\o M}{2M^2\dr{pl}}\dot\varPsi^2+\la_{c0}=
\frac12\o \d =\text{const.} &(2.61)\cr
&\o M\dot\varPsi^2=M^2\dr{pl}(\o\d-2\la_{c0}).&\text{(2.61a)}
}
$$
One gets from Eqs (2.14) and (2.16)
$$
\ealn{
&\frac{3\ddot R}R = -\frac1{M^2\dr{pl}}\cdot \frac{\rho_r}{e^{(n+2)
\varPsi}}-\frac12\(\frac{2\o M}{M^2\dr{pl}} \dot\varPsi^2-\la_{c0}\) &(2.62)\cr
&\dot R^2+k=\frac1{3M^2\dr{pl}}\cdot\frac{\rho_r}{e^{(n+2)\varPsi}}R^2
+\frac13\(\frac{\o M}{2M^2\dr{pl}}\dot\varPsi^2+\frac12 \la_{c0}\)R^2.&(2.63)
}
$$
Using Eq (2.61a) we get:
$$
\ealn{
&\frac{3\ddot R}R = -\frac1{M^2\dr{pl}}\cdot \frac{\rho_r}{e^{(n+2)
\varPsi}}-\o\d+\frac52\la_{c0} &(2.64)\cr
&\dot R^2+k=\frac1{3M^2\dr{pl}}\cdot\frac{\rho_r}{e^{(n+2)\varPsi}}R^2
+\frac16(\o\d- \la_{c0})R^2.&(2.65)
}
$$
One can derive from Eqs (2.64--65)
$$
\ealn{
\frac d{dt}(R\dot R)&=-\frac16\o\d R^2+\frac23 \la_{c0}R^2-k &(2.66)\cr
\la_{c0}&=-e^{n\varPsi}(\b e^{2\varPsi}+\g). &(2.67)
}
$$

Let us take $k=0$ and $\d=0$ and let us change independent and dependent
variables in (2.66) using (2.67) and (2.61a). One gets
$$
2y^2(y^2-1)\frac{d^2f}{dy^2}+y\((n+4)y^2-(n+1)y-2\)\frac{df}{dy}
=\frac{4\o M}3 (y^2-1)f(y) \eq68
$$
where
$$
y=\sqrt{\frac\b{|\g|}}e^\varPsi \eq69
$$
and
$$
f=R^2. \eq70
$$
It is easy to see that $y^2>1$ for
$$
\frac{d\varPsi}{dt}=\pm \frac{M\dr{pl}}{\sqrt{\o M}}\cdot\frac{|\g|^{\frac{n+2}4}}
{\b^\frac n4}\sqrt{y^n(y^2-1)}. \eq71
$$
Moreover according to our assumptions $\dot\varPsi\approx0$ and this can be
achieved only if $0<y-1<\e$, where $\e$ is sufficiently small.

Thus for further investigations we take $z=y-1$, $y=z+1$. One gets
$$
\al
2z(z+2)(z+1)^2\frac{d^2f}{dz^2}&+(z+1)\((n+4)z^2+(n+7)z
-(n+3)\)\frac{df}{dz}\\
&-\frac{4\o M}3 z(z+2)f(z)=0. 
\eal
\eq72
$$
Let us consider Eqs (2.64--65) supposing $k=0$ and $\o\d=0$.

After eliminating $\ddot R$ (via differentiation of Eq.~(2.65) with respect
to time $t$) and changing independent and dependent variables one gets:
$$
\frac d{dy}\biggl[\(8\pi G_N\frac{\t\rho_r}{y^{n+2}} + \frac12 y^n(y^2-1)\)f^2
\biggr]=-\frac d{dy}(f^2)y^n(y^2-1) \eq73
$$
where
$$
\o\rho_r=\rho_r \frac{|\g|^n}{\b^{n+1}}\,.  \eq74
$$
It is convenient for further investigations to consider a parameter
$$
\o r=\frac{4\o M}3\,. \eq75
$$
In such a way we get
$$
2y^2(y^2-1)\frac{d^2f}{dy^2}+y\((n+4)y^2-(n+1)y-2\)\frac{df}{dy}
- \o r(y^2-1)f(y) \eq76
$$
and
$$
\al
2z(z+2)(z+1)^2 \frac{d^2f}{dz^2}&+(z+1)\((n+4)z^2+(n+7)z-(n+3)\)
\frac{df}{dz}\\
&-\o rz(z+2)f(z)=0. 
\eal
\eq77
$$

One can transform Eq.~(2.73) into
$$
\frac d{dy}\left[\(
8\pi G_N \frac{\t\rho_r}{y^{n+2}}+\frac32 y^n(y^2-1)\)f^2\right]
=y^{n-1}\left[(n+2)y^2-n\right]f^2. \eq78
$$
Changing the independent variable from $y$ to $z=y-1$ one gets:
$$
\al
\frac d{dz}&\left[\(8\pi G_N\frac{\t\rho_r}{(z+1)^{n+2}}+
\frac32 (z+1)^nz(z+2)\)f^2\right]\\
&\qquad{}=(z+1)^{n-1}\left[(n+2)(z+1)^2-n\right]f^2. 
\eal
\eq79
$$
Let us come back to Eq.~(2.71) in order to find time-dependence of~$\varPsi$.
One gets
$$
\int\frac{d\varPsi}{\sqrt{\b e^{(n+2)\varPsi}-|\g|e^{n\varPsi}}}
=\pm \frac{t-t_1}{\sqrt{\o M}}\,M\dr{pl} \eq80
$$
or
$$
\int\frac{dx}{x\sqrt{\b x^{n+2}-|\g|x^n}}
=\pm \frac{t-t_1}{\sqrt{\o M}}\,M\dr{pl}. \eq80a
$$
If $n=2l$, where $l$ is a natural number, one gets for $\b>0$
$$
\al
\(\frac{|\g|}\b\)^{-\frac12}
&\Biggl\{ \sum_{k=1}^l\, \sum_{p=0}^k \frac{\binom kp(-2)^{k-p}}{2p}
\(\sqrt{\frac{\sqrt\b x-\sqrt{|\g|}}{\sqrt\b x+\sqrt{|\g|}}}\)^{2p+1}\\
&+\frac1{2\sqrt3} 
\ln\(\frac{\sqrt{\sqrt\b x-|\g|}-\sqrt{3(\sqrt\b x+|\g|)}}
{\sqrt{\sqrt\b x-|\g|}+\sqrt{3(\sqrt\b x+|\g|)}}\)\Biggr\}=
\pm \frac{(t-t_1)}{\sqrt{\o M}}\,M\dr{pl}. 
\eal
\eq81
$$

In this case it can be expressed in terms of elementary functions. For a
general case of $n$, $\b$ and~$\g$ one gets
$$
\frac{2x^{-\frac n2}\h12(\frac12,-\frac n4,(1-\frac n4);\frac{x^2\b}\g)}
{\sqrt\g n}=\pm\frac{(t-t_1)}{\sqrt{\o M}}\,M\dr{pl} \eq82
$$
where $\h12(a,b,c;z)$ is a hypergeometric function. For a small~$z$ (around
zero) one finds the following solution to Eq.~(2.77)
$$
f(z)=C_1z^\frac{(n+7)}4 \(1-\frac{(n+7)(5n+23)}{8(n+1)}z\)+
C_2\(1+\frac{\o r}{(n-1)}z^2\), \eq83
$$
$C_1,C_2=\text{const.}$ Thus
$$
\al
R(y)&=\Biggl[C_1(y-1)^\frac{(n+7)}4 \(1-\frac{(n+7)(5n+23)}{8(n+1)}(y-1)\)\\
&\qquad{}+C_2\(1+\frac{\o r}{(n-1)}(y-1)^2\)\Biggr]^{\frac12} 
\eal
\eq84
$$
where $y>1$ (but only a little).

Let us make some simplifications of the formulae taking under consideration
that $y-1$ is very small.

One gets
$$
\ealn{
f(y)&=C\(1+\frac{\o r}{(n-1)}(y-1)^2\), \quad C>0& (2.85)\cr
R(y)&=R_1\(1+\frac{\o r}{(n-1)}(y-1)^2\)^\frac12. &(2.86)
}
$$
After some calculations one gets
$$
\al
8\pi G_N\t\rho_r&=y^{2(n+1)}\biggl(-\frac{2(n+5)}{(n+4)(n-1)}y^4
+\frac{10\o r(n+2)}{(n-1)(n+4)}y^3\\
&+\frac{\(8\o r-(n-1)(n+4)\)}{2(n-1)(n+4)}y^2
-\frac{2\o r(n+3)}{(n^2-1)}y+ \frac{(\o r-n+1)}{(n-1)}\biggr)
\eal
\eq87
$$
(taking into account that $(y-1)$ is very small). Let us make some
simplifications in the formula (2.71) for $y$ close to~1. We find:
$$
\int \frac{dy}{\sqrt{y-1}}=\pm \frac{|\g|^\frac{n+4}4 \sqrt2}{\b^\frac n4
\sqrt{\o M}} (t-t_1)M\dr{pl} \eq88
$$
and finally
$$
y=\frac{|\g|^\frac{n+4}2}{2\o M \b^\frac n2}(t-t_1)^2+1 \eq89
$$
such that $y=1$ for $t=t_1$.

Let us pass to the minimum value for $\t\rho_r$ with respect to $y$ close
to~1. Thus we are looking for a minimum of a function
$$
\al
&\t\rho=\frac1{8\pi G_N} y^{2(n+1)}\biggl[
-\frac{2(n+5)}{(n+4)(n-1)}y^4
+\frac{10\o r(n+2)}{(n-1)(n+4)}y^3\\
&+\frac{\(8\o r-(n-1)(n+4)\)}{2(n-1)(n+4)}y^2
-\frac{2\o r(n+3)}{(n^2-1)}y+ \frac{(\o r-n+1)}{(n-1)}\biggr]
=\frac1{8\pi G_N}y^{2(n+1)}W_4(y)
\eal
\eq90
$$
for such a $y$ that is close to~1.

Let us write $y=1+\eta$, where $\eta$ is very small and develop $W_4(1+\eta)$
up to the second order in~$\eta$. One finds
$$
W_4(1+\eta)\simeq V_2(\eta) \eq91
$$
and
$$
V_2(\eta)=a\eta^2+b\eta+c \eq92
$$
where
$$
\ealn{
a&=\frac{\(-n^2+3n(20\o r-9)+4(32\o r-31)\)}{2(n+4)(n-1)} &(2.93)\cr
b&=\frac{\(-n^3-4n^2(7\o r-3)+5n(18\o r-11)+22(3\o r-2)\)}{(n+4)(n^2-1)} &(2.94)\cr
c&=\frac{\(-3n^3+2n^2(9\o r-8)+n(50\o r-29)+4(7\o r-4)\)}{2(n+4)(n^2-1)} &(2.95)
}
$$

Let us find minimum of $V_2(\eta)$ for $\eta>0$ such that $V_2(\eta)\ge0$.
For $a<0$, $b<0$, $c>0$ we get
$$
V_2(\eta\dr{min})=0 \eq96
$$
and
$$
\eta\dr{min}=\frac{b+\sqrt{b^2+4|a|c}}{2|a|}\,. \eq97
$$
In this case
$$
(1+\eta\dr{min})^{2(n+1)} W_4(1+\eta\dr{min})>0 \eq98
$$
and will be close to the real minimum of the function (2.90).
Simultaneously 
$$
\eta\dr{min}<1.
$$
For
$$
\t\rho_r=\frac{|\g|^n}{\b^{n+1}}\sigma T^4 \eq99
$$
(where $\sigma$ is a Stefan-Boltzmann constant) we get $T\dr{min}$ and we
call it a~$T_d$ (a decoupling temperature-decoupling of matter and
radiation). Thus $y_d=1+\eta\dr{min}$. $y_d$ is reached at a time $t_d$
(a~decoupling time)
$$
t_d=t_1 + \frac{\eta\dr{min}\o M\b^\frac n2}{2|\g|^{(\frac{n+4}2)}}\,M\dr{pl}. \eq100
$$

From that time ($t_d$) the evolution of the Universe will be driven by a
matter (with good approximation a dust matter) and the scalar field~$\varPsi$.
The radius of the Universe at $t=t_d$ is equal to
$$
R(t_d)=R_1\(1+\frac{\o r}{(n-1)}\eta^2\dr{min}\)^\frac12. \eq101
$$
The full field equations (generalized Friedmann equations) are as follows:
$$
\ealn{
\(\frac{\dot R}R\)^2 &= \frac{8\pi G\dr{eff}}3\rho_m+ \frac13\(\frac12
\frac{\o M}{M^2\dr{pl}}{\dot\varPsi}^2+\frac12 \la_{c0}\) &(2.102)\cr
\frac{3\ddot R}R &= \frac{8\pi G\dr{eff}}3\rho_m-\frac12\(2\frac{\o M}
{M^2\dr{pl}}{\dot\varPsi}^2-\la_{c0}\) &(2.103)\cr
-\frac{2\o M}{M^2\dr{pl}}\ddot \varPsi &-\frac{6\o M}{M^2\dr{pl}}
\cdot \frac{\dot R}R \dot\varPsi - \frac{d\la_{c0}}{d\varPsi}+(n+2)8\pi \rho_m
G\dr{eff}=0. &(2.104)
}
$$

Before we pass to those equations we answer the question what is a mass of
the scalar field~$\varPsi$ during the de Sitter phases and the radiation era. One
gets:
$$
\ealn{
-m_1^2&=\frac{M^2\dr{pl}}{2\o M}\frac{d^2\la_{c1}}{d\varPsi^2}(\varPsi_1)
&(2.105)\cr
-m_0^2&=\frac{M^2\dr{pl}}{2\o M}\frac{d^2\la_{c0}}{d\varPsi^2}(\varPsi_0).
&(2.106)
}
$$
The second question we answer is an evolution of the Universe during a period
from $t_r$ to $t_1$, i.e.\  when $y$ is changing from  $y=\sqrt{\frac n{n+2}}$
to $y=1$.

This will be simply the de Sitter evolution
$$
R(t)=R e^{H_1t} \eq107
$$
where
$$
H_1=\sqrt{\frac{\la_{c0}(x_0)}6}\,. \eq108
$$
This will be unstable evolution and will end at $t=t_1$, i.e.\ for
$\dot\varPsi\simeq0$ and value of the field~$\varPsi$ will be changed to the value
corresponding to $y=1$, i.e.\ 
$$
\varPsi(t_1)=\frac12 \ln\frac\b{|\g|}\,. \eq109
$$
Thus we have to do with two de Sitter phases of the evolution with two
different Hubble constants $H_1$ and~$H_0$. In the third phase (a radiation
era) the radius of the Universe is given by the formula
$$
R(t)=R_0e^{H_0t_r+H_1(t_1-t_r)} \cdot \(1+
\frac{|\g|^{n+4}}{3\o M(n-1)\b^n}(t-t_1)^4\)^\frac12
\eq110
$$
and such an evolution ends for $t=t_d$,
$$
R(t_d)=R_0\exp(H_0t_r+H_1(t_1-t_r)) \cdot \(1+\frac{\o r}{(n-1)}\eta\dr{min}^2\)
^\frac12. \eq111
$$
In both de Sitter phases the equation of state for the matter described by
the scalar field~$\varPsi$ is the same:
$$
\rho_\varPsi=-p_\varPsi. \eq112
$$
In the radiation era we get
$$
p_\varPsi=\tfrac43 \rho_\varPsi. \eq113
$$

Now we go to the matter dominated Universe and we change the notation
introducing a scalar field~$Q$ such that
$$
Q=\sqrt{\o M} \frac\varPsi{M\dr{pl}}\,. \eq114
$$
In this case we have
$$
\ealn{
\rho_Q&=\frac12{\dot Q}^2+U(Q) &\text{(2.115a)}\cr
p_Q&=\frac12{\dot Q}^2-U(Q) &\text{(2.115b)}
}
$$
where
$$
\ealn{
U(Q)&=\frac12\la_{c0}\(\frac {QM\dr{pl}}{\sqrt{\o M}}\)=-\frac\b2 e^{aQ}+
\frac12|\g|e^{bQ} &(2.116)\cr
&a=\frac{(n+2)}{\sqrt{\o M}}\,M\dr{pl}, \ b=\frac n{\sqrt{\o M}}\,M\dr{pl}. &(2.117)
}
$$

Let us write Eqs (2.102--104) in terms of $Q$:
$$
\ddot Q+3H\dot Q+U'(Q)-8\pi(n+2)G\dr{eff}\rho_m=0 \eq118
$$
where
$$
\ealn{
G\dr{eff}&=G_N e^{-\frac{(n+2)}{\sqrt{\o M}}M\dr{pl}Q} &(2.119)\cr
H^2&=\frac13\(8\pi G\dr{eff}\rho_m+\frac12{\dot Q}^2+U\)&(2.120)\cr
\frac{3\ddot R}R&= \frac12\(8\pi G\dr{eff}\rho_m+2({\dot Q}^2-U)\).&(2.121)
}
$$

Let us define an equation of state
$$
W_Q=\frac{p_Q}{\rho_Q}=\frac{\frac12{\dot Q}^2-U}{\frac12{\dot Q}^2+U}
\eq122
$$
and a variable
$$
x_Q=\frac{1+W_Q}{1-W_Q}=\frac{\frac12{\dot Q}^2}U, \eq123
$$
which is a ratio of a kinetic energy to a potential energy density for~$Q$.
It is interesting to combine (2.118) and (2.120) using (2.122)
and~(2.123). 

Let us define
$$
\ealn{
\O_Q&=\frac{\frac12 {\dot Q}^2+U}{3H^2} &(2.124)\cr
\O_m&=\frac{8\pi G\dr{eff}\rho_m}{3H^2}. &(2.125)\cr
}
$$
One gets
$$
\O_m+\O_Q=1 \eq126
$$
or
$$
8\pi G\dr{eff}\rho_m=3(1-\O_Q)H^2. \eq127
$$
We call $\rho_m$ an energy density of a background and for such a matter the
equation of state is
$$
W_m=\frac{p_m}{\rho_m}=0. \eq128
$$
It is natural to define
$$
\t\rho_m=8\pi G\dr{eff}\rho_m \eq129
$$
as an effective energy density of a background with the same equation of a
state as~(2.128).

Under an assumption of slow roll for $Q$, i.e.\ 
$$
\frac12{\dot Q}^2 \ll U(Q) \eq130
$$
one gets
$$
W_Q\simeq -1, \eq131
$$
i.e.\ an equation of state of a \co\ constant evolving in time. In this case
$x_Q \simeq 0$. This can give in principle an account for an acceleration of
an evolution of the Universe if we suppose those conditions for our
contemporary epoch.

\def\eq#1 {\eqno(\text{\rm3.#1})}
\section{3. Evolution of Higgs Field and Quintessential-Cosmological Models}
Let us come back to the inflationary era in our model. For our Higgs' field
is multicomponent, i.e.\ it is a multiplet of Higgs' fields, we have to do
with so called multicomponent inflation (see Ref.~[7]). In this case we can
define slow-roll parameter for our model in equations for Higgs' fields.

One gets from the first point of Ref.~[1] (see Eq.~(5.7.7), p.~385)
$$
\frac d{dt}\(L^{d0}_{\u b}\)_{av} - 3H\(L^{d0}_{\u b}\)_{av}=
-\frac{e^{n\varPsi_1}}{2r^2} l^{db}\left\{\frac{\d V'}{\d \varPhi^b_{\u n}}
g_{\u b\u n}\right\}_{av} \eq1
$$
where $L^d_{0\u a}=L^d_{t\u a}$ (a time component of $L^d_{\mu\u a}$) is
defined by
$$
l_{dc}L^d_{0\u a}+l_{cd}g_{\u a\u m}g^{\u m\u c}L^d_{0\u c}=
2l_{cd} g_{\u a\u m}g^{\u m\u c}\frac d{dt}\(\varPhi^d_{\u c}\),
\eq2
$$
$V'$, $\dfrac{\d V'}{\d \varPhi^b_{\u n}}$ and 
$\left\{\dfrac{\d V'}{\d \varPhi^b_{\u n}}\right\}_{av}$ are defined 
in [1] (Eq.~(5.6.9), (5.6.10), (5.424)).
$$
\ealn{
3H^2&=8\pi G_N \(\frac{e^{(n-2)\varPsi_1}}{r^4}V'(\varPhi)
-\frac{2e^{-2\varPsi_1}}{r^2}\cL\dr{kin}(\dot \varPhi)
+\frac12 \la_{c1}(\varPsi_1)\) & (3.3)\cr
\dot H&=8\pi G_N \(-\frac{2e^{-2\varPsi_1}}{r^2}\cL\dr{kin}(\dot \varPhi)\)
&(3.4)
}
$$
where $H$ is a Hubble constant and
$$
\cL\dr{kin}(\dot\varPhi)=l_{ab}\left\{g^{\u b\u n}L^a_{0\u b}
\frac d{dt}\varPhi^b_{\u n}\right\}_{av}. \eq5
$$
$\varPsi_1$ is a constant value for a field $\varPsi$. The slow-roll parameters are
defined by
$$
\ealn{
\e&=\frac{\l_{ab}\kl g^{\u b\u n}L^a_{0\u b}\ddt \varPhi^b_{\u n}\kr_{av}}
{H^2} &(3.6)\cr
\d&=\frac{\l_{ab}\kl g^{\u b\u n}\ddt\(L^a_{0\u b}\)\ddt \(\varPhi^b_{\u n}\)\kr_{av}}
{H^2\(\l_{ab}\kl g^{\u b\u n}L^a_{0\u b}\ddt \(\varPhi^b_{\u n}\)\kr_{av}\)}
&(3.7)
}
$$
with the assumptions of the slow roll
$$
\e=O(\eta),\ \d=O(\eta) \eq8
$$
for some small parameter $\eta$.

Under slow-roll conditions we have
$$
\ealn{
&3H\(L^d_{0\u b}\)_{av}+\frac{e^{n\varPsi_1}}{2r^2} l^{db}\(\frac{\d V'}{\d
\varPhi^b_{\u n}}g_{\u b\u n}\)_{av}\cong 0 &(3.9)\cr
&H^2\cong \frac{8\pi}{3M^2\dr{pl}}\kl\frac{e^{(n-2)\varPsi_1}}{r^4}V'(\varPhi)\kr
+\frac12\la_{c1}(\varPsi_1) &(3.10)
}
$$
and with a standard extra assumption
$$
\frac{\dot \d}{H}=O(\eta^2). \eq11
$$
We remind that the scalar field $\varPsi$ is the more important agent to drive
the inflation. However, the full amount of time the inflation takes place
depends on an evolution of Higgs' fields (under slow-roll approximation). The
evolution must be slow and starting for $\varPhi=\varPhi^1\dr{crt}$. It ends at
$\varPhi=\varPhi^0\dr{crt}$. 

Thus we have according to Eq.~(2.26)
$$
\o N=\ln\frac{R(t\dr{end})}{R(t\dr{initial})}=
H_0(t\dr{end}-t\dr{initial})
$$
where
$$
\ealn{
\varPhi(t\dr{end})&=\varPhi^0\dr{crt} &(3.12)\cr
\varPhi(t\dr{initial})&=\varPhi^1\dr{crt} &(3.13)\cr
t\dr{end}&=t_r
}
$$
(we omit indices for $\varPhi$).

Thus we should solve Eq.~(3.9) for initial condition (3.12) and with
an assumption $H=H_0$. We have of course a short period of an inflation with
$H=H_1$ from $t\dr{end}=t_r$ to $t=t_1$, i.e.\ 
$$
\o N\dr{tot}=\o N+\o N_1=H_0(t_r-t\dr{initial})+H_1(t_1-t_r). \eq14
$$

Let us remind to the reader that $\o N$ is called an amount of inflation. Let us
consider inflationary fluctuations of Higgs' fields in our theory. We are
ignoring \gr\ backreaction for the \co\ \ev\ is driven by the scalar field~$\varPsi$.
We write the Higgs field $\varPhi^a_{\u m}$ as a sum 
$$
\varPhi^a_{\u m}(\vec r,t)=\varPhi^a_{\u m}(t)+\d \varPhi^a_{\u m}(\vec r,t) \eq15
$$
where $\varPhi^a_{\u m}(t)$ is a solution of Eq.~(3.9) with boundary condition
(3.12--13) and $\d\varPhi^a_{\u m}(\vec r,t)$ is a small fluctuation which
will be written in terms of Fourier modes
$$
\d\varPhi^a_{\u m}(\vec r,t)=\sum_{\vec K} \d\varPhi^a_{\vec K\u m}(t)
e^{i\vec K\vec r}. \eq16
$$
The full field equation for Higgs' field in the de Sitter background
linearized for $\d\varPhi^a_{\vec K\u m}(t)$ can be written
$$
\al
\frac{d^2}{dt^2}\(M^d_{\vec K\u b}\)_{av}
&+3H_0\ddt \(M^d_{\vec K\u b}\)_{av}+\frac1{R^2}{\vec K}^2\(M^d_{\vec K}\)
_{av}\\
&=-\frac{e^{n\varPsi_1}}{2r^2} l^{db}\kl\frac{\d^2V'\(\varPhi^a_{\u b}(t)\)}
{\d\varPhi^b_{\u n}\d \varPhi^c_{\u m}}g_{\u b\u n}\kr
\d\varPhi^c_{\vec K\vec m}(t)
\eal
\eq17
$$
where
$$
l_{dc}M^d_{\vec K\u b}+l_{cd}g_{\u a\u m}g^{\u m\u c}M^d_{\vec K\u c}
=2l_{cd}g_{\u a\u m}g^{\u m\u c}\d\varPhi^d_{\vec K\u c}. \eq18
$$
Let us change \dt\ and in\dt\ variables.
$$
\ealn{
&d\tau=\frac{dt}{\o R(t)},\quad \o R(t)=R_0e^{H_0t} &(3.19)\cr
&R(\tau)=-\frac1{H_0\tau} &(3.20)\cr
&-\infty<\tau<0 \quad\hbox{(a conformal time)}&(3.21)
}
$$
and
$$
\d\varPhi^a_{\vec K\u m}=\frac{\chi^a_{\vec K\u m}}R\,. \eq22
$$
Simultaneously we neglect the term with the second derivative of the
potential. Eventually we get
$$
\frac{d^2}{d\tau^2}\(\t M^d_{\vec K\u b}\)_{av} - \frac2{\tau^2}
\frac d{d\tau}\(\t M^d_{\vec K\u b}\)_{av}+{\vec K}^2\(\t M^d_{\vec K\u
b}\)_{av}=0 
\eq23
$$
where
$$
l_{dc}\t M^d_{\vec K\u b}+l_{cd}g_{\u a\u m}g^{\u m\u c}\t M^d_{\vec K\u c}
=2l_{cd}g_{\u a\u m}g^{\u m\u c}\chi^d_{\vec K\u c}. \eq24
$$

Let us notice that $|\vec K\tau|$ is the ratio of the proper wavenumber
$|K|R^{-1}$ to the Hubble radius $\dfrac1{H_0}$. At early times $|K\tau|\gg1$,
the wavelength is small in comparison to Hubble radius and the mode
oscillates as in Minkowski space. However, if $\tau$ goes to zero,
$|K\tau|$~goes to zero too. The wavelength of the mode is stretched far
beyond the Hubble radius. It means the mode freezes. Thus the analyzes
of the modes are similar to those in classical symmetric inflationary \pt\ 
theory neglecting the fact that we have to do with multicomponent inflation
and that the structure of representation space of Higgs' fields should be
taken into account.

However, the form of Eq.~(3.23) is exactly the same as in one component
symmetric theory if we take $\t M^d_{\vec K\u a}$ a field to be considered.
The relation (3.24) between $\t M^d_{\vec K\u a}$ and $\chi^d_{\vec K\u a}$
is linear and under some simple conditions unambigous.

It is possible to find an exact solution to Eq.~(3.23) and Eq.~(3.24).
One gets
$$
\al
\chi^a_{\vec K\u m}&=C^a_{1\vec K\u m}\(\frac
{\tau|\vec K|\cos(\tau|\vec K|)-\sin(\tau|\vec K|)}\tau\)\\
&+C^a_{2\vec K\u m}\(\frac
{\tau|\vec K|\sin(\tau|\vec K|)+\cos(\tau|\vec K|)}\tau\) 
\eal
\eqno(*)
$$
and
$$
\al
M^d_{\vec K\u b}&=N^d_{1\vec K\u b}\(\frac
{\tau|\vec K|\cos(\tau|\vec K|)-\sin(\tau|\vec K|)}\tau\)\\
&+N^d_{2\vec K\u b}\(\frac
{\tau|\vec K|\sin(\tau|\vec K|)+\cos(\tau|\vec K|)}\tau\) 
\eal
\eqno(**)
$$
where
$$
l_{dc}N^d_{i\vec K\u b}+l_{cd}g_{\u a\u m}g^{\u m\u c}N^d_{i\vec K\u c}
=2l_{cd}g_{\u a\u m}g^{\u m\u c}C^d_{i\vec K\u c}, \quad i=1,2.
\eqno(*{*}*)
$$

In this way one obtains
$$
\d\varPhi^a_{\vec K\u m}=\frac1{R_0} e^{-H_0t}\chi^a_{\vec K\u m}
\(-\frac{e^{-H_0t}}{R_0H_0}\) \eqno(*{*}{*}{*})
$$
and using $(*)$ one easily gets
$$
\al
\d\varPhi^a_{\vec K\u m}&=\t C^a_{1\vec K\u m}
\(-\frac{|\vec K|}{R_0H_0}e^{-H_0t}\cos\(\frac{|\vec K|}{R_0H_0}e^{-H_0t}\)
+\sin\(\frac{|\vec K|}{R_0H_0}e^{-H_0t}\)\)\\
&+\t C^a_{2\vec K\u m}
\(\frac{|\vec K|}{R_0H_0}\sin\(\frac{|\vec K|}{R_0H_0}e^{-H_0t}\)
+\cos\(\frac{|\vec K|}{R_0H_0}e^{-H_0t}\)\) 
\eal
\eqno\text{(V$*$)}
$$
where
$$
\t C^a_{i\vec K\u m}=-H_0C^a_{i\vec K\u m}\,. \eqno\text{(VI$*$)}
$$
However, it is not necessary to use a full exact solution (V$*$) to
proceed an analysis which we gave before.

Thus the primordial value of $R_{\vec K}(t)$ is equal to
$$
R_{\vec K}=-\[\frac{H_0}{\ddt(\varPhi^d_{\u m})}\d\varPhi^d_{\vec K\u m}\]
_{\big| t=t^\ast} \eq25
$$
where $R_{\vec K}(t)$ is defined in \co\ \pt\ theory as
$$
R^{(3)}(t)=4\frac{\vec K^2}{R^2} R_{\vec K}(t) \eq26
$$
(see Ref.~[8]) and $t^\ast$ is such that $|\vec K|=R(t^\ast)H_0$, i.e.
$$
t^\ast=\frac1{H_0} \ln\biggl[\frac{|\vec K|}{R_0H_0}\biggr], 
\quad |\vec K|=K. \eq27
$$
The primordial value is of course time-in\dt. We suppose that fluctuations of
Higgs' fields (the initial values of them) are in\dt\ and Gaussian.

Thus we can repeat some classical results concerning a power spectrum of
primordial \pt s. One gets
$$
\[\bigl| \d\varPhi^d_{\vec K\u m}\bigr|^2\] = \(\frac{H_0}{2\pi}\)^2. \eq28
$$
Thus
$$
P_R(K) \cong \(\frac{H_0}{2\pi}\)^2 \sum_{\u m,d}
\(\frac{H_0}{\ddt(\varPhi^d_{\u m})}\)_{\big| t=t^\ast}^2
\eq29
$$
for $t^\ast$ given by Eq.~(3.27).

Taking a specific situation with concrete groups $H,G,G_0$ and constants 
$\xi,\zeta,r$ we can in principle calculate exactly $P_R(K)$ and a power
index~$n_s$. Thus using some specific software packages (i.e.\ CMBFAST code)
we can obtain theoretical curves of CMB (Cosmic Microwave Background)
anisotropy including polarization effects.

Let us give the following remark. The condition for slow-roll evolution
for~$\varPhi$ can be expressed in terms of the potential~$V'$. If those are
impossible to be satisfied for some models (depending on $G,G_0,H,G_0'$ and
parameters $\xi,\zeta$), we can employ a scenario with a tunnel effect from
``false'' to ``true'' vacuum and bubbles coalescence. In the last case the
time of an inflation (in the first de Sitter phase) depends on the
characteristic time of the coalescence.

Let us consider a more general model of the Universe filled with ordinary
(dust) matter, a radiation and a quintessence.

From Bianchi identity we have:
$$
\ddt \rho\dr{tot}+(\rho\dr{tot}+p\dr{tot})\frac{3\dot R}R=0 \eq30
$$
where
$$
\ealn{
\rho\dr{tot}&=\rho_Q + \t\rho_m+\t \rho_r &(3.31)\cr
p\dr{tot}&=p_Q+\t p_r. &(3.32)
}
$$
Let us suppose that the evolution of radiation, ordinary matter (barionic +
cold dark matter) and a \q\ are in\dt. In this way we get in\dt\ conservation
laws 
$$
\ealn{
&\ddt(R^3\t \rho_m)=0&(3.33)\cr
&\ddt(R^4\t \rho_r)=0&(3.34)
}
$$
and
$$
\dot\rho_Q+(\rho_Q+p_Q)\frac{3\dot R}R=0. \eq35
$$

One gets
$$
\ealn{
&\t \rho_m=\frac A{R^3},\quad A=\text{const.} &(3.36)\cr
&\t \rho_r=\frac B{R^4},\quad B=\text{const.} &(3.37)
}
$$
From Eq.~(3.35) we get
$$
\dot\rho_Q+(1+W_Q)\rho_Q\cdot\frac{3\dot R}R=0. \eq38
$$
Substituting Eqs (3.36--37) into Eqs (2.14) and (2.16), remembering Eqs
(2.114) and (2.116--117) one gets
$$
\ealn{
\(\frac{\dot R}R\)^2+k&= \frac1{3M^2\dr{pl}}\(\frac AR+\frac B{R^2}\)
+\frac1{3M^2\dr{pl}}\rho_Q R^2 &(3.39)\cr
\frac{3\ddot R}R&= -\frac1{M^2\dr{pl}}\(\frac A{2R^3}+\frac {2B}{3R^4}\)
-\frac1{2M^2\dr{pl}}(1+3W_Q)\rho_Q &(3.40)
}
$$
and from Eq.\ (2.118)
$$
\ddot Q+3H\dot Q-\frac{(n+2)A}{M^2\dr{pl}R^3}=0. \eq41
$$
However, now $Q$ and $U(Q)$ are redefined in such a way that
$\dfrac1{M^2\dr{pl}}$ factor appears before~$\rho_Q$. This is a simple
rescaling. We have of course
$$
\ealn{
\t\rho_m=e^{-(n+2)\varPsi}\rho_m&=e^{-aQ}\rho_m &(3.42)\cr
\t\rho_r=e^{-(n+2)\varPsi}\rho_r&=e^{-aQ}\rho_r. &(3.43)
}
$$
Thus we get
$$
\ealn{
&\rho_m=\frac A{R^3}e^{aQ} &(3.44)\cr
&\rho_r=\frac B{R^4}e^{aQ}. &(3.45)
}
$$

Let us consider Eqs (3.39--41) and Eq.~(3.35). Combining these equations
we get the following formula
$$
\frac{(n+2)A}{M^2\dr{pl}R^3}\sqrt{(1+W_Q)\rho_Q}=0 . \eq46
$$
The one way to satisfy this equation is to put $W_Q=-1$. It is easy to
see that if $A=0$ then Eq.~(3.46) is satisfied trivially. The condition
$A=0$ does not imply $B=0$ and we can still have $B\ne0$.
For such a solution $Q=Q_0=\text{const.}$ and $Q_0=\dfrac{\sqrt{\o M}}{M\dr{pl}}\varPsi_0$ found
earlier 
$$
e^{\varPsi_0}=x_0=\sqrt{\frac{n|\g|}{(n+2)\b}}=
\frac1{\a_s}\(\frac{m_{\u A}}{m\dr{pl}}\)\sqrt{\frac{n|\ul{\t P}|}
{(n+2)\t R(\t\G)}}. \eq47
$$
Thus
$$
e^{-aQ_0}=e^{-(n+2)\varPsi_0}=
\a_s^{n+2}\(\frac{m\dr{pl}}{m_{\u A}}\)^{n+2}\(\frac{n+2}n\)^\frac{n+2}2
\(\frac{\t R(\t\G)}{|\ul{\t P}|}\)^\frac{n+2}2 \eq48
$$
or
$$
e^{-aQ_0}=
\a_s^{n+2}\(\frac r{l\dr{pl}}\)\(\frac{n+2}n\)^\frac{n+2}2
\(\frac{\t R(\t\G)}{|\ul{\t P}|}\)^\frac{n+2}2. \eq49
$$

Let us put $W_Q=-1$ into Eqs (3.39--40). One gets
$$
\ealn{
{\dot R}^2+k&= \frac1{3M^2\dr{pl}}\(\frac AR+\frac B{R^2}
+\rho_Q R^2\) &(3.50)\cr
\frac{3\ddot R}R&= -\frac1{M^2\dr{pl}}\(\frac A{2R^3}+\frac {2B}{3R^4}
+\rho_Q\). &(3.51)
}
$$

From Eq.\ (3.50) one gets
$$
\frac{\pm R\,dR}{\sqrt{\frac{\rho_Q}{3M^2\dr{pl}}R^4
-kR^2+\frac{A}{3M^2\dr{pl}}R+\frac{B}{3M^2\dr{pl}}}}=dt \eq52
$$
if $A=0$,
$$
\frac{\pm R\,dR}{\sqrt{\frac{\rho_Q}{3M^2\dr{pl}}R^4
-kR^2+\frac{B}{3M^2\dr{pl}}}}=dt. \eq53
$$

Taking $x=R^2$ one gets
$$
\int\frac{dx}{\sqrt{\frac{\rho_Q}{3M^2\dr{pl}}x^2
-kx+\frac{B}{3M^2\dr{pl}}}}=\pm2(t-t_0) \eq54
$$
and finally
$$
\pm\int\frac{dy}{\sqrt{y^2-\frac{3kM^2\dr{pl}}{\sqrt{B\rho_Q}}y+1}}=
\frac{2\sqrt3\sqrt{\rho_Q}M\dr{pl}}B(t-t_0) \eq55
$$
where
$$
x=\sqrt{\frac B{\rho_Q}}y. \eq56
$$

For a flat case $k=0$ we find:
$$
\pm\int\frac{dy}{\sqrt{y^2+1}}=
\frac{2\sqrt3\sqrt{\rho_Q}M\dr{pl}}B(t-t_0). \eq57
$$
The last formula can be easily integrated and one gets
$$
R=\root 4 \of{\frac B{\rho_Q}} \sqrt{\Sh\(\frac{
2\sqrt3\sqrt{\rho_Q}M\dr{pl}}B(t-t_0)\)} \eq58
$$
where we take a sign $+$ in the integral on the left hand side of Eq.~(3.57).

Let us calculate a Hubble parameter (constant) for our model. One gets
$$
H=\frac{\dot R}R=\sqrt3 M\dr{pl}\(\frac{\sqrt{\rho_Q}}B\)
\ctgh\(\frac{2\sqrt3\sqrt{\rho_Q}M\dr{pl}}B(t-t_0)\). \eq59
$$
For a deceleration parameter one obtains
$$
q=-\frac{\ddot RR}{R^2}=-\frac{\(2\Sh^2(u)-\Ch(u)\)}
{\Ch(u)\Sh^\frac12(u)} \eq60
$$
where
$$
u=\frac{2\sqrt3\sqrt{\rho_Q}M\dr{pl}}B(t-t_0). \eq61
$$
Thus we get a spatially flat model of a Universe dominated by a radiation
which is expanding and an expansion is accelerating.

Let us take $k=0$ and $B=0$ in Eq.~(3.52). One gets
$$
\ealn{
&\frac{\pm R\,dR}{\sqrt{\frac{\rho_Q}{3M^2\dr{pl}}R^4
+\frac{A}{3M^2\dr{pl}}R}}=dt &(3.62)\cr
&R=\root3\of{\frac A{\rho_Q}}x,\quad x=R\root3\of{\frac{\rho_Q}A}\,.&(3.63)
}
$$
One finally gets
$$
\int\frac{x\,dx}{\sqrt{x(x^3+1)}}=
\pm\frac1{3M\dr{pl}}\sqrt{\rho_Q}(t-t_0). \eq64
$$
Let us notice that
$$
\ealn{
&H^2=\(\frac{\dot R}R\)^2=\frac1{3M^2\dr{pl}}\(\frac A{R^3}+\rho_Q\)>
\frac{\rho_Q}{M\dr{pl}} &(3.65)\cr
&\frac{\ddot R R}{{\dot R}^2}= \frac{(2\rho_QR^4-A)R}{2(A+\rho_QR^3)}\,.
&(3.66)
}
$$
Thus the model of the Universe is expanding and accelerating.

Let us consider the integral on the left hand side of Eq.~(3.64). One gets
after some tedious algebra:
$$
\al
&\int\frac{x\,dx}{\sqrt{x(x^3+1)}}=
\frac{\sqrt{\sqrt3 +2}}{12}\(9-2\sqrt3\)\ln W+
\frac{9\sqrt2}{52}\sqrt{4+\sqrt3}\(6\sqrt3+11\)\mPi\\
&-\frac{\sqrt6}{598}\(92\sqrt{4\sqrt3+3}\(141\sqrt3+272\)
+\(9\sqrt3-15\)\sqrt{598}\sqrt{16+9\sqrt3}\)F
\eal
\eq67
$$
where
$$
\al
&W=\left(-20-\dfrac{9\sqrt3}2+\dfrac{(16-7\sqrt3)}2
\(\dfrac{2x+\sqrt3-1}{2x-\sqrt3-1}\)^2 \vphantom{\sqrt{\sqrt{\Bigg[}}}\right.\\
&\left.\vphantom{\sqrt{\sqrt{\Bigg[}}}
{}+2\sqrt{6-\sqrt3}
\sqrt{\[\(\dfrac{(2x+\sqrt3-1)}{(2x-\sqrt3-1)}\)^2+7-4\sqrt3\]
\[\(\dfrac{(2x+\sqrt3-1)}{(2x-\sqrt3-1)}\)^2-1+\dfrac{\sqrt3}2\]}
\right)\\
&\qquad{}\times\(\(\dfrac{(2x+\sqrt3-1)}{(2x-\sqrt3-1)}\)^2-1\)^{-1},
\eal
\eq68
$$
$$
\mPi=\mPi\(\ar\(\(\frac{(2x+(\sqrt3-1))}{(2x-(\sqrt3+1))}\)\sqrt{2(2+\sqrt3)}\),
\(\sqrt3-1\),\sqrt{\frac{(5-2\sqrt3)}{13}}\) \eq69
$$
$$
F=F\(\ar\(\(\frac{(2x+(\sqrt3-1))}{(2x-(\sqrt3+1))}\)\sqrt{2(2+\sqrt3)}\),
\sqrt{\frac{(5-2\sqrt3)}{13}}\) \eq70
$$
where $\mPi$ is an elliptic integral of the third kind and $F$ is an elliptic
integral of the first kind
$$
\ealn{
F(u,k)&=\intop_0^u \frac{d\f}{\sqrt{1-k^2\sin^2\f}} &(3.71)\cr
k&=\sqrt{\frac{(5-2\sqrt3)}{13}} &(3.72)\cr
\mPi(u,n,k)&=\intop_0^u \frac{d\f}{(1-n\sin^2\f)\sqrt{1-k^2\sin^2\f}}
&(3.73)\cr
n&=\sqrt3-1&(3.74)
}
$$
and with the same modulus as $F(u,k)$
$$
k^2=\frac{5-2\sqrt3}{13}\,.
$$
However, the most important condition for an existence of a physical solution
is coming from the first part of a sum in the right hand side of
Eq.~(3.67). 

We should have
$$
W>0. \eq75
$$
In order to analyze condition (3.75) let us write $W$ in the following way
$$
W=\frac{-20-\frac{9\sqrt3}2+\frac{(16-7\sqrt3)}2y
+2\sqrt{6-\sqrt3}\sqrt{(y+7-4\sqrt3)\(y-1+\frac{\sqrt3}2\)}}{y-1}
\eq76
$$
where
$$
y=\(\frac{2x+\sqrt3-1}{2x-\sqrt3-1}\)^2. \eq77
$$
Obviously $y\ge 0$.

We have two possibilities:
$$
\displaylines{
\text{I} \hfill y>1 \hfill (3.78)
}
$$
and
$$
-20-\tfrac{9\sqrt3}2+\tfrac{(16-7\sqrt3)}2y
+2\sqrt{6-\sqrt3}\sqrt{(y+7-4\sqrt3)\(y-1+\tfrac{\sqrt3}2\)}>0
\eq79
$$
$$
\displaylines{
\text{II} \hfill y<1 \hfill (3.80)
}
$$
and
$$
-20-\tfrac{9\sqrt3}2+\tfrac{(16-7\sqrt3)}2y
+2\sqrt{6-\sqrt3}\sqrt{(y+7-4\sqrt3)\(y-1+\tfrac{\sqrt3}2\)}<0.
\eq81
$$

The first possibility can easily be solved:

$$\displaylines{
\text{I} \hfill y>4.0612 \hfill (3.82)
}
$$
or for the second possibility
$$\displaylines{
\text{II} \hfill \frac{2-\sqrt3}2 \le y < 1. \hfill (3.83)
}
$$
We have finally
$$
\ealn{
0.7916 &< x<\frac{\sqrt3+1}2\,,&(3.84)\cr
\frac{3\sqrt3-5}2 &< x < \frac12. &(3.85)}
$$
For 
$$
x>\frac{\sqrt3+1}2 \eq86
$$
we get
$$
x<3.072. \eq87
$$

Let us notice the following: the function $W$ has a finite limit at
$x=\frac{\sqrt3+1}2$. Thus we can remove a singularity at
$x=\frac{\sqrt3+1}2$. In this way we get
$$
0.7916 < x< 3.072 \eq88
$$
or
$$
\frac{3\sqrt3-5}2 < x < \frac12. \eq89
$$
This simply means that the solution exists only for
$$
0.7916 \root3\of{\frac A{\rho_Q}}< R < 3.072 \root3\of{\frac A{\rho_Q}}
\eq90
$$
or
$$
\frac{3\sqrt3-5}2 \root3\of{\frac A{\rho_Q}} <R< \frac12 \root3\of{\frac A{\rho_Q}}\,. \eq91
$$

Let us calculate a density of a \q\ energy:
$$
\al
\rho_Q&=M^2\dr{pl}U(Q_0)=-\frac12e^{n\varPsi_0}\(\b e^{2\varPsi_0}-|\g|\)\\
&=\(\frac n{n+2}\)^\frac n2 \(\frac{m_{\u A}}{\a_s^{2(n+1)}}\)
\(\frac{m_{\u A}}{m\dr{pl}}\)^n \(\frac {|\ul{\t P}|}{\t R(\t \G)}\)^\frac n2 |\ul{\t P}|.
\eal
\eq92
$$
An energy density of a matter is equal to
$$
\rho_m=\frac A{R^3} e^{aQ_0} \simeq \rho_Q e^{aQ_0} \eq93
$$
and
$$
\frac{\rho_m}{\rho_Q}\cong e^{aQ_0}=
\(\frac{1}{\a_s}\)^{n+2} \(\frac{m_{\u A}}{m\dr{pl}}\)^{n +2}
\(\frac n{n+2}\)^\frac {n+2}2 \(\frac {\t R(\t \G)}{|\ul{\t P}|}\)^\frac {n+2}2 .
\eq94
$$

Let us consider the behaviour of the ``effective'' \gr\ constant in our model
$$
\al
G\dr{eff}&=G_N e^{-(n+2)\varPsi_0}=G_N e^{-aQ_0}= G_N\(\frac1{x_0}\)^{n+2}\\
&=G_N \a_s^{2(n+2)}
\(\frac{m\dr{pl}}{m_{\u A}}\)^\frac{n+2}2
\(\frac{\t R(\t \G)}{|\ul{\t P}|}\)^\frac{n+2}2 \(\frac{n+2}n\)^\frac{n+2}2
\eal
\eq95
$$
Thus $G\dr{eff}$ is a constant. For we are living in that model we should
rescale the constant and consider $G\dr{eff}(Q_0)$ (Eq.~(3.95)) a Newton
constant. Thus
$$
G_N=G_0\a_s^{2(n+2)}\(\frac{m\dr{pl}}{m_{\u A}}\)^\frac{n+2}2
\(\frac{n+2}n\)^\frac{n+2}2 \(\frac{\t R(\t \G)}{|\ul{\t P}|}\)^\frac{n+2}2
\eq96
$$
where $G_0$ is a different constant responsible for a strength of \gr\
interactions in earlier epochs of an evolution of the Universe. It means we
should write
$$
G\dr{eff}=
\(G_N \a_s^{-2(n+2)}
\(\frac{m_{\u A}}{m\dr{pl}}\)^\frac{n+2}2
\(\frac n{n+2}\)^\frac{n+2}2 \(\frac {|\ul{\t P}|}{\t R(\t \G)}\)^\frac{n+2}2 \)\cdot
e^{-(n+2)\varPsi}. \eq97
$$
The obtained solution (i.e.\ (3.67)) is for $W_Q=-1$. However, this is not an
attractor of the dynamical equations. This is similar to the tracker solution
of Steinhardt (see Ref.~[9]). Moreover we cannot apply his method because
our equation for a scalar field~$\varPsi$ (or~$Q$) is different. It contains an
additional term with a trace of the energy-momentum tensor (matter +
radiation). Only with a radiation filled Universe our equation is the same (if
we identify $\t\rho_r$ with Steinhardt density of a matter). In this case we
could apply Steinhardt tracker solution method and apply his criterion for a
potential $U(Q)$. In our case $U(Q)$ is given by Eq.~(2.116)
$$
\ealn{
U'(Q)&=-\frac{M\dr{pl}}{2\sqrt{\o M}}\((n+2)\b e^{aQ} - n|\g|e^{bQ}\) &(3.98)\cr
U''(Q)&=-\frac{M\dr{pl}^2}{4\sqrt{\o M}}\((n+2)^2\b e^{aQ} - n^2|\g|e^{bQ}\). &(3.99)
}
$$

The Steinhardt criterion consists in finding
$$
\G=\frac{U''U}{(U')^2}=\frac
{\((n+2)^2\b e^{aQ} - n^2|\g|e^{bQ}\)\(\b e^{aQ}-|\g|e^{bQ}\)}
{\((n+2)\b e^{aQ} - n|\g|e^{bQ}\)^2}\,.
\eq100
$$
To get a tracker solution we need
$$
\G\ge 1. \eq101
$$
In our case
$$
\ealn{
\G&=1-\frac{4\b|\g|e^{(a+b)Q}}{\((n+2)\b e^{aQ}-n|\g|e^{bQ}\)^2} &(3.102)\cr
\G&\simeq 1- \frac{4|\g|}{(n+2)\b}e^{(b-a)Q}= 1-\frac{4|\g|}{\b(n+2)}
e^{-\frac{2M\dr{pl}}{\sqrt{\o M}}Q}=1-\frac{4|\g|}{\b(n+2)}
e^{-2\varPsi}.\qquad &(3.103)
}
$$

Thus asymptotically we are close to $\G=1$. Moreover the Steinhardt
criterion is only a sufficient condition. Our radiation filled Universe seems
to be unstable due to appearing of a matter (a~dust matter). Moreover a
quintessential period of a Universe model has a severe restrictions of a
value of a radius. Thus the solution with $W_Q=-1$ and a matter in\dt ly
evolving cannot evolve forever.

In order to answer a question what is a further evolution of the Universe we
come back to Einstein equations with a scalar field~$\varPsi$ and matter
sources. Supposing as usual a Robertson-Walker metric and a spatial flatness of
the metric we write once again a Bianchi identity
$$
\(\frac{\br{scal}T^{\mu \nu }}{M^2\dr{pl}}+\frac1{M^2\dr{pl}}
e^{-(n+2)\varPsi}\br{matter}T^{\mu \nu }\)_{;\nu }=0 \eq104
$$
where
$$
\br{matter}T^{\mu \nu }=\rho_m u^\mu u^\nu \eq105
$$
is an energy-momentum tensor for a dust matter. One gets
$$
\al
\frac1{M^2\dr{pl}} e^{-(n+2)\varPsi}&\(\rho_m+\dot \rho_m+\rho_m\frac{3\dot R}R
-(n+2)\rho_m\dot\varPsi\)\\
&+\frac{d\la_{c0}}{d\varPsi}\dot\varPsi
-\frac{\o M}{M^2\dr{pl}}\dot\varPsi \ddot\varPsi-\frac{\o M}{M^2\dr{pl}}{\dot\varPsi}^2
\frac{3\dot R}R=0.
\eal
\eq106
$$
We make the following ansatz concerning an evolution of a matter density
$$
\ealn{
\rho_m&=\rho_0e^{(n+2)\varPsi}&(3.107)\cr
\rho_0&=\text{const.} &(3.108)
}
$$
In that moment the scalar field $\varPsi$ and the matter $\rho_m$ interact
nontrivially. 

Using Eqs (3.107--108), (3.106) and an equation for a scalar field $\varPsi$
one obtains
$$
\frac{\rho_0}{M^2\dr{pl}}-\frac{(n+2)}{M^2\dr{pl}}\rho_0+
\frac{\rho_0}{M^2\dr{pl}}\frac{3\dot R}R=0. \eq109
$$
Using Eq.~(3.109) and Eq.~(2.102) one gets
$$
\ealn{
&\frac{d\varPsi}{dt}=\pm\frac{\sqrt2 M\dr{pl}}{\sqrt{3\o M}}
\sqrt{\(\frac{n+1}{M\dr{pl}}\)^2-\frac{3\rho_0}{M^2\dr{pl}}
-\frac32\la_{c0}} &(3.110)\cr
&\int \frac{dx}{x\sqrt{\d+3\b x^{n+2}+3\g x^n}}=\pm
\frac{M\dr{pl}}{\sqrt{\o M}}(t-t_0) &(3.111)
}
$$
where
$$
\ealn{
\d&=2\(\frac{n+1}{M\dr{pl}}\)^2 - \frac{3\rho_0}{M^2\dr{pl}} &(3.112)\cr
x&=e^\varPsi. &(3.113)
}
$$

Using Eq.~(3.111) and Eq.~(2.102) one gets
$$
\ddt \ln R=\frac{(n+1)}{M\dr{pl}} \eq114
$$
or
$$
R=R_0 e^{(\frac{n+1}{M\dr{pl}})(t-t_0)}.\eq114a
$$
Let us consider Eq.~(3.111) and let us suppose that $x<1$ (and practically
small). In this case we can simplify
$$
\int \frac{dx}{x\sqrt{\d+3\b x^{n+2}+3\g x^n}}\simeq
\int \frac{dx}{x\sqrt{\d+3\g x^n}} \eq115
$$
and finally we get
$$
\varPsi=\frac1{2n} \ln\(\frac\d{3|\g|}\)-\frac1n\sqrt{\frac{2\d}{3\o M}}M\dr{pl}
(t-t_0). \eq116
$$
Thus our prediction that $x<1$ and is small has been justified. Using
Eq.~(3.116) one easily gets
$$
\ealn{
p_\varPsi &= \frac{\d M^2\dr{pl}}{3n^2}+\frac{\sqrt{|\g|\d}}{2\sqrt3}
\exp\(-\sqrt{\frac{2\d}{3\o M}}M\dr{pl}(t-t_0)\) &(3.117)\cr
\rho_\varPsi &= \frac{\d M^2\dr{pl}}{3n^2}-\frac{\sqrt{|\g|\d}}{2\sqrt3}
\exp\(-\sqrt{\frac{2\d}{3\o M}}M\dr{pl}(t-t_0)\). &(3.118)
}
$$
It is easy to see that
$$
\frac{p_\varPsi}{\rho_\varPsi}\underset{t\to\infty}\to\longrightarrow 1. \eq119
$$
Thus in this model we have to do asymptotically (practically very quickly)
with a stiff matter equation of state:
$$
p=\rho. \eq120
$$

Simultaneously an energy density for a scalar field~$\varPsi$ is dominated by a
kinetic energy which is practically constant and
$$
\frac12{\dot Q}^2 \cong \frac{\d M^2\dr{pl}}{3n^2}\,. \eq121
$$
In this sense a dark matter generated by $\varPsi$ in the model is in some sense
$K$-essence (not a \q).

Let us calculate an energy density for a matter
$$
\rho_m=\rho_0\(\frac\d{3|\g|}\)^\frac{(n+2)}{2n}
\exp\(-\frac{(n+2)}n \sqrt{\frac{2\d}{3\o M}} M\dr{pl}(t-t_0)\). \eq122
$$
The effective \gr\ constant reads
$$
G\dr{eff}=G_N \(\frac{3|\g|}\d\)^\frac{(n+2)}n \exp\(
\(\frac{n+2}n\)\sqrt{\frac{2\d}{3\o M}}M\dr{pl}(t-t_0)\). \eq123
$$
It is easy to write $R$ as a function of $\varPsi$:
$$
R=R_0 \exp\(\frac{n+1}{M^2\dr{pl}}\sqrt {\o M} \intop_{x_0}^{e^\varPsi}
\frac{dx}{x\sqrt{\d+3\b x^{n+2}+3\g x^n}}\) \eq124
$$
where
$$
x_0=\(\frac{3|\g|}\d\)^{2n}. \eq125
$$
Eq.\ (3.124) can be easily reduced to the Eq.\ (3.114) using Eq.~(3.111).

Thus an energy density of matter goes to zero in an exponential way. The
effective \gr\ constant is growing exponentially. What does it mean for the
future of the Universe? First of all the \U\ will be very diluted after some
time of such an evolution. Secondly, a relative strength of \gr\ interactions
will be stronger. This means that if we take substrat particles as cluster of
galaxies then in a quite short time any cluster will be a lonely island in
the \U\ without any communication with other clusters. All the clusters will
be beyond a horizon. On the level of a single cluster the strength of \gr\
interactions will be stronger and eventually they collapse to form a black
hole. In an intermediate time a \gr\ dynamics on the level of galaxies'
clusters will be governed by a Newtonian dynamics (nonrelativistic) with a
\gr\ potential corrected by a \co\ constant. The \co\ constant induces a
positive pressure (a~stiff matter) and will be important up to a moment of
sufficiently large \gr\ constant. After a sufficiently long time only a
classical Newtonian gravity will drive a dynamics of galaxies' clusters. Let
us roughly estimate this effect. In order to do this let us suppose that
$$
\rho_\varPsi=p_\varPsi=\frac{\d M^2\dr{pl}}{3n^2}=\text{const.} \eq126
$$
In an energy momentum we get
$$
\br{scal}T^{\mu \nu }=2\cdot \frac{\d M^2\dr{pl}}{3n^2} u^\mu u^\nu - 
\frac{\d M^2\dr{pl}}{3n^2} g^{\mu \nu }. \eq127
$$

However, we have in the place of Newtonian constant $G_N$ a $G\dr{eff}$ which
is a function of \co\ time and in the equation for \gr\ field for~$g_{\mu \nu }$
we should put
$$
\frac1{(M\dr{pl}\ur{eff})^2}\br{scal}T^{\mu \nu }=
\frac\d{3n^2}\(\frac{M\dr{pl}}{M\dr{pl}\ur{eff}}\)^2
\(2u^\mu u^\nu -g^{\mu \nu }\) \eq128
$$
where
$$
\al
M\dr{pl}\ur{eff}=\frac1{\sqrt{8\pi G\dr{eff}}}&=
M\dr{pl}\(\frac{\a_s}{m\dr{pl}m_{\t A}}\)^{n+2}
\(\frac{2(n+1)^2-3\rho_0}{3|\uP|}\)^\frac{n+2}2\\
&\times\exp\[\(\frac12\(\frac{n+2}2\)\sqrt{\frac{2\d}{3\o M}}\)M\dr{pl}(t-t_0)\].
\eal
\eq129
$$
Thus
$$
\frac1{(M\dr{pl}\ur{eff})^2}\br{scal}T^{\mu \nu }=
\t Ae^{-\k (t-t_0)}\(2u^\mu u^\nu -g^{\mu \nu }\) \eq130
$$
where
$$
\ealn{
&\k=\frac12\(\(\frac{n+2}2\)\sqrt{\frac{2\d}{3\o M}}\) &(3.131)\cr
&\t A=\frac{3n^2}\d \(\frac{\a_s}{m\dr{pl}m_{\t A}}\)^{-(n+2)}
\(\frac{2(n+1)^2-3\rho_0}{3|\uP|}\)^{-(\frac{n+2}2)}. &(3.132)
}
$$

Consider the following case (the justification will be given below):
$$
\frac1{(M\dr{pl}\ur{eff})^2}\br{scal}T^{\mu \nu }=
-\t A e^{-\k(t-t_0)}g^{\mu \nu }. \eq133
$$
Now let us solve the following problem. Let $\o \La$ be a value 
$\t Ae^{-\k(\bar t-t_0)}$ in an established \co\ time $t=\o t$ and the same
for $\o G=G\dr{eff}(\o t)$. Let us consider a \gr\ field of a point
mass~$M_0$ static and spherically symmetric. Consider a metric
$$
ds^2=B(r)dt^2-\frac{dr^2}{B(r)}-r^2\(d\theta^2+\sin^2\theta \,d\theta^2\)
\eq134
$$
in spherical coordinates and Einstein equations
$$
R_{\mu \nu }-\frac12g_{\mu \nu }R=-\o\La g_{\mu \nu } \eq135
$$
for the metric (3.134) where
$$
\o\La=\t Ae^{-\k(\bar t-t_0)}. \eq136
$$
Via a standard procedure one obtains
$$
B(r)=1-\frac{2\o GM_0}r -\frac{\o \La r^2}3. \eq137
$$
Thus an effective nonrelativistic Newton-like \gr\ potential has a shape
$$
V(r)=\frac{\o GM_0}r + \frac{\o\La r^2}6\,.\eq138
$$
Remembering that $\o G$ is growing exponentially in time and $\o\La$ is
decaying in time exponentially we see that the second term is important only
for a short time. For a contemporary time
$$
\o\La\simeq 10^{-52} \text{m}^{-2} \eq139
$$
according to the Perlmutter data and it could be important on the level of a
size of galaxies' cluster. However, in a short time it will be negligible
even on that level. Let us calculate a Schwarzschild radius for a mass $M_0$
with an effective \gr\ constant $G\dr{eff}$. One gets
$$
\al
r\ur{eff}_g&=\sqrt{2M_0\o G}=\sqrt{2M_0 G\dr{eff}(\o t)}
=r_g\(\frac{m\dr{pl}m_{\u A}}{\a_s}\)^{n+2}
\\
&\times \(\frac{3|\uP|}{2(n+1)^2-3\rho_0}\)^\frac{n+2}2\exp
\[\(\frac12\(\frac{n+2}2\)\sqrt{\frac{2\d}{3\o M}}\)M\dr{pl}(t-t_0)\]
\eal
\eq140
$$

Thus $r\ur{eff}_g$ (calculated for an effective \gr\ constant $G\dr{eff}(\o
t)$) is growing exponentially. It means that after some time it will be of an
order of size of galaxies' cluster. Then galaxies' clusters collapse to form
black holes. The time of a collapse is easy to calculate.
$$
\ealn{
r\ur{eff}_g &= r\dr{cluster} &(3.141)\cr
t\dr{collapse} &=t_0+(n+2)\sqrt{6\o M}
\Biggl[\ln\(\dfrac{r\dr{cluster}}{r_g}\)
+(n+2)\ln\(\dfrac{m\dr{pl}m_{\t A}}{\a_s}\)\cr
&\quad{}+
\(\dfrac{n+2}2\)\ln\(\dfrac{3|\uP|}{2(n+1)^2-3\rho_0}\)\Biggr]
\cdot\[\sqrt\d (n+2)M\dr{pl}\]^{-1} &(3.142)
}
$$
where $r_g$ is a \gr\ radius of a mass of a galaxies' cluster with $G=G_N$. 

Let us calculate a Hubble constant and a deceleration parameter for our
model. One gets
$$
\ealn{
H&=\frac{\dot R}R =(n+1)M\dr{pl} &(3.143)\cr
q&=-\frac{\ddot RR}{{\dot R}^2}=-1.&(3.144)
}
$$
Summing up we get the following scenario of an evolution. The \U\ starts in a
``false'' vacuum state and its stable evolution is driven by a ``\co\
constant'' formed from $\t R(\t \G)$, $\o P$, $V(\varPhi^1\dr{crt})$ and $\varPsi_1$
being a minimal solution to an effective selfinteracting potential
for~$\varPsi$. This evolution is stable against small \pt\ for~$\varPsi$ and~$R$.
The evolution is \ex\ and a Hubble constant is calculable in terms of our
theory (the nonsymmetric Jordan--Thiry theory). This evolution ends at a
moment the ``false'' vacuum state changes into a ``true'' vacuum state. In
that moment an energy of a vacuum is released into radiation (and a matter).
The \U\ is reheating and a big-bang scenario begins at a very hot stage.
Moreover the evolution is still governed by a scalar field~$\varPsi$ which
attains a new minimum of an effective potential. The effective potential is
different and a new value of~$\varPsi$, a~$\varPsi_0$, is different too. The
evolution is \ex\ and a new Hubble constant~$H_1$ is also calculable in terms
of the underlying theory. In this background we have an evolution of a
multiplet of scalar fields~$\varPhi^a_{\u m}$ from $\varPhi^1\dr{crt}$ to
$\varPhi^0\dr{crt}$. This goes to a spectrum of fluctuations calculable in the
theory. After that time the field~$\varPsi$ is slowly changing $\dot\varPsi\approx
0$ and the temperature of the \U\ is going down. The evolution of a radiation
is in the second inflationary stage adiabatic and afterwards during next phase
($\dot\varPsi\simeq0$) nonadiabatic. The last means there is an energy exchange
between radiation and a scalar field~$\varPsi$. The total amount of an inflation
$\o N\dr{tot}$ is calculable in the theory. The evolution of a factor $R(t)$ is
governed by a simple elementary function of a small rise. After this the
radiation condenses into a matter (a~dust with zero pressure) and a matter
(a~dust), a radiation and a scalar field evolve adiabatically without
interaction among them. The evolution of a scalar field is governed by a \q,
i.e.\ $p_Q=W_Q\rho_Q$, where $W_Q\cong-1$. 

The \q\ field $Q$ is a normalized scalar field~$\varPsi$. During a radiation era
($\dot\varPsi\simeq0$) an equation of state for a scalar field~$\varPsi$ is
different: $p_\varPsi=\frac43\rho_\varPsi$ (similar to stiff matter equation of
state). The scale factor $R(t)$ evolves according to some elementary function
built from hyperbolic function in radiation + \q\ scenario and according to
complicated function formed from the first and the second elliptic integrals
(after reversing of those functions). It expands and accelerates. The
solution with radiation and \q\ has a restricted behaviour (Eq.~(3.90--91)).
The energy density of radiation is going to zero. Only a \q\ energy density
is constant. The energy of a \q\ is a potential energy dominated. However, the
solution cannot be continued up to the end for an interaction between a
matter and a scalar field starts to be important. The effective \gr\ constant
remains constant during the period. Let us remind to the reader that during
\il ary epochs the effective \gr\ constant was constant. Only during a
radiation era it was slightly changing. In order to keep the \U\ to evolve
forever it is necessary to find a solution with an interaction between a
matter and a scalar field~$\varPsi$. Using an appropriate ansatz for an \ev\ of
an energy density of a matter we get such a solution. The scale factor $R(t)$
is \ex ly expanding. 

An energy density of a matter is going to zero and an energy density of a
scalar field is approaching (\ex ly) a constant. However, now a scalar
field~$\varPsi$ forms a different form of a matter---a stiff matter with an
equation of a state $\rho_\varPsi\simeq p_\varPsi$ ($W_\varPsi\simeq1$). The effective
\gr\ constant is growing \ex ly in time. In contradistinction with the
previous period of an \ev\ the scalar field~$\varPsi$ forms a matter with a
kinetic energy dominance (i.e.\ $K$-essence). In the previous period we have
to do with a potential energy dominance (i.e.\ \q). 

Let us give some remarks
on a spatial curvature of a model of the \U. In the first period (\il ary
scenario) the spatial curvature has to be zero and any \pt s of initial
conditions cannot change it. In next periods it is natural to suppose that it
is not changed. Thus our model of a \U\ is spatially flat. In the moment of
change of a phase of an \ev\ we should apply a matching conditions for an
energy density and for a scale factor (a~radius of a model of a \U). However,
an \ev\ in such moments undergoes phase transitions for a law of an \ev\ of a
scalar field and a scale factor, which can be considered as a second order or
even a first order phase transition. It is worth to notice that in formulae
concerning Hubble constants, deceleration parameters, \gr\ constants
(effective) meet some parameters from unification theory of fundamental
interactions and from cosmology. It seems that we are on a right track to
give an account for large number hypothesis by P.~Dirac. This demands of
course some investigations especially in developing nonsymmetric Kaluza--Klein
(Jordan--Thiry) theory. It is also interesting to mention that the scalar
field~$\varPsi$ plays many r\^oles in our theory. It works as an \il\ field during
an \il\ epoch. (However Higgs' fields $\varPhi^a_{\u b}$ play an important
r\^ole in creating fluctuations spectrum.) It plays also a r\^ole of a \q\
and a $K$-essence, except its influence on an effective \gr\ constant and on
an effective scale of masses in unification of fundamental interactions. This
field is massive getting masses from several mechanisms. During \il\ phases
it acquires mass due to spontaneous symmetry breaking, different in both
epochs. It gets a mass from a \co\ background.
Thus except its \co\ importance the fluctuations of a scalar field
around its \co\ value could be detected (in principle) as massive scalar
particles of a large mass. In this way an effective \gr\ potential in
nonrelativistic limit takes a form
$$
V(r)=\frac{G_NM_0}r \(1-\frac\a r e^{-(\frac r{r_0})}\)e^{\k(t-t_0)}
+\frac{\t A r^2}6e^{-\k(t-t_0)}
\eq145
$$
where $\k,\t A$ are given by Eqs (3.131--132), $r_0$ is a range of massive
scalar $\d\varPsi$ (a~fluctuation of a scalar field~$\varPsi$ in a \co\
background). This could be detected as a tiny change of a Newton constant in
\gr\ law (e.g.\ the fifth force) on short distances. Let us notice that on
short distances and on short time scales an \ef\ \gr\ potential reads:
$$
V(r)=\frac{G_NM_0}r\(1-\frac \a r \,e^{-(\frac r{r_0})}\). \eq146
$$
On large distances and short time scales 
$$
V(r)=\frac{G_NM_0}r+\frac{\o \La r^2}6. \eq147
$$
On short distances and large time scales 
$$
V(r)=\frac{G_NM_0}r\(1-\frac\a r\,e^{-(\frac r{r_0})}\)e^{\k(t-t_0)}. \eq148
$$
On intermediate distances and large time scales 
$$
V(r)=\frac{G_NM_0}r e^{\k(t-t_0)}. \eq149
$$
The latest being a cause to collapse a cluster of galaxies into a black hole.
It seems also possible to consider an \ef\ coupling \ct~$\a_s\ur{eff}$
for~$\a_s$ enters~$m_{\u A}$ and~$m_{\u A}\ur{eff}$. However
$$
m\ur{eff}_{\u A} = e^{-\varPsi}\(\frac{\a_s}r\), \eq150
$$
$r$ is a radius of a manifold $G/G_0$ (e.g.\ a radius of a sphere~$S^2$). If
we rewrite Eq.~(3.150) as
$$
\frac{m\ur{eff}_{\u A}}{\a_s\ur{eff}} = \frac1r e^{-\varPsi}, \eq151
$$
it would be quite difficult to distinguish a drift of $m\ur{eff}_{\u A}$ from
$\a_s\ur{eff}$ drift. Only a quotient has a definite dependence on the scalar
field~$\varPsi$. Especially it is interesting to consider a drift of $\a\dr{em}$
(a~fine structure \ct) reported from several sources. However, without
extending our theory to include fermion (spinor) fields this is impossible.
Thus it is necessary to construct a nonsymmetric-geometric version with
supersymmetry and supergravity including noncommutative (anticommutative)
coordinates to settle the problem. The scalar field~$\varPsi$ can form an
infinite tower of massive scalar fields which could have important
astrophysical and high-energy consequences.

Let us consider Eqs (5.2.51) and (5.2.56) from the first point of Ref.~[1],
p.~318--319, and changing a base in the Lie algebra
$\gH$ we get
$$
\mu^a_{\hi}=\d^a_\hi \eq152
$$
and
$$
\a^c_i=\d^c_i. \eq153
$$
Simultaneously for $G\subset H$ ($\gG\subset \gH$) we get
$$
\ealn{
f^\hi_{\u a\u b}&= C^\hi_{\u a\u b}&\text{(3.154a)}\cr
f^{\u d}_{\u a\u b}&= C^{\u d}_{\u a\u b}&\text{(3.154b)}
}
$$
where $\u a,\u b,\u c$ refer to the complement $\gM$ in $\gG$ ($\gG=\gG_0
\dot+ \gM$), $\hi,\hj,\hat k$ to the subalgebra $\gG_0$ of~$\gG$. Let us
suppose a symmetry requirement
$$
[\gM,\gM]\subset\gG_0. \eq155
$$
From Eq.\ (3.155) it follows that
$$
C^{\u a}_{\u b\u c}=f^{\u a}_{\u b\u c}=0. \eq156
$$
Eq.\ (2.5.51) looks like
$$
\varPhi^c_{\u b}C^{\u b}_{\hi\u a}=\varPhi^b_{\u a}C^c_{\hi b}. \eq157
$$
Eq.\ (2.5.55)
$$
H^c_{\u a\u b}=C^c_{ab}\varPhi^a_{\u a}\varPhi^b_{\u b}- C^c_{\u a\u b}
-\varPhi^c_{\u d}C^{\u d}_{\u a\u b}. \eq158
$$
Let us come back to Eq.\ (3.1).

For our \il\ driving agent is a scalar field~$\varPsi$ we do not carry any
backreaction of Higgs' fields and we can consider some nonstandard scenarios
of an \ev\ of those fields. Thus we can consider an \ev\ which is not a
slow-roll \ev. Let us suppose that
$$
\ddt\(\varPhi^a_{\u m}\)\simeq0. \eq159
$$
In this way we get from (3.159)
$$
\ddt\({L^d}{}^0_{\u b}\)_{av}+\frac{e^{n\varPsi_1}}{2r^2}l^{db}\kl
\frac{\d V'}{\d\varPhi^b_{\u n}}g_{\u b\u n}\kr. \eq160
$$
The next step in simplification is such that we collapse all the degrees of
freedom of Higgs' fields into one
$$
\varPhi^a_{\u m}=\d^a_{\u m}\varPhi. \eq161
$$
In this way constraints (3.157) are satisfied trivially and Eq.~(3.158) is
as follows:
$$
H^c_{\u a\u b}=C^c_{\u a\u b}\(\varPhi^2-1-\varPhi\). \eq162
$$
In this way $V=V'$ (no constraints) and
$$
V=\frac A{\a^2_s}\(\a^2_s\varPhi^2-1-\a_s\varPhi\)^2 \eq163
$$
where
$$
A=\[l_{ab}\(2g^{[\u m\u n]}g^{[\u a\u b]}C^c_{\u m\u n}C^b_{\u a\u b}
-C^b_{\u m\u n}g^{\u a\u n}g^{\u b\u m}\t L^a_{\u a\u b}\)\]_{av} \eq164
$$
and
$$
\ealn{
&L^a_{\u a\u b}=\t L^a_{\u a\u b}\(\a_s\varPhi^2-\frac1{\a_s}-\varPhi\) &(3.165)\cr
&l_{dc}g_{\u m\u b}g^{\u c\u m}\t L^d_{\u c\u a}+
l_{cd}g_{\u a\u m}g^{\u m\u c}\t L^d_{\u b\u c}=
2l_{cd}g_{\u a\u m}g^{\u m\u c}C^d_{\u b\u c}. &(3.166)
}
$$

We can get the following identity
$$
l_{dc}g^{\u c\u q}g^{\u a\u p}\t L^d_{\u c\u a}C^c_{\u p\u q}+
l_{cd}g^{\u p\u c}g^{\u q\u b}\t L^d_{\u b\u c}C^c_{\u p\u q}=
2l_{cd}g^{\u p\u c}g^{\u q\u b}C^d_{\u b\u c}C^c_{\u p\u q}. \eq167
$$
Using Eq.\ (3.160) we get the following equation for $\varPhi$:
$$
\frac{d^2\varPhi}{dt^2} + \frac{e^{n\varPsi_1}}{2r^2}\frac{\d V}{d\varPhi}=0 \eq168
$$
if the following constraints are satisfied
$$
\int \sqrt{|\t g|}d^nx \(g_{\u q\u c}\t l^{\u p\u e}+l^{\u e\u n}
g_{\u c\u n}\d^{\u p}_{\u q}\)=2V_2\d^e_{\u c}\d^{\u p}_{\u q}. \eq169
$$
Thus from Eq.\ (3.168) one gets a first integral of motion
$$
\frac{{\dot\varPhi}^2}2+\frac{e^{n\varPsi_1}}{2r^2}V=\frac B2=\text{const.} 
\eq170
$$

From Eq.\ (3.170) one gets
$$
\intop_{\varPhi_0}^\varPhi \frac{d\varPhi}
{\sqrt{-\dfrac{e^{n\varPsi_1}}{r^2\a_s^2} A\(\a_s^2\varPhi^2-\a_s\varPhi-1\)^2+B}}
=\pm(t-t_0) \eq171
$$
and finally
$$
\intop_{\f_0}^\f \frac{d\f}{\sqrt{(a-\f^2+\f+1)(a+\f^2-\f-1)}}
=\pm\sqrt{\frac{Ae^{n\varPsi_1}}{r^2}}(t-t_0) \eq172
$$
where
$$
\ealn{
a&=\a_s\sqrt{\frac{Br^2}{Ae^{n\varPsi_1}}} &(3.173)\cr
\varPhi&=\frac1{\a_s}\f. &(3.174)
}
$$

In order to find an integral on the left hand side of Eq.~(3.172) and its
inverse function we use a uniformization theory of algebraic curves. Let
$$
f(\f)=(a-\f^2+\f+1)(a+\f^2-\f-1)=-\f^4+2\f^3+\f^2-2\f+(a^2-1). \eq175
$$
One easily notices that
$$
\f_0=\frac{1+\sqrt{5+4a}}2 \eq176
$$
is a root of the polynomial $f(\f)$ (a real root).

In this way one gets
$$
\f=\frac{f'(\f_0)}{4\(P(z)-\frac1{24}f''(\f_0)\)}+\f_0 \eq177
$$
where
$$
P(z)=P(z;g_2,g_3) \eq178
$$
is a $P$ Weierstrass function with invariants:
$$
\ealn{
&g_2=\frac{5-3a^2}{3} &(3.179)\cr
&g_3=\frac{a^2}{12}+\frac{10}{27}\,. &(3.180)
}
$$

One easily finds
$$
\ealn{
f'(\f_0)&=\tfrac12 (5+3a)\sqrt{5+4a} &(3.181)\cr
f''(\f_0)&=-2(5+6a) &(3.182)
}
$$
and
$$
z=\int_{\f_0}^\f [f(x)]^{-\frac12}\,dx. \eq183
$$
In this way one gets
$$
\varPhi(t)=\frac{(5+3a)\sqrt{5+4a}}{8\a_s\(P(z)+\frac1{12}(5+6a)\)}
+\frac{1+\sqrt{5+4a}}{2\a_s} \eq184
$$
where
$$
z=\pm\sqrt{\frac{Ae^{n\varPsi_1}}{r^2}}(t-t_0). \eq185
$$

Let us come back to the potential $V$ Eq.\ (3.163) and let us look for its
critical points. One gets
$$
\frac{dV}{d\varPhi}=\frac{2A}{\a_s}\(\a_s^2\varPhi^2-\a_s\varPhi-1\)(2\a_s\varPhi-1) \eq186
$$
and from
$$
\frac{dV}{d\varPhi}=0 \eq187
$$
easily finds
$$
\ealn{
\varPhi_1&=\frac{1}{2\a_s} &\text{(3.188a)}\cr
\varPhi_2&=\frac{1-\sqrt5}{2\a_s} &\text{(3.188b)}\cr
\varPhi_3&=\frac{1+\sqrt5}{2\a_s} &\text{(3.188c)}
}
$$
One obtains
$$
\ealn{
V(\varPhi_1)&=\frac{25A}{16\a^2_s} &\text{(3.189a)}\cr
V(\varPhi_2)&=0 &\text{(3.189b)}\cr
V(\varPhi_3)&=0. &\text{(3.189c)}
}
$$
It is easy to see that $\varPhi_1$ is a maximum and $\varPhi_2$ and~$\varPhi_3$ are
minima of the potential~$V$.

However, in our simplified model of an \ev\ of Higgs' fields we have not 
a second (local)
minimum. Thus in order to mimic a real situation we start an \ev\ of a Higgs
field from the value
$$
\varPhi=0. \eq190
$$
In this case
$$
V(0)=\frac A{\a_s^2}\,.\eq191
$$
Thus
$$
\ealn{
\varPhi(t\dr{initial})&=0 &(3.192)\cr
\varPhi(t\dr{end})&=\frac{1+\sqrt5}{2\a_s}\,. &(3.193)
}
$$
Coming back to Eq.\ (3.172) we get
$$
\intop_{0}^\frac{1+\sqrt5}{2} \frac{d\f}{\sqrt{(a-\f^2+\f+1)(a+\f^2-\f-1)}}
=\sqrt{\frac{Ae^{n\varPsi_1}}{r^2}}(t\dr{end}-t\dr{initial}). \eq194
$$

Thus the amount of \il\ obtained in our \ev\ is equal to
$$
\o N_0=H_0\sqrt{\frac {r^2}{Ae^{n\varPsi_1}}}
\intop_{0}^\frac{1+\sqrt5}{2} \frac{d\f}{\sqrt{(a-\f^2+\f+1)(a+\f^2-\f-1)}}\,.
\eq195
$$
Supposing simply
$$
a<\frac54 \eq196
$$
one gets
$$
\o N_0= H_0\sqrt{\frac {2r^2}{5Aae^{n\varPsi_1}}}
\intop_{\f_2}^{\f_1}\frac{d\theta}{\sqrt{1-\(\frac{4a+5}{10}\)\sin^2\theta}}
\eq197
$$
where
$$
\ealn{
\cos^2\f_1&=\frac1{5+4a} &(3.198)\cr
\cos^2\f_2&=\frac5{5+4a} &(3.199)
}
$$
or
$$
\o N_0=H_0\sqrt{\frac {2r^2}{5Aae^{n\varPsi_1}}}
\intop_{\f_2}^{\f_1}\frac{d\theta}{\sqrt{1-\(\frac{4a+5}{10}\)\sin^2\theta}}
\eq200
$$
or
$$
\o N_0=\frac{H_0}{\sqrt{5\a_s}}\root4\of{\frac {4r^2}{ABe^{n\varPsi_1}}}
\intop_{\f_2}^{\f_1}\frac{d\theta}{\sqrt{1-\(\frac{4a+5}{10}\)\sin^2\theta}}\,.
\eq201
$$
In order to get 60-fold \il
$$
\o N_0\simeq 60 \eq202
$$
we should play with parameters in our theory. Let us notice that in our
simple model the constant $\a_1$ equals
$$
\a_1=m^2_{\u A}\(\frac{m_{\u A}}{m\dr{pl}}\)^2\(\frac A{\a^6_s}\)=
\frac1{r^2}\(\frac{l\dr{pl}}r\)^2\(\frac A{\a^6_s}\) \eq203
$$
(see Eqs (2.8--9a).

Let us consider a slow-roll dynamic of Higgs' field in this simplified model.
From Eq.~(3.9) one gets
$$
\ddt\(\varPhi^f_{\u p}\)=-\frac{e^{n\varPsi_1}}{12V_2r^2H_0}
\(l^{fb}g_{\u p\u n}+l^{bf}g_{\u n\u p}\)\frac{\d V'}{\d \varPhi^b_{\u n}}\,.
\eq204
$$
Using our simplifications from Eqs (3.152--162) and Eqs (3.143--149) one
gets 
$$
\frac{d\varPhi}{dt}=-\frac{e^{n\varPsi_1}}{12r^2H_0}\frac{dV}{d\varPhi} \eq205
$$
supposing a condition
$$
\d^f_{\u p}=l^{f\u n}g_{\u p\u n}+l^{\u nf}g_{\u n\u p} \eq206
$$
or
$$
\frac{d\varPhi}{dt}+\frac{e^{n\varPsi_1}A}{6\a_sr^2H_0}
\(\a_s^2\varPhi^2-\a_s\varPhi-1\)\(2\a_s\varPhi-1\)=0.
\eq207
$$
Changing the \dt\ variable into
$$
\f=\a_s\varPhi \eq208
$$
one gets
$$
\frac{d\f}{dt}+b\(\f^2-\f-1\)(2\f-1)=0 \eq209
$$
where
$$
b=\frac{e^{n\varPsi_1}A}{6r^2H_0}>0. \eq210
$$
From Eq.\ (3.209) one finds
$$
\frac{\f^2-\f-1}{(\f-\frac12)^2}=e^{-5b(t-t_0)} \eq211
$$
and finally
$$
\varPhi(t)=\frac1{2\a_s}\[1\pm\sqrt{\frac5
{1-e^{-5b(t-t_0)}}}\].
\eq212
$$

One easily notices that for $(t-t_0)\to\infty$
$$
\varPhi(t)\to \frac1{\a_s}\(\frac{1\pm\sqrt5}2\). \eq213
$$
It simply means that a slow-roll \il\ never ends.

If we suppose that an \il\ starts at $\f=3$, i.e.\ for
$$
t\dr{initial}=t_0+\frac1{5b}\ln\frac54\,, \eq214
$$
it is evident that to complete it we need an eternity.

Summing up we get in our simplified model a finite amount of an \il\ for a
nonstandard dynamic of a Higgs field and an infinite amount of \il\ for a
slow-roll dynamic. The truth probably is in a middle.

Probably it would be necessary to consider a full equation for a Higgs field
in a simplified model
$$
\frac{d^2\varPhi}{dt^2}-3H_0\frac{d\varPhi}{dt}-3bH_0
\(\a^2_s\varPhi^2-\a_s\varPhi-1\)\(2\a_s\varPhi-1\)=0 \eq215
$$
or
$$
\frac{d^2\f}{dt^2}-3H_0\frac{d\f}{dt}-3b\a_s H_0
\(\f^2-\f-1\)\(2\f-1\)=0. \eq215a
$$
Changing an in\dt\ variable from $t$ to $\tau=H_0t$ one gets
$$
\frac{d^2\f}{d\tau^2}-3\frac{d\f}{d\tau}-3\frac {b\a_s}{H_0}
\(\f^2-\f-1\)\(2\f-1\)=0. \eq215b
$$
Let
$$
\frac{3b\a_s}{H_0}=\o a \eq216
$$
and let us change a \dt\ variable $\f$ into
$$
\ealn{
z&=2\f-1 &(3.217)\cr
\f&=\frac{z+1}2\,. &(3.218)
}
$$
One gets
$$
\frac{d^2z}{d\tau^2}-3\frac{dz}{d\tau}-\frac{\o a}4 z^3-\frac{(15\o a)}2z=0. 
\eq219
$$

Now let us take a special value for $\o a=-\frac4{15}$ and let us change $z$ into~$r$:
$$
z=i\sqrt5r. \eq220
$$
One gets
$$
\frac{d^2r}{d\tau^2}+3\frac{dr}{d\tau}-2r^3+2r=0. \eq221
$$
This equation can be solved in general [10]
$$
r(\tau)=iC_1e^\tau \sn(C_1e^\tau+C_2,-1). \eq222
$$
Thus
$$
z(\tau)=-\sqrt5C_1e^\tau \sn(C_1e^\tau+C_2,-1). \eq223
$$

$\sn(u,-1)$ is a Jacobi elliptic function with a modulus $K^2=-1$, $C_1$
and~$C_2$ are constants. In this way
$$
\varPhi(t)=\frac1{2\a_s}\(1-\sqrt5C_1e^{H_0t}\sn(C_1e^{H_0t}+C_2,-1)\). \eq224
$$
However, we are forced to put $\o a=-\frac4{15}$. Let us remind that
$$
\o a=\frac{e^{n\varPsi_1}}{2r^2H_0^2}A \eq225
$$
and we are supposing $A>0$. However (see Eq.~(3.164) for a definition
of~$A$), it is possible to consider $A<0$ only for a pleasure to play. Thus we
use formula (3.224) to consider a dynamics of a Higgs field. 
Let us start an \il\ for $t=0$ with $\varPhi=0$.
$$
\varPhi(0)=0. \eq226
$$
One finds
$$
\frac1{C_1\sqrt5}=\sn(C_1+C_2,-1) \eq227
$$
(i.e.\ $t\dr{initial}=0$).

Let
$$
\varPhi(t\dr{end})=\frac1{\a_s}\(\frac{1+\sqrt5}2\) \eq228
$$
(a true minimum---a ``true'' vacuum). One gets
$$
\frac{1}{2}=-C_1e^{H_0t\dr{end}}\sn\(C_1e^{H_0t\dr{end}}
+C_2,-1\). \eq229
$$

Let us calculate the derivative of $\varPhi(t)$. One gets
$$
\al
\frac{d\varPhi}{dt}&=-\frac{\sqrt5}{2\a_s}C_1H_0e^{H_0t}
\Bigl(\sn\(C_1e^{H_0t}+C_2,-1\)\\
&+\cn\(C_1e^{H_0t}+C_2,-1\)\cdot\dn\(C_1e^{H_0t}+C_2,-1\)\Bigr).
\eal
\eq230
$$

Let us suppose a slow movement of $\varPhi$ such that
$$
\frac{d\varPhi}{dt}(0)=0. \eq231
$$
From (3.231) one gets
$$
\sc(C_1+C_2,-1)=-\dn(C_1+C_2,-1). \eq232
$$

This gives us an equation for a sum of integration \ct s. Thus we can get
from Eq.~(3.232) and Eq.~(3.227) integration \ct s $C_1$ and~$C_2$ and from
Eq.~(3.229) $t\dr{end}$, which can give us an amount of \il
$$
\o N_0=H_0t\dr{end}. \eq233
$$
In some sense Eq.~(3.229) is an equation for an amount of \il
$$
\frac{1}{2}=-C_1e^{\bar N_0}\sn\(C_1e^{\bar N_0}+C_2,-1\). \eq234
$$

Using Eq.~(3.232) and Eq.~(3.227) one easily gets
$$
C_1=-\sqrt{\frac{5-\sqrt5}{10}} \simeq -0.5256. \eq235
$$
One gets
$$
C_1+C_2=\frac1{\sqrt2}\(F(90^\circ\backslash 45^\circ)
-F(32^\circ\backslash 45^\circ)\)=0.9431 \eq236
$$
where $F$ is an elliptic integral of the first kind and
$$
\arcsin\(\sqrt{\frac{5-\sqrt5}{10}}\)\simeq\arcsin(0.5256)\simeq32^\circ. 
\eq237
$$

Let us come back to Eq.~(3.234) and let us denote
$$
e^{\bar N_0}=x_0. \eq238
$$
One gets using (3.236) and (3.235)
$$
C_2\simeq 1.4687 \eq239
$$
and finally
$$
1.4866x_0-5.0354=F\(\f/45^\circ\),
\eq240
$$
where 
$$
\cos\f=\frac{1.9026}{x_0}\,. \eq241
$$

From Eq.~(3.240) one finds
$$
\frac{2.8284}{\cos\f}-5.0354=\intop_0^\f \frac{d\theta}
{\sqrt{1-\frac12\sin^2\theta}}\,. \eq242
$$
We come back to this equation later.

Thus we get the following results. The Higgs field evolves from a ``false''
vacuum value ($\varPhi=0$) to the ``true'' vacuum value ($\varPhi=\varPhi_0$)
completing a second order phase transition. The potential of selfinteracting
$\varPsi$~field changes and a new equilibrium $\varPsi_0$ is attained. However, the
field~$\varPsi$ evolves from~$\varPsi_0$ to new value a little different
from~$\varPsi_0$ and afterwards a radiation era starts. The change of the value
of a scalar field~$\varPsi$ is small in terms of a variable~$y$ (see Eq.~(2.69)
for a definition) only from $\sqrt{\frac n{n+2}}$ to~1 ($n$~is a natural number
$n=\dim H$, and $H$ is at least~$G2$, $\dim G2=14$). Let us consider an
\ev\ of a field~$\varPsi$ in this epoch. From Eq.~(2.59) we get
$$
\ddot\f+\om^2_0 \f=0 \eq243
$$
where
$$
\ealn{
\om^2_0&=\frac{M^2\dr{pl}}{2\o M}(n+2)\(\frac n{n+2}\)^\frac n2
|\g|\(\frac{|\g|}\b\)^\frac n2 &(3.244)\cr
\varPsi&=\varPsi_0+\f &(3.245)
}
$$
and we linearize Eq.~(2.59) around~$\varPsi_0$ neglecting a term with a first
derivative of~$\f$. Thus a scalar field~$\varPsi$ undergoes small oscillations
around the equilibrium~$\varPsi_0$. These oscillations cannot be too long for a
field should change from
$$
\varPsi_0=\ln\(\sqrt{\frac{n|\g|}{(n+2)\b}}\) \eq246
$$
to
$$
\o\varPsi=\ln\(\sqrt{\frac{|\g|}{\b}}\). \eq247
$$
In terms of a field $\f$ from $\f=0$ to $\f=\frac12\ln(1+\frac2n)\simeq
\frac1n$. Thus we have
$$
\f=\f_0\sin(\om_0t). \eq248
$$

For $t=0$, $\f=0$ the time $\D t$ to go from $\f=0$ to $\f=\frac1n$ is simply
equal to
$$
\D t=\frac{\left|\arcsin(\frac1{\f_0n})\right|}{\om_0}
\le \frac\pi{2\om_0}\,. \eq249
$$
Thus the full amount of an \il\ in this epoch is equal to
$$
N_1=\D tH_1=\frac{H_1}{\om_0}\left|\arcsin\(\frac1{\f_0n}\)\right|
\le \frac{\pi H_1}{2\om_0}=\frac{\pi M\dr{pl}}{2\sqrt{6\o M}(n+2)}\,.
\eq250
$$
In our notation
$$
t_1=t_r+\D t \eq251
$$
(see Eq.~(2.89)).

After a time $t_1$ the \U\ goes to the phase of radiation domination with a
strong interaction between a radiation and a scalar field~$\varPsi$ up to a
minimal temperature where an ordinary (a~dust) matter appears. This matter
evolves afterwards in\dt ly of a radiation and of a scalar field~$\varPsi$. The
scalar field now is evolving as a \q.

In order to get some comparison let us consider an \ev\ of a scalar
field~$\varPsi$ during the second de Sitter phase in a slow-roll approximation.
Using Eq.~(2.59) one gets
$$
\frac{dy}{dt}=\frac{\o By^\frac{(n-2)}2(y^2-(\frac n{n+2}))}
{\sqrt{1-y^2}} \eq252
$$
or
$$
\int \frac{\sqrt{1-y^2}}{y^{(\frac{n-2}2)}(y^2-(\frac n{n+2}))}\,dy
=\o B(t-t_0) \eq253
$$
where
$$
\o B=2(n+2)|\g| \(\frac n{n+2}\)^\frac n2\(\frac{|\g|}\b\)^\frac n2 \eq254
$$
and
$$
y=\sqrt{\frac\b{|\g|}}\cdot e^\varPsi. \eq255
$$

Let
$$
I=\int \frac{\sqrt{1-y^2}\,dy}{y^{(\frac{n-2}2)}(y^2-(\frac n{n+2}))}
\simeq \int \frac{\sqrt{1-y^2}\,dy}{(y^2-(\frac n{n+2}))}
\eq256
$$
for
$$
\ealn{
&\sqrt{\frac n{n+2}} \le y \le 1 &(3.257)\cr
&n>\dim G2=14. &(3.258)
}
$$
For the integral $I$ one finds:
$$
\al
I&=-\arcsin(y)+\frac1{\sqrt{n(n+2)}}\sqrt{1-y^2}\\
&+\(\sqrt{\frac 2n}+\sqrt{\frac n2}\)\frac1{n+2} \cdot
\ln\[\frac{2\[\(1+y\sqrt{\frac n{n+2}}\)\(y+\sqrt{\frac n{n+2}}\)
+\sqrt{\frac 2{n+2}}\sqrt{1-y^2}\]}{\(y+\sqrt{\frac n{n+2}}\)}\]\\
&-\frac1{n+2}\(\sqrt{\frac 2n}+\sqrt{\frac n2}\) \cdot
\ln\[\frac{2\[\(1-y\sqrt{\frac n{n+2}}\)\(y-\sqrt{\frac n{n+2}}\)
+\sqrt{\frac 2{n+2}}\sqrt{1-y^2}\]}{\(y-\sqrt{\frac n{n+2}}\)}\].
\eal
\eq259
$$
In order to find an amount of an \il\ let us find limits of~$I$ for $y=1$ and
$y=\sqrt{\frac n{n+2}}$. One finds
$$
\ealn{
\lim_{y\to1}I &= -\frac\pi2+\(\sqrt{\frac 2n}+\sqrt{\frac n2}\)
\frac{\ln\(2\(1+\sqrt{\frac n{n+2}}\)\)}{(n+2)}\cr
&\hphantom{{}=-\frac\pi2}-\(\sqrt{\frac 2n}+\sqrt{\frac n2}\)
\frac{\ln\(2\(1-\sqrt{\frac n{n+2}}\)\)}{(n+2)} &(3.260)\cr
\lim_{y\to\sqrt{\frac n{n+2}}} I &=-\infty. &(3.261)
}
$$
Thus we see that a slow-roll approximation offers an infinite in time
\ev\ of a field~$y$ from $\sqrt{\frac n{n+2}}$ to~1. It is similar to an
\ev\ of a Higgs field in some slow-roll approximation schemes. The \il\ never
ends. Let us come back to the Eq.~(2.110) and let us calculate a Hubble \ct\
and a deceleration parameter for this model
$$
\ealn{
H&=\frac{\dot R}R=\(\frac{2|\g|^{n+4}}{3\o M\b^n(n-1)}\)
\frac{(t-t_1)^3}{\(1+\frac{|\g|^{n+4}}{3\o M\b^n(n-1)}
(t-t_1)^4\)} &(3.262)\cr
q&=-\frac{\ddot R R}{{\dot R}^2}= -\frac23\(\frac1{(t-t_1)^{4}}
+\frac{|\g|^{n+4}}{9\o M\b^n}\). &(3.263)
}
$$

Let us compare Eq.~(3.262) with the similar equation for radiation dominated
\U\ in General Relativity
$$
H\dr{GR}=\frac2{3(t-t_1)} \eq264
$$
and let us calculate a speed up factor
$$
\D_1=\frac H{H\dr{GR}}=\frac{|\g|^{n+4}}{\o M\b^n(n-1)}
\frac{(t-t_1)^4}{\(1+\frac{|\g|^{n+4}}{3\o M\b^n(n-1)}
(t-t_1)^4\)} \eq265
$$
and let us do the same for a model Eq.~(3.58) (i.e.\ radiation dominated
with a \q). Using Eq.~(3.59) one gets
$$
\D_2=\frac H{H\dr{GR}} = \frac{3\sqrt3 m\dr{pl}\rho_Q(t-t_1)}
{2B\tgh\(\frac{2\sqrt3\rho_Q}B m\dr{pl}(t-t_1)\)}\,. \eq266
$$
Let us notice that for a $(t-t_1)\sim0$
$$
\D_1\simeq 0 \eq267
$$
and
$$
\lim_{t\to\infty}\D_1=+\infty. \eq268
$$

Thus at early stages model Eq.~(2.110) is slower than that in General
Relativity. For large time the behaviour depends on details of the theory. In
the case of model Eq.~(3.58) we have
$$
\lim_{t\to t_1}\D_2=\tfrac34 \eq269
$$
and
$$
\lim_{t\to\infty}\D_2=+\infty. \eq270
$$
Thus in the first case $\D$ is monotonically going from~0 to
$+\infty$ and in the second case from $\frac34$ to infinity.
According to modern ideas an expansion rate in early \U\ has an important
influence on a production of light elements, the so called primordial
abundance of light elements (see Ref.~[11]). If at a beginning of primordial
nucleosynthesis the \U\ expansion rate is slower than in~GR, then we have
${}^4$He underproduction. This can be balanced by considering a larger ratio
of a number density of barions to number density of photons (remember the
barion number is a conserved quantity)
$$
\eta=\frac{n_B}{n_\g} \eq271
$$
where $n_B,n_\g$---barion and photon density numbers for a high redshift
$z=10^{10}$. In contrast to the latest if an expansion rate became faster
than in~GR during nucleosynthesis process, those bigger~$\eta$ (traditionally
one uses the so called) $\eta_{10}$
$$
\eta_{10}=10^{10}\eta \eq272
$$
do not result in excessive burning of deuterium because this happens in a
shorter time. The standard model of big-bang nucleosynthesis (SBBN) demands
$$
3\le \eta_{10}\le 5.6. \eq273
$$
It seems in the light of observational data from cosmic microwave background
(CMB) (BOOMERANG and MAXIMA) and from the Lyman $\a$-forest that
$\eta_{10}=8.8\pm1.4$, significantly higher than the SBBM value (for CMB) and
$8.2\pm2$ or $12.3\pm2$ for Lyman $\a$-forest. Thus the second model could
help in principle to explain the data without modification of nuclear
reaction rates due to neutrino degeneracy or introducing new decaying
particles. As the \U\ expands, cools and becomes more dilute, the nuclear
reactions cease to create and destroy nuclei. The abundances of the light
nuclei formed during this epoch are determined by the competition between a
time available (an expansion rate) and a density of reactans: neutrons and
protons. 

The abundances of D, ${}^3$He, ${}^7$Li are limited by an expansion rate and
are determined by the competition between the nuclear production/destruction
rates and an (universal) expansion of the \U. The SBBN theory is based on a
flat radiation dominated \U\ model in General Relativity and found in a
laboratory nuclear reaction rates. Thus it is strongly constrained. Any
significant discrepancy between observed and calculated value of~$\eta_{10}$
(known as baryometry) could be very dangerous. Thus changing the ratio of an
universal expansion relative to the model of \GR\ can give some margin in a
primordial alchemy.

In our theory after two \il ary phases we have in principle two radiation
dominated phases described by Eq.~(2.110) and Eq.~(3.58) with speed up
factors (3.265) and~(3.266). However, the important point is to match those
models. We get
$$
\al
R(t_d)&=R_0\exp\(H_0t_r+H_1(t_1-t_r)\)\(1+\frac{\o r}{(n-1)}\eta^2\dr{min}\)^\frac12\\
&=\root4\of{\frac B{\rho_Q}}\(\sh\(\frac{2\sqrt3\sqrt{\rho_Q}}B
m\dr{pl}\(t_1-\o t_0+\frac{\sqrt{2\o M}\b^\frac n4}{|\g|^{\frac{n+4}4}}\eta\dr{min}^\frac12
\)\)\)^\frac12 
\eal
\eq274
$$
and
$$
\al
B&=\rho_r(t_d)R^3(t_d)=\rho_r\(t_1+
\frac{\sqrt{2\o M}\b^\frac n4}{|\g|^{\frac{n+4}4}}\eta\dr{min}^\frac12\)R^3\(t_1+
\frac{\sqrt{2\o M}\b^\frac n4}{|\g|^{\frac{n+4}4}}\eta\dr{min}^\frac12\)\\
&=
\rho\dr{min}R_0^3\exp\(3H_0t_r+3H_1(t_1-t_r)\)\(1+\frac{\o r}{(n+1)}
\eta\dr{min}^2\)^\frac32 
\eal
\eq275
$$
where $\rho\dr{min}$ is given by
$$
\rho\dr{min}=M^2\dr{pl}\frac{\b^{n+1}}{|\g|^n}\(1+\eta\dr{min}\)^{2(n+1)}
W_4(1+\eta\dr{min}), \eq276
$$
$W_4(y)$ is given by formula (2.90) and $\eta\dr{min}$ by formula (2.97).

The time $t_r=t\dr{end}$ in any \il\ model for an \ev\ of the Higgs field in
the first de Sitter phase. The time $t_1=t_r+\D t$ can be calculated from
Eq.~(3.249) 
$$
\D t=\frac{\arcsin(\frac1{\f_0n})}{\om_0} \eq277
$$
and is bounded by
$$
\D t\le \frac1{\om_0}=\frac1{M\dr{pl}|\g|^\frac12}
\sqrt{\frac{2\o M}{(n+2)}}\(\frac{n+2}n\)^\frac n4\(\frac\b{|\g|}\)^\frac n4.
\eq278
$$
From Eq.\ (3.274) we can get the constant $\o t_0$
$$
\o t_0=t_1+\frac{\sqrt{2\o M}\b^\frac n4}{|\g|^{\frac{n+4}4}}\eta\dr{min}
^\frac12 -T \eq279
$$
where
$$
T=\frac{B\ars\[R_0^2\sqrt{\frac{\rho_Q}B}\exp\(2H_0t_r+2H_1(t_1-t_r)\)
\(1+\frac{\o r}{(n-1)}\eta^2\dr{min}\)\]}
{2\sqrt3\sqrt{\rho_Q}M\dr{pl}} \eq280
$$
and $B$ is given by Eq.~(3.275).

Let us notice that for $t=t_d$ the \U\ undergoes a phase transition due to a
change in an equation of state for a scalar field~$\varPsi$ from
$p_\varPsi=\frac43\rho_\varPsi$ in a first radiation dominated phase to
$p_\varPsi=-\rho_\varPsi$ in a second radiation dominated phase (a~\q\ phase). 

A matter which appears in the second phase does not play any r\^ole in an
\ev\ of the \U. For $t=t_d$ we have a discontinuity in Hubble constants
(parameters) for both phases
$$
H(t_d)=\frac{2|\g|^{n+4}}{3\o M(n-1)\b^n} \cdot 
\frac{(t_d-t_1)^\frac32}{\(1+\frac{\o r}{(n-1)}\eta\dr{min}
^2\)} \eq281
$$
in the first phase and
$$
\o H(t_d)=\sqrt3 m\dr{pl} \(\frac{\sqrt{\rho_Q}}B\)
\ctgh\(\frac{2\sqrt3}B \sqrt{\rho_Q} m\dr{pl} T\) 
\eq282
$$
for the second one,
$$
H(t_d)\ne \o H(t_d) \eq283
$$
$$
\rho_Q=\frac12\la_{c0}(\varPsi_0)=
\frac{|\g|^{\frac n2+1}n^\frac n2}{\b^\frac n2(n+2)^{\frac n2+1}} . \eq284
$$

Let us notice that a speed up factor $\D_1$ changes from
$$
\D_1(t_1)=0 \eq285
$$
to
$$
\D_1(t_d)=\frac{|\g|^{n+4}}{\o M(n-1)\b^n}\cdot\frac
{(t_d-t_1)^4}{\(1+\frac{\o r}{(n-1)}
\eta^2\dr{min}\)} \eq286
$$
and an expansion rate can be slower than in \GR.

The speed up factor $\D_2$ has a more complicated behaviour than~$\D_1$. It
is natural to expect a rapid speed up of a rate of expansion resulting in
non-equilibrium nuclear synthesis of light elements. Thus the presented
scenario offers interesting possibilities for an \ev\ of early \U. Recently
some papers have appeared on scalar-tensor theories of gravitation exploiting
the idea of a speed up factor in primordial nucleosynthesis~[12], [13].

Finally let us come back to the model (3.64), (3.67).
We cannot invert analytically the formula~(3.67). Moreover taking under
consideration Eqs (3.85--88) and making some natural simplifications we
come to the following approximative formulae for $R(t)$.

For $x<1.1969$ (Eq.\ (3.86))
$$
R(t)=\root3\of{\frac A{\rho_Q}}\(0.44 + 0.085 N\) \eq287
$$
where
$$
N=\exp\(0.374 \frac{\sqrt{\rho_Q}}{M\dr{pl}}(t-t_0)+157.93\). \eq288
$$
One can calculate a Hubble parameter and a deceleration parameter
and gets:
$$
\ealn{
H&=\frac{\dot R}R = \frac{\sqrt{\rho_Q}}{M\dr{pl}}
\frac{3.17N}{44+8.5N} & (3.289)\cr
-q&=\frac{\ddot R R}{{\dot R}^2}=\frac{5.21}{N}+1. &(3.290)
}
$$
We have $H>0$, and $q<0$. Thus the model expands and
accelerates. First of all we consider
$$
\rho_m=\frac A{R^3} e^{aQ} 
$$
and we take boundary for $R$. In this way one gets
$$
\rho_m=a_i \cdot \rho_Q e^{aQ}, \quad i=1,2,3,4, \eq291
$$
$$
\al
a_1&=2.015\\
a_2&=0.034\\
a_3&=1060\\
a_4&=8
\eal
\eq292
$$
If we reconsider a contemporary \gr\ \ct~$G_N$ as $G_Ne^{-(n+2)\varPsi_0}$ we
would get interesting ratios. Thus we get
$$
\frac{\rho_m}{\rho_Q}=0.034 \div 2.015 \hbox{\quad or\quad }8\div 1060 \eq293
$$
which for the first is an excellent agreement with recent data concerning an acceleration
of the \U. Solving Eq.~(3.288) for $N$ if $R=1.33\sqrt{\frac A{\rho_Q}}$
gives us
$$
\ealn{
N&=10.47 &(3.294)\cr
H&=\frac{\sqrt{\rho_Q}}{M\dr{pl}}\cdot 0.25 &(3.295)\cr
-q&=1.5. &(3.296)
}
$$

Using Eq.\ (3.288) and Eq.\ (3.295) we can estimate a time of our
contemporary epoch
$$
t=-\frac 1h \cdot10^{15}\text{yr} - t_0 \eq297
$$
where $h$ is a dimensionless Hubble parameter, $H=h\cdot H_0$ (we take for a
contemporary Hubble parameter $H_0=100\frac{\text{km/s}}{\text{Mps}}$), 
$0.7<h<1$, which seems to be too much.

We can also estimate a density of a \q
$$
\rho_Q=\frac{H^2}{G_N}\cdot 0.571=h^2\cdot 0.08\cdot 10^{-23}
\frac{\text{kg}}{\text{m}^3}=8h^2\cdot 10^{-29}
\frac{\text{g}}{\text{cm}^3}\,. \eq298
$$

Let us come back to the Eqs (3.177) and (3.184). In Eq.~(3.184) a Higgs
field is given by a $P$~Weierstrass function. This function can be expressed
by Jacobi elliptic function
$$
P(z)=e_3+(e_1-e_2)\ns^2\(z(e_1-e_2)^\frac12\) \eq299
$$
with
$$
K^2=\frac{e_2-e_3}{e_1-e_3} \eq300
$$
where $e_1,e_2,e_3$ are roots of the polynomial
$$
4x^3-g_2x-g_3. \eq301
$$
Thus we should solve a cubic equation
$$
x^3-\frac1{12} \(5-3a^2\)-\frac1{12} \(\frac{a^2}{4}+\frac{10}9\)=0.
\eq302
$$
We want Eq.\ (3.302) to have all real roots. Thus we need
$$
p=-\frac1{12} \(5-3a^2\)<0 \eq303
$$
and
$$
D=\frac{p^3}{27}+\frac{q^2}4<0 \eq304
$$
where
$$
q=-\frac1{12}\(\frac{a^2}4+\frac{10}9\).\eq305
$$
Both conditions (3.303) and (3.304) are satisfied if
$$
0<a<0.3115\,. \eq306
$$
In this case we get
$$
\ealn{
e_1&=\frac13 \sqrt{5-3a^2} \cos\frac\f3 &(3.307)\cr
e_2&=\frac13 \sqrt{5-3a^2} \cos\frac{\f+2\pi}3 &(3.308)\cr
e_3&=\frac13 \sqrt{5-3a^2} \cos\frac{\f+4\pi}3 &(3.309)
}
$$
where
$$
\cos\f=-\frac{(9a^2+40)}{12(5-3a^2)^{3/2}}\,. \eq310
$$
From (3.307--309) one gets
$$
\ealn{
&K^2=\frac{\sin\frac\f3}{\sin\frac{\f+2\pi}3} &(3.311)\cr
&e_1>e_2>e_3 &\text{(3.312)}
}
$$
Thus
$$
\varPhi(t)=\frac{(5+3a)\sqrt{5+4a}}{8\a_s(A_1\ns^2(u)+A_2)}
+\frac{\(1+\sqrt{5+4a}\)}{2\a_s} \eq313
$$
where
$$
\ealn{
u&=\root4\of{\frac{Ae^{n\varPsi_1}\(5Ae^{n\varPsi_1}-3\a_s^2r^2\)}{r^4}}
\sqrt{\sin\frac{\f+\pi}3}(t-t_0) &(3.314)\cr
A_1&=e_1-e_2=\sqrt{\frac53-a^2}\,\sin\frac{\f+\pi}3 &(3.315)\cr
A_2&=e_3+\frac1{12}(5+6a)=\sqrt{\frac53-a^2}\,
\cos\frac{\f+4\pi}3+\frac1{12}(5+6a)&(3.316)
}
$$
and
$$
\ealn{
\cos\f&=-\frac{\sqrt A e^{\frac n2\varPsi_1}
\(9\a_s^2Br^2+40Ae^{n\varPsi_1}\)}{12\(5Ae^{n\varPsi_1}-3\a_s^2Br^2\)^{3/2}}
&(3.317)\cr
&0<\sqrt{\frac{\a_s^2Br^2}{Ae^{n\varPsi_1}}}<0.3115 &(3.318)
}
$$
and $a$ is given by Eq.\ (3.173).

The function $\ns(u,K)$ is an elliptic Jacobi function with a modulus~$K$
given by Eq.~(3.311)
$$
\ealn{
\ns(u,K)&=\frac1{\sn(u,K)} &(3.319)\cr
\sn(u,K)&=\operatorname{sinam}(u,K). &(3.320)
}
$$
In order to justify our treatment of Eqs (3.130) and (3.133) we consider
the full field equations
$$
R_{\mu \nu }-\frac12g_{\mu \nu }R=2\o\La u_\mu u_\nu - \o\La g_{\mu \nu } \eq321
$$
(where $\o\La$ is given by Eq.\ (3.136)).

However, in this case we consider $\o\La$ arbitrary. Let us consider static and
spherically symmetric metric
$$
ds^2=e^v\, dt^2-e^\la\, dr^2-r^2\(d\theta^2+\sin^2\theta\,d\f^2\) \eq322
$$
where $v=v(r)$ and $\la=\la(r)$.

From Eqs (3.321--322) one gets
$$
\ealn{
r^2\o\La &= e^{-\la}(rv'+1)-1 &\text{(3.323)} \cr
r^2\o\La &= e^{-\la}(r\la'-1)+1 &\text{(3.324)} \cr
\o\La' &= -\o\La v' &\text{(3.325)}
}
$$
where $'$ means derivation with respect to~$r$. Summing up (3.323) and
(3.324) one gets
$$
e^{-\la}(\la'+v')=2\o\La r. \eq326
$$
Moreover in our case $\o\La$ is very small. Thus we get approximately
$$
\frac{d\o\La}{dr}\simeq 0 \eq327
$$
and
$$
\la' \cong -v'. \eq328
$$
In this way we go to the solution (3.137) for a small $\o\La$ which is our
case. In this way $\la=-v$ and
$$
B(r)=e^v=e^{-\la}. \eq329
$$

Let us check a consistency of our solution. In order to do this we consider
Eqs (3.323--325) in full. One gets from Eq.~(3.325)
$$
\o\La=e^{-\mu_0}e^{-v}. \eq330
$$
For $\o\La$ is very small we should suppose that $\mu_0$ is positive and very
large ($\mu_0\to\infty$). From Eqs~(3.323--324) one gets
$$
\frac d{dr}(\la+v)=2e^{-\mu_0}\cdot r \cdot e^{(\la-v)}. \eq331
$$
Supposing that
$$
\la(r_0)+v(r_0)=0 \eq332
$$
for an established $r_0>0$ ($r_0$ is greater than a Schwarzschild radius of
the mass~$M_0$ from the solution (3.137)). Taking sufficiently big~$\mu_0$
we get approximately
$$
\frac d{dr}(\la+v)\simeq0 \eq333
$$
and of course
$$
\la\cong -v. \eq334
$$

Let us come back to the Eq.\ (3.242) and consider it for
$$
\f=k\pi+\tfrac\pi2 -\e, \quad k=0,\pm1,\pm2,\ldots, \eq335
$$
where $\e$ is considered to be small. One gets
$$
\frac{2.8284(-1)^k}{\sin\e}-5.0354=
\intop_0^{\frac\pi2-\e} \frac{d\theta}{\sqrt{1-\frac12\sin^2\theta}}
+2kK\(\tfrac12\), \eq336
$$
where
$$
K\(\tfrac12\)=\intop_0^{\frac\pi2}\frac{d\theta}{\sqrt{1-\frac12\sin^2\theta}}
=1.8541 \eq337
$$
is a full elliptic integral of the first kind for a modulus equal $\frac12$.
For a small~$\e$ we can write 
$$
\ealn{
\sin\e &\simeq \e &(3.338)\cr
\intop_0^{\frac\pi2-\e} \frac{d\theta}{\sqrt{1-\frac12\sin^2\theta}}
&\cong K\(\tfrac12\) - \sqrt2\, \e.&(3.339)}
$$

Thus one gets for $k=2l$ 
$$
\frac{2.8284}{\e}-5.0354=-\sqrt2\,\e+(4l+1)1.8541. \eq340
$$
It is easy to notice that we can realize a 60-fold \il\ in our model for
$$
x_0=\frac{1.9026}{|\e|} \eq341
$$
can be arbitrary big for sufficiently big $l$.

Let us take $l=25\cdot 10^{24}$. In this case we have for $\e$ an equation
$$
\e^2-\bigl((4l+1)1.3110+5.0354\bigr)\e+2=0 \eq342
$$
or
$$
\e^2-\(10^{26}+6.0354\)\e+2=0, \eq342a
$$
i.e.
$$
\e^2-10^{26}\e+2=0, \eq342b
$$
and finally
$$
\ealn{
|\e|&\simeq 10^{-26} &(3.343)\cr
x_0&\simeq 1.9026\cdot 10^{26} &(3.344)\cr
\ln x_0&\simeq 60.51 &(3.345)
}
$$
which gives us a 60-fold \il.

Moreover the Eq.\ (3.242) has an infinite number of solutions for $0\leq \psi
\leq\frac\pi2$, $\f=k\pi+\psi$,
$$
\frac{2.8284}{\cos\psi}-5.0354=\intop_0^\psi \frac{d\theta}
{\sqrt{1-\frac12\sin^2\theta}}+7.5164\cdot l \eq346
$$
with
$$
x_0=\frac{1.9026}{\cos\psi},\quad
l=0,1,2,3,\ldots,\quad k=2l. \eq347
$$
One can find roots of Eq.\ (3.346) for 

\noindent
$l=5$
$$
\al
x_0&=16.83\\
\ln x_0&=2.82,
\eal \eq348a
$$
$l=50$
$$
\al
x_0&=0.24\cdot 10^3\\
\ln x_0&=5.48
\eal \eq348b
$$
and $\ln x_0$ for $l=5\cdot10^{24}$. In the last case one finds using
70-digit arithmetics
$$
\ln x_0\simeq 58.479\ldots \eq349
$$
which gives us an almost 60-fold \il. In general an amount of \il
$$
\o N_0=\ln x_0 \eq350
$$
is a function of $l=0,1,2,\ldots$ and probably can be connected to the Dirac
large number hypothesis. 

Finally let us take $l=10^n$ where $n>10$. In this case one can solve Eq.\
(3.342a) and get
$$
|\e|\simeq \frac2{[(4\cdot10^n+1)1.3110+5.0354]}\,. \eq351
$$
Thus for $x_0$ we find
$$
x_0\simeq 5\cdot 10^n \eq352
$$
and
$$
\o N_0(n) \simeq 1.60+2.30n. \eq353
$$
In this way, for large $l$, $\o N_0$ is a linear function of a logarithm
of~$l$, 
$$
\al
&\o N_0(10)\simeq 24.6, \ \o N_0(20)\simeq 47.6,\ \o N_0(24)\simeq 56.8,\\
&\o N_0(25)\simeq 59.1, \ \o N_0(26)\simeq 61.4
\eal \eq354
$$
or
$$
\o N_0(\log_{10} l)=1.6+\ln l. \eq355
$$
It is easy to see that Eq.\ (3.355) is an excellent approximation even for
$l=50$. 

Let us come back to the Eq.\ (3.29) in order to find a power spectrum for
our simplified model of \il. Using Eqs (3.161), (3.230), (3.235) and
(3.239) one gets after some algebra
$$
P_R(K) \cong 2.8944 \(\frac{H_0}{2\pi}\)^2 n_1\a_s^2
f\(\frac K{R_0H_0}\) \eq356
$$
where
$$\al
f(x)&=x^{-2}\Bigl(\sn(1.4687-0.5256x,-1)\\
&\qquad{}+\cn(1.4687-0.5256x,-1)
\dn(1.4687-0.5256x,-1)\Bigr)^{-2}. 
\eal \eq357
$$
Using some relations among elliptic functions one gets
$$
f(x)=\frac1{x^2} \cdot \frac{2\dn^4(u,\frac12)}
{\(\sn(u,\frac12)\dn(u,\frac12)+\sqrt2\,\cn(u,\frac12)\)^2}
\eq358
$$
where
$$
u=2.0770-0.7433x. \eq359
$$

Let us consider a more general situation for the Eq.\ (3.231), i.e.
$$
\frac{d\varPhi}{dt}(0)=h \eq360
$$
where $h\ne0$. In this way one gets
$$
h=-\frac{\sqrt5}{\a_s} C_1H_0\bigl(\sn(C_1+C_2,-1)
+\cn(C_1+C_2,-1)\dn(C_1+C_2,-1)\bigr). \eq361
$$
Using Eq.\ (3.227) one gets
$$
\al
C_2&=\frac{K(\frac12)}{\sqrt2}-\frac{\sqrt{10}}
{\sqrt{\sqrt{\bigl(1+\frac{\a_sh}{H_0}\bigr)^4+4} - 
\bigl(1+\frac{\a_sh}{H_0}\bigr)^2}}\\
&-\frac1{\sqrt2}\int_0^
{\arccos\sqrt{\frac{\sqrt{\(1+\frac{\a_sh}{H_0}\)^4+4} -
\(1+\frac{\a_sh}{H_0}\)^2}2}}
\frac{d\f}{\sqrt{1-\frac12\sin^2\f}}
\eal \eq362
$$
and
$$
|C_1|=\frac{\sqrt{10}}{\sqrt{\sqrt{\bigl(1+\frac{\a_sh}{H_0}\bigr)^4+4} -
\bigl(1+\frac{\a_sh}{H_0}\bigr)^2}}\,. \eq363
$$

Let us come back to the Eq.\ (3.229) to find an amount of \il\ for $C_2$
and~$C_1$ given by Eqs (3.362--363). One gets
$$
\sqrt2 C_2 - K\(\tfrac12\) = \intop_0^{\arccos\bigl(\frac1{2C_1e^{N_0}}\bigr)}
\frac{d\theta}{\sqrt{1-\frac12\sin^2\theta}} - \sqrt2 C_1 e^{N_0}, \eq364
$$
or using a natural substitution
$$
\cos\f=\frac1{2C_1e^{N_0}}\,, \eq365
$$
$$
\sqrt2 C_2 - K\(\tfrac12\) = \intop_0^\f
\frac{d\theta}{\sqrt{1-\frac12\sin^2\theta}} - \frac{\sqrt2}{2\cos\f}\,. \eq366
$$
Taking as usual
$$
\f=2l\pi+\tfrac\pi2-\e, \quad l=0,1,2,\ldots, \eq367
$$
one gets for small $\e$
$$
\e\simeq \frac1{\sqrt2 (4l+2)K(\frac12)-2C_2} \simeq 
\frac1{(4l+2)\sqrt2 K(\frac12)}\eq368
$$
(for large $l$) and eventually
$$
\al
&e^{N_0}\simeq \frac{4\sqrt2 K(\frac12)}{C_1}l\\
&N_0=\ln\(\frac{4\sqrt2 K(\frac12)}{C_1}\)+\ln l.
\eal \eq369
$$

Let us calculate $\frac{d\varPhi}{dt}$ for $t=t^\ast=\frac1{H_0}
\ln\bigl(\frac K{R_0H_0}\bigr)$. One gets
$$
\al
\frac{d\varPhi}{dt}(t^\ast)&=
-\frac{\sqrt5}{2\a_s}C_1H_0 \(\frac K{R_0H_0}\)
\biggl(\sn\Bigl(C_1\Bigl(\frac K{R_0H_0}\Bigr)+C_2,-1\Bigr)\\
&+\cn\Bigl(C_1\Bigl(\frac K{R_0H_0}\Bigr)+C_2,-1\Bigr)
\dn\Bigl(C_1\Bigl(\frac K{R_0H_0}\Bigr)+C_2,-1\Bigr)\biggr).
\eal \eq370
$$
Thus we can write down a $P_R(K)$ function.
$$
P_R(K)=\(\frac{H_0}{2\pi}\)^2 \cdot \frac{4\a_s^2n_1}{5C_1^2}\,
f\(\frac K{R_0H_0}\) \eq371
$$
where
$$
f(x)=\frac1{x^2}\bigl(\sn(C_1x+C_2,-1)+\cn(C_1x+C_2,-1)\dn(C_1x+C_2,-1)\bigr)
^{-2}.
\eq372
$$
Using some relation among elliptic functions one finds
$$
P_R(K)=\(\frac{H_0}{2\pi}\)^2 \cdot \frac{4\a_s^2n_1}{5C_1^2}\,
g\(\frac K{R_0H_0}\) \eq373
$$
where
$$
g(x)=\frac1{x^2} \cdot \frac2
{\(\sd(u,\frac12)+\sqrt2 \cd(u,\frac12)\nd(u,\frac12)\)^2} 
\eq374
$$
and
$$
u=\sqrt2 C_1x+\sqrt2 C_2\,. \eq375
$$

Moreover we can reparametrize (3.374--375) in the following way
$$
u=\sqrt2 C_1x+K(\tfrac12)-C_1\sqrt2-
\intop_0^{\arccos\(\frac{\sqrt5}{C_1}\)}
\frac{d\f}{\sqrt{1-\frac12\sin^2\f}} \eq376
$$
where
$$
\ealn{
\frac{\a_sh}{H_0}&=-1\pm \frac1{|C_1|}\sqrt{\frac{C_1^4-25}5} &(3.377)\cr
|C_1|&>\sqrt5. &(3.378)
}
$$
For large $C_1$ one gets
$$
\ealn{
&u\cong \sqrt2 C_1(x-1) &(3.379)\cr
&\frac{\a_sh}{H_0}\simeq \frac{C_1}{\sqrt5}\,. &(3.380)
}
$$

Using (3.369),
$$
u=\frac{4\sqrt2 K(\frac12)l}{e^{N_0}}\,(x-1). \eq381
$$
If we take large $C_1$ (it means, a large $h$)
$$
h=\frac{C_1H_0}{\sqrt5 \a_s} \eq382
$$
and simultaneously sufficiently large $l$ we can achieve a 60-fold \il\ with
a function $P_R(K)$ given by the formula (3.374). Large $C_1$ means here
$C_1\simeq 100$, large $l$ means $l\simeq 10^{25}$.

Let us calculate the spectral index for our $P_R(K)$ function, i.e.
$$
n_s(K)-1 = \frac{d\ln P_R(K)}{d\ln K}\,.\eq383
$$
One gets
$$
n_s(K)-1 = -2-2C_1x\,\frac{\sqrt2 \cd(u,\frac12)-\sd^3(u,\frac12)}
{\sd(u,\frac12)+\sqrt2 \cd(u,\frac12)\nd(u,\frac12)} \eq384
$$
where
$$
\al
u=\sqrt2 (C_1x+C_2)&=\frac{\sqrt2}{R_0H_0}\,(C_1K+C_2R_0H_0)\\
x&=\frac K{R_0H_0}
\eal
\eq385
$$
A very interesting characteristic of $P_R(K)$ is also $\dfrac{dn_s(K)}{d\ln
K}$. One gets
$$
\al
\frac{dn_s(K)}{d\ln K}&=
-2C_1x\,\frac{\sqrt2 \cd(u,\frac12)-\sd^3(u,\frac12)}
{\sd(u,\frac12)+\sqrt2 \cd(u,\frac12)\nd(u,\frac12)}\\
&{}-2C_1^2x^2\sd^2(u,\tfrac12)-2C_1^2x^2
\frac{\(\sqrt2 \cd(u,\frac12)-\sd^3(u,\frac12)\)^2}
{\(\sd(u,\frac12)+\sqrt2 \cd(u,\frac12)\nd(u,\frac12)\)^2}
\eal
\eq386
$$
where $u$ and $x$ are given by Eq.\ (3.385).

Let us take for a trial $C_1=-0.5256$ and $C_2=1.4687$. In this case one
finds 
$$
n_s(K)-1 = -2+1.0512x\,\frac{\sqrt2 \cd(u,\frac12)-\sd^3(u,\frac12)}
{\sd(u,\frac12)+\sqrt2 \cd(u,\frac12)\nd(u,\frac12)} \eq387
$$
$$
\al
\frac{dn_s(K)}{d\ln K}&=
1.0512x\,\frac{\sqrt2 \cd(u,\frac12)-\sd^3(u,\frac12)}
{\sd(u,\frac12)+\sqrt2 \cd(u,\frac12)\nd(u,\frac12)}\\
&{}-0.5525x^2\sd^2(u,\tfrac12)-0.5525x^2
\frac{\(\sqrt2 \cd(u,\frac12)-\sd^3(u,\frac12)\)^2}
{\(\sd(u,\frac12)+ \sqrt2\cd(u,\frac12)\nd(u,\frac12)\)^2}
\eal
\eq388
$$
$$
\al
u&=2.0771-0.7433x\\
x&=\frac K{R_0H_0}
\eal
\eq388a
$$

The interesting point is to find $n_s(K)\simeq1$ (a~flat power spectrum). One
gets
$$
\sd(u,\tfrac12)+ \sqrt2\cd(u,\tfrac12)\nd(u,\tfrac12)=
C_1x\(\sd^3(u,\tfrac12)-\sqrt2 \cd(u,\tfrac12)\) \eq389
$$
and
$$
\sd(u,\tfrac12)+ \sqrt2\cd(u,\tfrac12)\nd(u,\tfrac12)\ne0. \eq390
$$
Using (3.389--390) one gets
$$
\frac{dn_s}{d\ln K} = -2C_1^2x^2\sd^2(u,\tfrac12) \eq391
$$
if Eqs (3.389--390) are satisfied.

In the case of special $C_1$ and $C_2$ one gets
$$
\frac{dn_s}{d\ln K} = -0.5525x^2\sd^2(u,\tfrac12) \eq392
$$
where $x,u$ are given by Eq.\ (3.388a).

Let us reparametrize the Eq.\ (3.389). One gets
$$
\sd(u,\tfrac12)+ \sqrt2\cd(u,\tfrac12)\nd(u,\tfrac12)=
\frac{u-\sqrt2 C_2}{\sqrt2}\(\sd^3(u,\tfrac12)-\sqrt2 \cd(u,\tfrac12)\). \eq393
$$
For sufficiently big $u$ one gets (the equation has infinite number of roots)
$$
u=u_1+2lK(\tfrac12),\qquad l=0,\pm1,\pm2,\ldots \eq394
$$
where $u_1$ satisfies the equation
$$
\sn(u_1,\tfrac12)=\frac13\(1+\root3\of5
\(\root3\of{65+\sqrt{20357}}-\root3\of{\sqrt{20357}-65}\)\)^\frac12
\simeq 0.65\ldots \eq395
$$
Moreover for $x=\frac K{R_0H_0}>0$ we have the condition
$$
\frac{u_1+2lK(\frac12)-\sqrt2 C_2}{\sqrt2C_1}>0. \eq396
$$

One finds
$$
\sd^2(u_1,\tfrac12)=0.53\ldots \eq397
$$
Thus
$$
\al
\frac{dn_s}{d\ln K}\(u_1+2lK(\tfrac12)\) &\simeq
-0.53\(\frac{u_1+2lK(\frac12)-\sqrt2 C_2}{\sqrt2C_1}\)^2,\\
 l&=0,\pm1,\pm2,\ldots 
\eal
\eq398
$$
One gets
$$
\ealn{
&\arcsin0.65 \simeq 40^\circ.54 &(3.399)\cr
&u_1\cong \intop_0^{40^\circ.54}\frac{d\theta}{\sqrt{1-\frac12 \sin^2\theta}}
=F(40^\circ.54/45^\circ)\cong 0.73\ldots &(3.400)
}
$$
Taking special value of $C_1$ and $C_2$ one gets
$$
x\cong 1.81 - 4.68l \eq401
$$
where $l=0,\pm1,\pm2,\ldots$. For $x>0$ it is easy to see that $l$~should be
nonpositive. Practically $l$~should be a negative integer
$$
l<-3 \eq402
$$
and
$$
\frac{dn_s}{d\ln K} \cong -0.29 (1.81-4.68l)^2. \eq403
$$

Thus for 
$$
K_l=R_0H_0(1.81-4.68l)
$$
we get
$$
n_s(K_l)\cong1. \eq404
$$
The important range of $\ln K$, $\D\ln K$ is about 10.

Thus
$$
n_s(K)\simeq 1\pm2.9(1.81-4.68l)^2. \eq405
$$
Taking $l=-3$ one gets
$$
-300 < n_s(K) < 300 \eq406
$$
and
$$
P_R(K) \sim K^{1-n_s(K)}
$$
is not flat in the range considered.

Moreover we can improve the results considering Eq.~(3.398) for large~$C_1$.
For large $C_1$ one gets
$$
\frac{C_2}{C_1}\simeq -1. \eq407
$$
Thus we find
$$
\frac{dn_s}{d\ln K} \simeq -1.06\(u_1+2lK(\tfrac12)+C_1\)^2. \eq408
$$
Taking large value of $C_1$ in such a way that
$$
C_1=-2lK(\tfrac12)-u_1+\e \eq409
$$
where $\e$ is a small number,
$$
\e\simeq 10^{-n}, 
$$
one gets
$$
\frac{dn_s}{d\ln K}\simeq -1.06\cdot 10^{-2n}. \eq410
$$
The last condition means that we should take
$$
h\cong \frac{H_0|C_1|}{\sqrt5 \a_s}= 
\frac{H_0}{\sqrt5 \a_s}\,\(0.73-2lK(\tfrac12)-10^{-n}\) \eq411
$$
where $l$ is an integer, $l\simeq-100$.

Thus for some special values of $h$ we can get arbitrarily small
$\dfrac{dn_s}{d\ln K}$ which means we can achieve a flat power spectrum in
the range considered ($\D\ln K\sim 10$),
$$
n_s(K)=1\pm10^{-(2n-1)}, \eq412
$$
$n$ arbitrarily big.

Let us consider the value of $K$ coresponding to our value of $n_s(K)=1$. One
gets 
$$
x=\frac{0.73+2lK(\frac12)-\sqrt2 C_2}{\sqrt2 C_1}\,. \eq413
$$
Using our assumption on a large $C_1$ and Eq.\ (3.409) one finds
$$
x\cong \(1-\frac1{\sqrt2}\)-\frac \e {2lK(\frac12)} \eq414
$$
or (for $\e$ is small and $l$ quite big)
$$
x\simeq 0.293 \eq415
$$
and
$$
K\simeq 0.3R_0H_0 \eq416
$$
with the range $\D\ln K\sim10$. It means that
$$
0.3e^{-10} \le \frac K{R_0H_0} \le 0.3\cdot e^{10} \eq417
$$
which gives us a full \co ly interesting region
$$
10^{-4} \le \frac K{R_0H_0} \le 10^4. \eq418
$$

\smallskip
Finally let us consider in this section two problems. First we match our
solutions of an \ev\ of the \U, i.e.\ both de Sitter phases, radiation
dominated \U, matter dominated \U\ and K-essence \U\ (kinetic energy
dominated \q). In order to simplify calculations we match the first de Sitter
phase to radiation dominated \U\ using some issues from the second de Sitter
phase. Secondly we write down solution to the \ev\ of \U\ with a \co\ \ct\
(from the second de Sitter phase), with a radiation and with a matter
(a~dust). All of these material ingredients are evolving in\dt ly in this
case, similarly as a matter (dust) and a \q\ for the model \rf3.67 . 

Let us consider Eq.\ \rf2.26 \ and let us match it to Eq.~\rf3.58 . One gets
$$
\ealn{
R_1&=\root4\of{\frac B{\rho_Q}} \sqrt{\sh\(\frac{2\sqrt3\sqrt{\rho_Q}M\dr{pl}}
{B} t\dr{end}\)} &\rf3.419 \cr
R_1&=R_0e^{H_0t\dr{end}} &\rf3.420 
}
$$
where
$$
t\dr{end}=\frac{N_0}{H_0} \eq421
$$
is the time of the end of the first de Sitter phase. ($N_0$ is the amount of
an \il.) 

Simultaneously we need a local conservation of an energy, i.e.
$$
\t \rho_r(R_1)=\frac B{R_1^4} \eq422
$$
where
$$
\t \rho_r(R_1)=6\(H_0^2-H_1^2\); \eq423
$$
see the formula
$$
\rho_r=\la_{c1}(\P_0)-\la_{c0}(\P_1)=
\la_{c0}(\P_1)\(\frac{H_0^2}{H_1^2}-1\). 
$$

After some calculations one gets
$$
B=\frac{2\sqrt3 M\dr{pl}\sqrt{\rho_Q}t\dr{end}}{\ars\(\sqrt{\frac{\rho_Q}
{\u\rho_r}}\)}\,. \eq424
$$

Now we need match a solution with a radiation dominated period to the
solution with matter dominated period (with the same \co\ \ct---\q\ energy
density). One gets
$$
\frac A{R_2^3}+\rho_Q= \frac B{R_2^4}+\rho_Q \eq425
$$
and finally
$$
A=\frac B{R_2} \eq426
$$
where
$$
R_2=0.7916\root3\of{\frac A{\rho_Q}}=\root4\of{\frac B{\rho_0}}
\sqrt{\sh\(\frac{2\sqrt3\sqrt{\rho_Q}M\dr{pl}}{B}\,t_1\)}\,, \eq427
$$
$t_1$ is a time to match (see \rf3.90 ). Let us consider
$$
N_1=H_1\(t_1-t\dr{end}\) \eq428
$$
as an amount of an \il\ during the second de Sitter phase. In this way one
gets 
$$
t_1=\frac{N_1}{H_1}+\frac{N_0}{H_0} \eq429
$$
and finally
$$
\ealn{
A&=2.015\cdot \frac{M\dr{pl}^{3/4}\rho_Q^{5/8}\(\frac{N_0}{H_0}\)^{3/4}}
{\ars^{4/3}\sqrt{\frac{\rho_Q}{\u \rho_r}}}\,
\sh^{3/2}\(\sqrt{\frac{\rho_Q}{\t\rho_r}}\(\frac{N_1}{N_0}\,
\frac{H_0}{H_1}+1\)\) &\rf3.430 \cr
B&=\frac{2\sqrt3\,M\dr{pl}\sqrt{\rho_Q}}{\ars\(\sqrt{\frac{\rho_Q}{\u
\rho_r}}\)} \,\frac{N_0}{H_0}\,. &\rf3.431
}
$$
However, we should match also a time. In this way we get from Eqs \rf3.287--288
$$
t_0=\frac{418.48}{\sqrt{\rho_Q}}\,M\dr{pl}+\frac{N_1}{H_1} \eq432
$$
or using the value $\rho_Q=\la_{c0}=10^{-52}\frac1{{\text m}^2}$ we get
$$
t_0=0.447\cdot 10^{14}\text{yr}+\frac{N_1}{H_1}\,. \eq433
$$

If $R_0$ is an initial value of ``a radius'' of the \U\ we get
$$
\al
R_0=R_1e^{-N_0}
&=\root4\of{\frac
{2\sqrt3\,M\dr{pl}\sqrt{\rho_Q}}{\t\rho_r\ars\(\sqrt{\frac{\rho_Q}{\u
\rho_r}}\)}}
\cdot \exp\(-\(N_0+\frac{n+2}4 \P_0\)\)\cr
&=\root4\of{\frac
{2\sqrt3\,M\dr{pl}\sqrt{\rho_Q}}{\t\rho_r\ars\(\sqrt{\frac{\rho_Q}{\u
\rho_r}}\)}}\cdot\(\frac{(n+2)\b}{n|\g|}\)^{(n+2)/8}e^{-N_0}. 
\eal
\eq434
$$
The only one ingredient from the second de Sitter phase is using the formula
\rf3.428 . Now we should match the solution \rf3.287--288 \ with conditions
\rf3.90 \ to the solution \rf3.124 , \rf3.116 , \rf3.122 . We should match a
field~$\P$, a density of an energy, a radius and a time. One gets
$$
\ealn{
&R_3=\o R_0 \exp\(\frac{n+1}{M\dr{pl}}(t_2-\o t_0)\)=
3.07\root3\of {\frac A{\rho_Q}} &\rf3.435 \cr
&\P(t_2)=\frac1{2n}\ln\(\frac\d{3|\g|}\)-\frac1n \sqrt{\frac{2\d}{3|\o M|}}
\,M\dr{pl}(t_2-\o t_0)=\P_0 &\rf3.436 \cr
&\rho_Q+\frac A{R_3^3}=\rho_0e^{(n+2)\P_0}+\rho_\P &\rf3.437 
}
$$
where
$$
\ealn{
&\rho_\P=\frac{\d M^2\dr{pl}}{3n^2}-\frac{\sqrt{|\g|\d}}{2\sqrt3}
\exp\(-\sqrt{\frac{2\d}{3\o M}}\, M\dr{pl}(t_2-\o t_0)\) &\rf3.438 \cr
&\d=2\,\frac{n+1}{M^2\dr{pl}}-\frac{3\rho_0}{M^2\dr{pl}} &\rf3.439 \cr
&\o t_0\ne0. &\rf3.440
}
$$

One finds:
$$
t_2=\sqrt{\frac{3\o M}{2\d}}\,\frac1{M\dr{pl}}\ln\(\sqrt{
\frac{\d(n+2)^n\b^n}{3|\g|^{n+1}n^n}}\)+\o t_0 \eq441
$$
and $t_2$ is large,
$$
\ealn{
&\rho_\P\cong \frac{\d M^2\dr{pl}}{3n^2}=\frac{M^2\dr{pl}}{3n^2}
\(2\,\frac{n+1}{M^2\dr{pl}}-\frac{3\rho_0}{M^2\dr{pl}}\) &\rf3.442 \cr
&\rho_0=\frac{n^2\la_{c0}-3(n+1)}{n^2x_0^{n+2}-1}\,. &\rf3.443
}
$$
From
$$
\al
\frac A{R^3_3}&=-\rho_Q+\rho_0\(x_0^{n+2}-\frac1{n^2}\)+\frac{3(n+1)}{n^2}\\
R_3&=3.072\root3\of{\frac A{\rho_Q}}
\eal
\eq444
$$
one obtains
$$
\d=\frac1{M^2\dr{pl}}\(2(n+1)-3\,\frac{\la_{c0}n^2-3(n+1)}{n^2x_0^{n+2}-1}\).
\eq445
$$
From Eqs \rf3.90 , \rf3.287 \ and \rf3.288 \ we get
$$
\ealn{
&t_2=\frac{N_1}{H_1}+\frac{5.38}{\sqrt{\rho_Q}}\,M\dr{pl} &\rf3.446 \cr
&\o t_0=\frac{N_1}{H_1}+\frac{5.38}{\sqrt{\rho_Q}}\,M\dr{pl}
-\sqrt{\frac{3\o M}{2\d}} \, \frac1{M\dr{pl}} \ln\(\sqrt{
\frac{\d(n+2)^n\b^n}{3|\g|^{n+1}n^n}}\)
&\rf3.447 \cr
&M\dr{pl}\frac{5.38}{\sqrt{\rho_Q}}=5.64\cdot 10^{11}\,\text{yr}. &\rf3.448
}
$$
Finally we find $\o R_0$:
$$
\o R_0=R_3\exp\(-\frac{n+1}{M\dr{pl}}(t_2-\o t_0)\)=
3.072\root3\of{\frac A{\rho_Q}} \exp\(-\frac{n+1}{M\dr{pl}}(t_2-\o t_0)\).
\eq449
$$

In this way we match an \ev\ of the \U\ from the very beginning up to
\hbox{K-essence} dominance. We find all the \ct s. Let us notice that the solution
\rf3.90 , \rf3.287--288 \ has a quite known behaviour in the theory of ordinary
nonlinear differential equations. The solution cannot be extended after some
value of the \dt\ variable (and also in\dt). One says the solution dies. Due
to this interesting behaviour we obtain physical consequences. Matter and a \q\
cannot evolve in\dt ly. We get also a ratio between a density of matter and a
\q\ which agrees with observational data and helps (in principle) to
calculate a time of our contemporary epoch.

If we use Eqs\ \rf3.287 \ and \rf3.288 \ we can calculate a time of our
contemporary epoch, i.e.\ a time when the ratio of a density of a matter to a
\q\ density
is~$\frac37$ (this is this ratio measured now). One gets
$$
\o t\dr{contemporary}=8.97\,\frac{M\dr{pl}}{\sqrt{\rho_Q}}+\frac{N_1}{H_1}
=9.42\cdot 10^{10}\,\text{yr}+\frac{N_1}{H_1}\,.\eq450
$$
In the same way using Eqs \rf3.287 , \rf3.288 \ and \rf3.289 \ we calculate a
Hubble parameter for our contemporary epoch.
$$
H\dr{contemporary}=h\dr{contemporary}\frac{100\frac{\text km}{\text s}}
{\text{Mpc}} \eq451
$$
where
$$
h\dr{contemporary}=1.21. \eq452
$$
Both values are a little too big. 

Moreover one can improve the results. However, the age of the \U\ is a little
longer,
$$
t\dr{age of the \U}=t\dr{contemporary}+t_1=
9.42\cdot 10^{10}\,\text{yr}+2\,\frac{N_1}{H_1}+\frac{N_0}{H_0}\,.
\eq453
$$
In general one gets
$$
H=\frac{\sqrt{\rho_Q}}{M\dr{pl}}\,\frac{32.27\root3\of{\frac{\rho_Q}{\rho_m}}
-16.38}{100\root3\of{\frac{\rho_Q}{\rho_m}}-227.48}
$$
and
$$
-q=\frac{5.21}{11.76\root3\of{\frac{\rho_Q}{\rho_m}}-5.17}+1
$$
where $\frac{\rho_Q}{\rho_m}$ is a ratio of a \q\ energy density to a matter
(dust) energy density, or
$$
h=0.9115\,\frac{32.27\root3\of{\frac{\rho_Q}{\rho_m}}
-16.38}{14.96\root3\of{\frac{\rho_Q}{\rho_m}}-0.03}\,.
$$

Finally, let us express an age of the \U\ in terms of the ratio
$\frac{\rho_Q}{\rho_m}$. One gets
$$
t\(\frac{\rho_Q}{\rho_m}\)=2.667 \ln\(31.87\root3\of{\frac{\rho_Q}{\rho_m}}
-14.03\)\frac{M\dr{pl}}{\rho_Q} + 2\(\frac{N_1}{H_1}\)+\(\frac{N_0}{H_0}\)\,
$$
or
$$
t\(\frac{\rho_Q}{\rho_m}\)=2.82\cdot 10^{10}\,\text{yr}\cdot
 \ln\(31.87\root3\of{\frac{\rho_Q}{\rho_m}}
-14.03\)\frac{M\dr{pl}}{\rho_Q} + 2\(\frac{N_1}{H_1}\)+\(\frac{N_0}{H_0}\)\,.
$$
For $\frac{\rho_m}{\rho_Q}=0.0034$ one gets
$$
t(294.12)=14.91\cdot 10^{10}\,\text{yr}+
2\(\frac{N_1}{H_1}\)+\(\frac{N_0}{H_0}\)\,. 
$$
For $\frac{\rho_m}{\rho_Q}=2.015$ 
$$
t(0.496)=6.76\cdot 10^{10}\,\text{yr}+
2\(\frac{N_1}{H_1}\)+\(\frac{N_0}{H_0}\)\,. 
$$
$$
\D t=t(294.12)-t(0.496)=8.15\cdot 10^{10}\,\text{yr}.
$$
The last number is a duration time of the epoch for $\rho_Q$ is equal to the
contemporary value of a \co\ \ct.

Let us come back to the Eq.\ \rf3.52 \ and consider it in the flat case $K=0$.
One gets
$$
\frac{R\,dR}{\sqrt{\frac{\rho_Q}{3\Mp^2}\cdot R^4+\frac A{3\Mp^2}\cdot R
+\frac B{3\Mp^2}}}=\pm dt. \eq454
$$
We change $R$ into $x$:
$$
R=\root3\of{\frac A{\rho_Q}}\,x \eq455
$$
and finally get
$$
\pm\int\frac{x\,dx}{\sqrt{x^4+x+a}}=\sqrt{\frac{\rho_Q}{3\Mp^2}}\,(t-t_0)
\eq456
$$
where
$$
0<a=\frac BA \root3\of{\frac{\rho_Q}A}\,. \eq457
$$

This is an \ev\ of a flat \U\ with radiation, matter (dust) and a density of
a \q\ (this is a \co\ \ct\ $\la_{c0}$). The integral on LHS of Eq.~\rf3.456 \
is an elliptic integral and can be calculated. The properties of the result
of calculations strongly depend on the value of~$a$. Let us notice that the
integral we have calculated here has $a=0$ (no radiation).

The \ev\ of a \q, a matter and a radiation is here in\dt. One gets the
following results for
$$
I=\int\frac{x\,dx}{\sqrt{x^4+x+a}}\,. \eq458
$$

In general there are two cases:
$$
\displaylines{
\text{I\hphantom{I}} \hfill B_2>0 \hfill \rf3.459 \cr
\text{II} \hfill B_2<0 \hfill \rf3.460
}
$$
where
$$
B_2=\frac{\sqrt{12b_1^4-a}-2b_1^2-\sqrt{4b_1^4-a}}
{2\sqrt{12b_1^4-a}} \eq461
$$
and $b_1>0$ is a solution of the cubic equation
$$
\displaylines{
\hfill y^3-ay-\tfrac18=0 \hfill \rf3.462 \cr
\hfill y=2b_1^4 \hfill \rf3.463
}
$$
Eq.\ \rf3.462 \ can be solved by using the Cardano formulae in two cases

\item{1)} $D>0$,
\item{2)} $D<0$,

$$
D=\frac{27-256a^3}{27\cdot256}\,. \eq464
$$
In case 1) one gets
$$
\al
b_1&=\frac12 \root4\of{\root3\of{4+\tfrac12\sqrt{27-256a^3}}
+ \root3\of{4-\tfrac12\sqrt{27-256a^3}}}>0\\
0&<a<\frac3{4\root3\of4}=0.472470393\ldots
\eal \eq465
$$
In case 2) one gets
$$
b_1=\root4\of{\frac a3}\sqrt{\cos\frac\f3} \eq466
$$
where
$$
\displaylines{
\hfill \cos\f=\frac3{16a}\sqrt{\frac3a} \hfill \rf3.467 \cr
\hfill a>\frac3{4\root3\of4}\,, \quad \f\in\langle0,\tfrac\pi2\rangle. \hfill 
\rf3.468
}
$$
Condition I can be transformed into
$$
4b_1^2<a+1 \eq469
$$
for both cases 1) and 2). In case 1) this condition has no solution. It means
we have always $B_2<0$. In case~2) we have always \rf3.469 . It means
$$
B_2>0 \eq470
$$
for an equation
$$
\frac1{\root4\of4}\,\frac{\cos^2\(\frac{\f}3\)}{\cos^{4/3}\f}=
\frac3{4\root3\of4 \cos^{2/3}\f}+1. \eq471
$$
has no solution in $(0,\frac\pi2)$ and in
$(\frac{3\pi}2, 2\pi)$.

The fact that in case 1) we have always
$$
4b_1^2>a+1 \eq472
$$
is caused by a nonexistence of real solutions of an equation
$$
4(a+1)^{1/2}=\root3\of{4+\tfrac12\sqrt{27-256a^3}}
+\root3\of{4-\tfrac12\sqrt{27-256a^3}} \eq473
$$
where
$$
0<a<\frac3{4\root3\of4}\,. \eq474
$$
Thus for sufficiently small $a$ we always have case~I (in a limit $a=0$, no
radiation also).

One can express $b_1$ and $a$ in terms of $\f$:
$$
\ealn{
b_1^4&=\frac3{4\root3\of4}\,\frac{\cos^2(\frac\f3)}{\cos^{2/3}\f} &\rf3.475 \cr
a&=\frac3{4\root3\of4}\cos^{-2/3}\f. &\rf3.476
}
$$
Thus one obtains in case I
$$
\ealn{
&I=\frac12 \ln \frac {e(\frac{x-\a}{x-\b})^2 +f
+2\sqrt{\bigl((\frac{x-\a}{x-\b})^2+c\bigr)\bigl((\frac{x-\a}{x-\b})^2+d\bigr)(1+c)(1+d)}}
{(\frac{x-\a}{x-\b})^2-1}\cr
&\quad{}+K_1(b_1,a)\mPi\(\arctg\[P(b_1,a)\,\frac{x-\a}{x-\b}\],n_1,q_1\)\cr
&\quad{}+L_1(b_1,a)F\(\arctg\[P(b_1,a)\,\frac{x-\a}{x-\b}\],q_1\)
&\rf3.477 \cr
\noalign{\vskip3pt plus2pt}
&P_1(b_1,a)=\(\frac{\sqrt{12b_1^4-a}-2b_1^2-\sqrt{4b_1^4-a}}
{\sqrt{12b_1^4-a}+2b_1^2+\sqrt{4b_1^4-a}}\)^{1/2} &\rf3.478 \cr
\noalign{\vskip3pt plus2pt}
&q_1=\frac{\sqrt6 |b_1|}{\(3b_1^2+\sqrt{12b_1^4-a}\)^{1/2}} &\rf3.479 \cr
\noalign{\vskip3pt plus2pt}
&K_1(b_1,a)=\(\frac{2b_1^2+\sqrt{4b_1^4-a}+\sqrt{12b_1^4-a}}
{\sqrt{12b_1^4-a}-2b_1^2-\sqrt{4b_1^4-a}}\)^{3/2} &\rf3.480 \cr
\noalign{\vskip3pt plus2pt}
&n_1=\frac{2\sqrt{12b_1^4-a}}{2b_1^2+\sqrt{4b_1^4-a}+\sqrt{12b_1^4-a}}
&\rf3.481 \cr
\noalign{\eject}
&L_1(b_1,a)=\(M(b_1,a)+\frac{2b_1^2+\sqrt{4b_1^4-a}+\sqrt{12b_1^4-a}}
{2\sqrt{12b_1^4-a}}\)\frac1{\(3b_1^2+\sqrt{12b_1^4-a}\)^{1/2}}\cr
&\quad{}\times\frac
{2(12b_1^4-a)^{1/2}\(\sqrt{4b_1^4-a}-b_1^2\)^{1/2}}
{\(2b_1^2+\sqrt{4b_1^4-a}-\sqrt{12b_1^4-a}\)
\(2b_1^2+\sqrt{4b_1^4-a}+\sqrt{12b_1^4-a}\)^{1/2}} &\rf3.482 \cr
\noalign{\vskip3pt plus2pt}
&M(b_1,a)=\frac\b{\a-\b}\cr
&\quad{}=\biggl((6b_1^4+a)\sqrt{4b_1^4-a}+2b_1^2a+
\sqrt{192b_1^{12}-112b_1^8a+20b_1^4a+a^3}\cr
&\qquad{}+\sqrt{576b_1^{12}-240b_1^8a+28b_1^4a-a^3}
+6b_1^2\sqrt{48b_1^8-16b_1^4a+a^2}\cr
&\qquad{}+(a-4b_1^4)\sqrt{12b_1^4-a}\biggr)^{1/2}\cr
&\quad{}\times\Biggl(\biggl((12b_1^4-a)\sqrt{4b_1^4-a}
+4b_1^4\sqrt{12b_1^4-a}-4b_1^2a+24b_1^6+4b_1^4\sqrt{12b_1^4-a}\cr
\noalign{\vskip3pt plus2pt}
&\qquad\quad{}-\sqrt{192b_1^{12}-112b_1^8a+20b_1^4a+a^3}
-\sqrt{576b_1^{12}-240b_1^8a+28b_1^4a-a^3}\biggr)^{1/2}\cr
&\qquad{}-\biggl(8b_1^2\sqrt{48b_1^8-16b_1^4a+a^2}+2(6b_1^4-a)\sqrt{4b_1^4-a}
-4b_1^4\sqrt{12b_1^4-a}\cr
&\qquad\quad{}+4b_1^6+6b_1^2a+\sqrt{192b_1^{12}-112b_1^8a+20b_1^4a+a^3}\cr
&\qquad\quad{}+\sqrt{576b_1^{12}-240b_1^8a+28b_1^4a-a^3}\biggr)^{1/2}\Biggr)^{-1}.
&\rf3.483 
}
$$
$$
\ealn{
\a&=\sqrt{\frac{\sqrt{48b_1^8-16b_1^4a+a^2}+6b_1^2\sqrt{4b_1^4-a}
+2b_1^2\sqrt{12b_1^4-a}+a}
{2b_1^2+\sqrt{4b_1^4-a}+\sqrt{12b_1^4-a}}}>0 &\rf3.484 \cr
\noalign{\vskip3pt plus2pt}
\b&=\sqrt{\frac{\sqrt{48b_1^8-16b_1^4a+a^2}+6b_1^2\sqrt{4b_1^4-a}
-2b_1^2\sqrt{12b_1^4-a}+a}
{2b_1^2-\sqrt{4b_1^4-a}-\sqrt{12b_1^4-a}}}>0 &\rf3.485 \cr
\noalign{\vskip3pt plus2pt}
c&=\frac{\(3b_1^2+\sqrt{12b_1^4-a}\)\(2b_1^2-\sqrt{4b_1^4-a}-\sqrt{12b_1^4-a}\)}
{\(\sqrt{4b_1^4-a}+2b_1^2\)\(\sqrt{12b_1^4-a}-3b_1^2\)} \cr
\noalign{\vskip3pt plus2pt}
d&=\frac{\sqrt{12b_1^4-a}+2b_1^2+\sqrt{4b_1^4-a}}
{\sqrt{12b_1^4-a}-2b_1^2-\sqrt{4b_1^4-a}}\,,\quad
|d|>1 &\rf3.486 \cr
\noalign{\eject}
e&=\frac{2\sqrt{12b_1^4-a}\(\sqrt{12b_1^4-a}+\sqrt{4b_1^4-a}-4b_1^2\)}
{\(\sqrt{4b_1^4-a}+2b_1^2+\sqrt{12b_1^4-a}\)\(\sqrt{12b_1^4-a}-3b_1^2\)}
&\rf3.487 \cr
\noalign{\vskip3pt plus2pt}
f&=\frac{2\sqrt{12b_1^4-a}}
{\(\sqrt{12b_1^4-a}-3b_1^2\)\(\sqrt{12b_1^4-a}-2b_1^2-\sqrt{4b_1^4-a}\)}\cr
&\quad{}\times\(2b_1^2+\sqrt{4b_1^4-a}+\sqrt{12b_1^4-a}\)^{-2}\cr
&\quad{}\times\Bigl(2(14b_1^4-a)\sqrt{4b_1^4-a}-4b_1^2a
+2\sqrt{192b_1^{12}-112b_1^8a+20b_1^4a+a^3}\cr
&\qquad{}+2(b_1^4-a)\sqrt{12b_1^4-a}-2\sqrt{576b_1^{12}-240b_1^8a+28b_1^4a-a^3}
\Bigr). &\rf3.488
}
$$

Case II
$$
\al
I&=\frac12 \ln \frac {e(\frac{x-\a}{x-\b})^2 +f
+2\sqrt{\bigl((\frac{x-\a}{x-\b})^2+c\bigr)\bigl((\frac{x-\a}{x-\b})^2+d\bigr)(1+c)(1+d)}}
{(\frac{x-\a}{x-\b})^2-1}\\
&+K_2(b_1,a)\mPi\(\arccos\[P_2(b_1,a)\,\frac{x-\a}{x-\b}\],n_2,q_2\)\\
&+L_2(b_1,a)F\(\arccos\[P_2(b_1,a)\,\frac{x-\a}{x-\b}\],q_2\)
\eal \eq489
$$
$$
\ealn{
&q_2=\frac1{q_1}=\frac{\(3b_1^2+\sqrt{12b_1^4-a}\)^{1/2}}{\sqrt6 |b_1|}
&\rf3.490 \cr 
\noalign{\vskip3pt}
&n_2=\frac{2b_1^2+\sqrt{4b_1^4-a}+\sqrt{12b_1^4-a}}{2\sqrt{12b_1^4-a}}
&\rf3.491 \cr
\noalign{\vskip3pt}
&K_2(b_1,a)=\frac
{\(2b_1^2+\sqrt{4b_1^4-a}-\sqrt{12b_1^4-a}\)\sqrt{12b_1^4-a}
\(\sqrt{4b_1^4-a}-b_1\)^{1/2}}
{\(\sqrt{12b_1^4-a}+2b_1^2+\sqrt{4b_1^4-a}\)\(\sqrt{12b_1^4-a}
-3b_1^2\)^{3/2}}\hskip30pt &\rf3.492 \cr
\noalign{\vskip3pt}
&P_2(b_1,a)=\(\frac{2b_1^2+\sqrt{4b_1^4-a}-\sqrt{12b_1^4-a}}
{\sqrt{12b_1^4-a}+2b_1^2+\sqrt{4b_1^4-a}}\)^{1/2} &\rf3.493 \cr
\noalign{\vskip3pt}
&L_2(b_1,a)=M(b_1,a)\cr
&\quad{}-\frac
{4(12b_1^4-a)\(\sqrt{4b_1^4-a}-b_1^2\)^{1/2}
\(\sqrt{12b_1^4-a}-2b_1^2-\sqrt{4b_1^4-a}\)}
{\(3b_1^2-\sqrt{12b_1^4-a}\)^{3/2}\(2b_1^2+\sqrt{4b_1^4-a}+\sqrt{12b_1^4-a}\)}\,.
&\rf3.494
}
$$

The function \rf3.456 \ cannot be inverted globally. It can be inverted only locally in
some intervals where the solution lives. The solution dies on the end of an
interval being reborning on beginning of a next interval. All the intervals
can be obtained from the condition
$$
\o W>0 \eq495
$$
where
$$
\o W=\frac {e(\frac{x-\a}{x-\b})^2 +f+2\sqrt{\bigl((\frac{x-\a}{x-\b})^2
+c\bigr)\bigl((\frac{x-\a}{x-\b})^2+d\bigr)(1+c)(1+d)}}
{(\frac{x-\a}{x-\b})^2-1} \eq496
$$
and
$$
\left|\frac{x-\a}{x-\b}\right|\ne1 \eq497
$$

 In the case I $c>0$, $d>0$.

In the case II $c>0$, $d<0$.

Thus one can get very interesting behaviour from the physical point of view,
because every end of an interval of living (existence) of the solution of an
\ev\ equation means a start of nontrivial
interaction between a matter, a radiation and a \q. In all of these intervals
one can find an approximate inverse function, i.e.\ one can write
$$
R=F(t) \eq498
$$
where
$$
F(t_1)=R_1<R<R_2=F(t_2) \eq499
$$
and
$$
t_1<t<t_2. \eq500
$$

Moreover we do not develop this project here in details. $F$ and $\mPi$ are
usual elliptic integrals of the first and of the third kind given by the
formulae \rf3.71 \ and \rf3.73 .

For future convenience of this project we find roots of the polynomial
$$
P(x)=x^4+x+a, \quad a>0. \eq501
$$
First of all according to the Ferrari method we should solve a resolvent
equation which is a cubic equation. One gets
$$
z^3-4az-1=0. \eq502
$$
The discriminant of Eq.\ \rf3.502 \ reads
$$
\o D=\frac{27-256a^3}{108}
$$
and we have two cases
$$
\ealn{
&\o D>0,\quad 0<a<\frac3{4\root3\of4}, &\rf3.503 \cr
&\o D<0, \quad a>\frac3{4\root3\of4}, &\rf3.504
}
$$

In the case \rf3.503 \ one gets
$$
\ealn{
z_1&=\root3\of{\frac{\sqrt{27-256a^3}+3\sqrt3}{6\sqrt3}}
-\root3\of{\frac{\sqrt{27-256a^3}-3\sqrt3}{6\sqrt3}} &\rf3.505 \cr
z_{2,3}&=\frac12 \(\root3\of{\frac{\sqrt{27-256a^3}-3\sqrt3}{6\sqrt3}}
-\root3\of{\frac{\sqrt{27-256a^3}+3\sqrt3}{6\sqrt3}}\)\cr
&\pm\frac{i\sqrt3}2 \(\root3\of{\frac{\sqrt{27-256a^3}-3\sqrt3}{6\sqrt3}}
+\root3\of{\frac{\sqrt{27-256a^3}+3\sqrt3}{6\sqrt3}}\) &\rf3.506
}
$$

In the second case \rf3.504 
$$
\ealn{
z_1&=4\sqrt{\frac a3}\cos\(\frac{\o\f}3\) &\rf3.507 \cr
z_2&=4\sqrt{\frac a3}\cos\(\frac{\o\f}3+\frac{2\pi}3\) &\rf3.508 \cr
z_3&=4\sqrt{\frac a3}\cos\(\frac{\o\f}3+\frac{4\pi}3\) &\rf3.509 
}
$$
where $\cos\o\f=\sqrt{\frac3{16a}}$.

We can also distinguish the special case $\o D=0$ ($a=\frac3{4\root3\of4}$):
$$
\ealn{
z_1&=\root3\of4 & \rf3.510 \cr
z_{2,3}&=-\tfrac12\root3\of4 & \rf3.511
}
$$

In the first case we get one real and two complex conjugated roots, in the
second three real roots, in the third case two different real roots (one of
them is double).

According to the Ferrari method one gets
$$
\ealn{
2x_1&=\sqrt{z_1}+\sqrt{z_2}+\sqrt{z_3} &\rf3.512a \cr
2x_2&=\sqrt{z_1}-\sqrt{z_2}-\sqrt{z_3} &\rf3.512b \cr
2x_3&=-\sqrt{z_1}+\sqrt{z_2}-\sqrt{z_3} &\rf3.512c \cr
2x_4&=-\sqrt{z_1}+\sqrt{z_2}+\sqrt{z_3} &\rf3.512d 
}
$$
$$
z_1z_2z_3=1. \eq513
$$

In the future analysis of an \ev\ of our model of the \U\ there are three
important cases for $z_1$, $z_2$ and~$z_3$:

\item{A.} All $z_1,z_2,z_3$ are real and positive. In this case all $x_1,x_2,
x_3$ and~$x_4$ are real.

\item{B.} One root (e.g.\ $z_1$) is positive, two remaining ($z_2,z_3$) are
negative. In this case we have two pairs of conjugate roots.

\item{C.} One root (e.g.\ $z_1$) is positive, two remaining ($z_2,z_3$) are
complex conjugate. In this case we have two real roots and one pair of
complex conjugate roots.

This analysis is quite important because only for positive real roots the
integral~$I$ can have singular points. It means in A and C case. This gives
us some restrictions on the parameter~$a$.

Let us notice that in the case of $a=0$ (considered before) we have to do
with the case~C. Fortunately one real root is zero and the second real root
is negative ($x_2=-1$).

Let us notice that our case with $a=0$ is in some sense exceptional from the
point of view of a general theory. In that case some of the formulae are
singular and because of this it was considered separately.

Let us notice that for sufficiently big $a$ the polynomial \rf501 \ has no real
roots. This is true for $a>\frac3{4\root3\of4}$. In this case the integral
\rf458 \ has no singular points. This is case~II.

\section{Conclusions}

In the paper we consider some \co\ consequences of the Nonsymmetric
Kaluza--Klein (Jordan--Thiry) Theory. Especially we use the scalar
field~$\P$ appearing there in order to get \co\ models with a \q\ and phase
transitions. We consider a dynamics of Higgs' fields with various
approximations and models of \il\ [14].

Eventually we develop a toy model of this dynamics to obtain an amount of
\il\ and $P_R(K)$ function (a~spectral function for fluctuations). We
calculate a spectral index $n_s(K)$ and $\dfrac{dn_s}{d\ln K}$ for this
model. We calculate a Hubble parameter and an age of the \U.

\def\ii#1 {\item{[#1]}}
\section{References}
\setbox0=\hbox{[11]\enspace}
\parindent\wd0

\ii1 {Kalinowski} M. W., 
{\it Nonsymmetric Fields Theory and its Applications\/},
World Scientific, Singapore, New Jersey, London, Hong Kong 1990.

\item{} {Kalinowski} M. W., 
{\it Nonsymmetric Kaluza--Klein (Jordan--Thiry) Theory in a general
nonabelian case}\/,
Int. Journal of Theor. Phys. {\bf30}, p.~281 (1991).

\item{} {Kalinowski} M. W., 
{\it Nonsymmetric Kaluza--Klein (Jordan--Thiry) Theory in the electromagnetic
case}\/,
Int. Journal of Theor. Phys. {\bf31}. p.~611 (1992).

\item{} {Kalinowski} M. W., 
{\it Can we get confinement from extra dimensions}\/,
in: Physics of Elementary Interactions (ed. Z.~Ajduk, S.~Pokorski,
A.~K.~Wr\'oblewski), World Scientific, Singapore, New
Jersey, London, Hong Kong 1991.

\ii2 {Bahcall} N. A., {Ostriker} J. P., {Perlmutter} S., 
{Steinhardt} P. J.,
{\it The cosmic triangle: revealing the state of the Universe}\/,
Science {\bf284}, p.~1481 (1999).

\item{} {Wang} L., {Caldwell} R. R., {Ostriker} J. P., 
{Steinhardt} P. J.,
{\it Cosmic concordance and quintessence}\/, 
astro-ph/9901388v2.

\item{} {Wiltshire} D. L., 
{\it Supernovae Ia, evolution and quintessence}\/, 
astro-ph/0010443.

\item{} {Primack} J. R.,
{\it Cosmological parameters},
Nucl. Phys.~B (Proc. Suppl.) {\bf87}, p.~3 (2000).

\item{} {Bennett} C. L. et al., 
{\it First Year Wilkinson Microwave Anisotropy Probe (WMAP) Observations:
Determination of Cosmological Parameters}\/,
astro-ph/0302209v2.

\ii3 {Caldwell} R. R., {Dave} R., {Steinhardt} P. J.,
{\it Cosmological imprint of an energy component with general equation
of state}\/,
Phys. Rev. Lett. {\bf80}, p.~1582 (1998).

\item{} {Armeendariz-Picon} C., {Mukhanov} V., 
{Steinhardt} P. J.,
{\it Dynamical solution to the problem of a small cosmological constant and
late-time cosmic acceleration}\/,
Phys. Rev. Lett. {\bf85}, p.~4438 (2000).

\item{} {Maor} J., {Brustein} R., {Steinhardt} P. J.,
{\it Limitations in using luminosity distance to determine the equation of
state of the Universe}\/,
Phys. Rev. Lett. {\bf86}, p.~6 (2001).

\ii4 {Axenides} M., {Floratos} E. G., {Perivolaropoulos} L.,
{\it Some dynamical effects of the cosmological constant}\/,
Modern Physics Letters {\bf A15}, p.~1541 (2000).

\ii5 {Vishwakarma} R. G.,
{\it Consequences on variable $\Lambda$-models from distant type Ia
supernovae and compact radio sources}\/,
Class Quantum Grav. {\bf18}, p.~1159 (2001).

\ii6 {Di{\'a}z-Rivera} L. M., {Pimentel} L. O.,
{\it Cosmological models with dynamical $\Lambda$ in scalar-tensor 
theories}\/,
Phys. Rev. {\bf D66}, p.~123501-1 (1999).

\item{} {Gonz{\'a}lez-Di{\'a}z} P. F.,
{\it Cosmological models from quintessence}\/,
Phys. Rev. {\bf D62}, p.~023513-1 (2000).

\item{} {Frampton} P. H.,
{\it Quintessence model and cosmic microwave background}\/,
astro-ph/0008412.

\item{} {Dimopoulos} K.,
{\it Towards a model of quintessential inflation}\/,
Nucl. Phys.~B (Proc. Suppl.) {\bf95}, p.~70 (2001).

\item{} {Charters} T. C., {Mimoso} J. P., {Nunes} A.,
{\it Slow-roll inflation without fine-tun\-ning}\/,
Phys. Lett. {\bf B472}, p.~21 (2000).

\item{} {Brax} Ph., {Martin} J.,
{\it Quintessence and supergravity}\/,
Phys. Lett. {\bf B468}, p.~40 (1999).

\item{} {Sahni} V.,
{\it The cosmological constant problem and quintessence}\/,
Class. Quantum Grav. {\bf19}, p.~3435 (2002).

\ii7 {Liddle} A. R.,
{\it The early Universe}\/,
astro-ph/9612093.

\item{} {Gong} J. O., {Stewart} E. D.,
{\it The power spectrum for multicomponent inflation to second-order
corrections in the slow-roll expansion}\/,
Phys. Lett. {\bf B538}, p.~213 (2002).

\item{} {Sasaki} M., {Stewart} E. D.,
{\it A general analytic formula for the spectral index of the density
perturbations produced during inflation}\/,
Progress of Theoretical Physics {\bf95}, p.~71 (1996).

\item{} {Nakamura} T. T., {Stewart} E. D.,
{\it The spectrum of cosmological perturbations produced by a
multicomponent inflation to second order in the slow-roll approximation}\/,
Phys. Lett. {\bf B381}, p.~413 (1996).

\item{} {Turok} N.,
{\it A critical review of inflation}\/,
Class. Quantum Grav. {\bf19}, p.~3449 (2002).

\ii8 {Liddle} A. R., {Lyth} D. H., 
{\it Cosmological Inflation and Large-Scale Structure}\/,
Cambridge Univ. Press, Cambridge 2000.

\ii9 {Steinhardt} P. J., {Wang} L., {Zlatev} I.,
{\it Cosmological tracking solutions}\/,
Phys. Rev. {\bf D59}, p.~123504-1 (1999).

\item{} {Steinhardt} P. J.,
{\it Quintessential Cosmology and Cosmic Acceleration}\/,
Web page\hfil\break {\tt http://feynman.princeton.edu/\~{}steinh}

\ii10 {Painlev{\'e}} P.,
{\it Sur les \'equations diff\'erentielles du second ordre et d'ordre
sup\'erieur dont l'int\'egrale g\'en\'erale est uniforme\/},
Acta Mathematica {\bf25}, p.~1 (1902).

\ii11 {Steigman} G., 
{\it Primordial alchemy: from the big-bang to the present Universe},
astro-ph/0208186.

\item{} {Steigman} G.,
{\it The baryon density though the (cosmological) ages}\/.

\item{} {Scott} J., {Bechtold} J., {Dobrzycki} A., 
{Kul-Karni} V.~P.,
{\it A uniform analysis of the LY-$\alpha$ forest of $Z=0-5$}\/,
astro-ph/0004155.

\item{} {Hui} L., {Haiman} Z., {Zaldarriaga} M., 
{Alexander} T.,
{\it The cosmic baryon fraction and the extragalactic ionizing background}\/,
astro-ph/0104442.

\item{} {Balbi} A., {Ade} P., {Bock} J., {Borrill} J., 
{Boscaleri}~A., {De\ Bernardis}~P.,
{\it Constraints on cosmological parameters from Maxima-$1$}\/,
astro-ph/0005124.

\item{} {Jeff} A. H. et al.,
{\it Cosmology from Maxima-$1$, BOOMERANG \& COBE/DMR CMB observations}\/,
astro-ph/0007333.

\ii12 {Serna} A., {Alimi} J. M.,
{\it Scalar-tensor cosmological models}\/,
astro-ph/9510139.

\item{} {Serna} A., {Alimi} J. M.,
{\it Constraints on the scalar-tensor theories of gravitation from
primordial nucleosynthesis}\/,
astro-ph/9510140.

\item{} {Navarro} A., {Serna} A., {Alimi} J. M.,
{\it Search for scalar-tensor gravity theories with a non-monotonic time
evolution of the speed-up factor}\/,
Class. Quantum Grav. {\bf19}, p.~4361 (2002).

\ii13 {Serna} A., {Alimi} J. M., {Navarro} A., 
{\it Convergence of scalar-tensor theories toward General Relativity and
primordial nucleosynthesis}\/,
gr-qc/0201049.

\ii14 {Kalinowski} M. W.,
{\it Dynamics of Higgs' field and a quintessence 
in the nonsymmetric Kaluza--Klein (Jordan--Thiry) Theory}\/,
hep-th/0306241.
\end